 \documentclass[final,3p,times]{elsarticle}

\usepackage{amssymb}

\usepackage{amsmath}

\usepackage{graphicx}
\usepackage{dcolumn}
\usepackage{bm}
\usepackage[usenames, dvipsnames]{xcolor} 
\usepackage{algorithm} 
\usepackage{tikz} 
\usepackage{setspace} 

\usetikzlibrary{math} 
\usetikzlibrary{decorations.markings} 

\usepackage{hyperref}

\usepackage{diagbox}   
\usepackage{hyperref}  
\usepackage{pifont}    
\usepackage{threeparttable} 

\usepackage{caption}    
\usepackage{subcaption} 

\usepackage{fontawesome} 

\usetikzlibrary{patterns}
\usetikzlibrary{decorations}
\usetikzlibrary{decorations.pathmorphing}
\usetikzlibrary{decorations.pathreplacing}
\usetikzlibrary{decorations.shapes}
\usetikzlibrary{decorations.text}
\usetikzlibrary{decorations.markings}
\usetikzlibrary{decorations.fractals}
\usetikzlibrary{decorations.footprints}
\usetikzlibrary{arrows,calc}
\usetikzlibrary{shapes}

\usetikzlibrary{matrix}

\usetikzlibrary{fit}

\begin{document}


\newcommand{\acP}{\begin{tikzpicture}
\draw[rotate=-90,scale=0.75,line width=0.3mm] (0,0) -- (0.23,0) -- (1/2*0.23, {sqrt(3)/2*0.23}) -- cycle[draw=blue,fill=white]{};
\end{tikzpicture}}

\newcommand{\acM}{\begin{tikzpicture}
\draw[rotate=90,scale=0.75,line width=0.3mm] (0,0) -- (0.23,0) -- (1/2*0.23, {sqrt(3)/2*0.23}) -- cycle[draw=black!40!green,fill=white]{};
\end{tikzpicture}}

\newcommand{\shear}{\begin{tikzpicture} 
\draw [color=white,line width=0.3mm] (-0.25,0) -- (0.25,0);
\draw [color=red,line width=0.3mm] (-0.1,-0.1) -- (0.1,-0.1) ;  
\draw [color=red,line width=0.3mm] (-0.1,-0.1) -- (-0.1,0.1) ;  
\draw [color=red,line width=0.3mm] (-0.1,0.1) -- (0.1,0.1) ; ;  
\draw [color=red,line width=0.3mm] (0.1,0.1) -- (0.1,-0.1) ; ;  
\end{tikzpicture}}

\newcommand{\entrop}{\begin{tikzpicture} 
\draw [color=white,line width=0.3mm] (-0.25,0) -- (0.25,0); 
\draw [color=black,line width=0.3mm] (0,0) circle (0.1);  
\end{tikzpicture}}

\newcommand{\macroSpu}{\begin{tikzpicture} 
\draw [color=black,line width=0.3mm, fill=black] (0,0) circle (0.06);  
\end{tikzpicture}}

\newcommand{\ghost}{\begin{tikzpicture} 
\node[draw, diamond, line width=0.3mm, aspect=0.6, scale=0.35, color=black!50!white] at (0,0) {};
\end{tikzpicture}}

\begin{frontmatter}



\author[label1]{Florian Renard\corref{cor1}}
\ead{renard@cerfacs.fr}
 
\author[label1]{Gauthier Wissocq}
\ead{gauthier.wissocq@cerfacs.fr}

\author[label1]{Jean-Fran\c cois Boussuge}
\ead{boussuge@cerfacs.fr}

\author[label2]{Pierre Sagaut}
\ead{pierre.sagaut@univ-amu.fr}

\address[label1]{CERFACS, 42 Avenue G. Coriolis, 31057 Toulouse cedex, France}
\address[label2]{Aix Marseille Univ, CNRS, Centrale Marseille, M2P2, 13451 Marseille, France.}

\title{A linear stability analysis of compressible hybrid lattice Boltzmann methods}


\author{}

\address{}

\begin{abstract}
An original spectral study of the compressible hybrid lattice Boltzmann method (HLBM) on standard lattice is proposed. 
In this framework, the mass and momentum equations are addressed using the lattice Boltzmann method (LBM), while finite difference (FD) schemes solve an energy equation. Both systems are coupled with each other thanks to an ideal gas equation of state.
This work aims at answering some questions regarding the numerical stability of such models, which strongly depends on the choice of numerical parameters. 
To this extent, several one- and two-dimensional HLBM classes based on different energy variables, formulation (primitive or conservative), collision terms and numerical schemes are scrutinized.
Once appropriate corrective terms introduced, it is shown that all continuous HLBM classes recover the Navier-Stokes Fourier behavior in the linear approximation. However, striking differences arise between HLBM classes when their discrete counterparts are analysed. 
Multiple instability mechanisms arising at relatively high Mach number are pointed out and two exhaustive stabilization strategies are introduced: (1) decreasing the time step by changing the reference temperature $T_{ref}$ and (2) introducing a controllable numerical dissipation $\sigma$ \textit{via} the collision operator.
A complete parametric study reveals that only HLBM classes based on the primitive and conservative entropy equations are found usable for compressible applications. 
Finally, an innovative study of the macroscopic modal composition of the entropy classes is conducted. Through this study, two original phenomena, referred to as \textit{shear-to-entropy} and \textit{entropy-to-shear} transfers, are highlighted and confirmed on standard two-dimensional test cases.
\end{abstract}


\end{frontmatter}



\section{Introduction}
\label{sec:Intro}

Based on a simple and efficient collide and stream algorithm, the lattice Boltzmann method (LBM) has proven in the last few years to be a valuable alternative to the standard computational fluid dynamics (CFD) solvers. Projecting the Boltzmann equation onto an appropriate discrete-velocity lattice, this Cartesian method allows simulating a large panel of fluid phenomena, ranging from magneto-hydrodynamics~\cite{dellar2002lattice,ahrar2017novel}, meteorological flows~\cite{feng2019hybrid}, multiphase flows~\cite{shan1993lattice,chen1998lattice}, turbulent flows~\cite{yu2005lattice, sagaut2010toward}, porous media~\cite{succi1989three} or even hemodynamics and biomedical applications~\cite{ye2016particle,li2018application}. The success of this method partly lies in its ease of implementation~\cite{latt2020palabos}, its advantages for massively parallel computing~\cite{schornbaum2016massively} as well as its suitability to handle complex geometries~\cite{succi1989three, martys1996simulation}, thus making it an alternative to standard CFD solvers notably in the aeronautical field~\cite{shmilovich2019practical,barad2017lattice,rodarte2019analysis}. In this particular domain, the LBM is especially promising for computational aeroacoustics~\cite{sengissen2015simulations,hou2019lattice} owing to its low-dissipation properties~\cite{marie2009comparison}.
Nevertheless, the standard LBM schemes, relying on lattices using only neighboring nodes, are restricted to low Mach number and isothermal flows, greatly reducing their applications in the aeronautical field. 

To circumvent this issue and make the simulation of compressible flows possible, three methods emerged: (1) the multispeed (MS) method~\cite{alexander1993lattice}, where higher-order lattices are considered to recover a thermal equation; (2) the double distribution function (DDF) method~\cite{he1998novel}, where an additional thermal population is introduced to compute the energy; (3) the hybrid lattice Boltzmann method (HLBM)~\cite{lallemand2003hybrid}, where an extra energy equation, discretized using finite differences (FD) schemes, is coupled to the LBM. Despite recent promising advances of DDF and MS methods~\cite{saadat2019lattice, latt2019efficient}, they remain limited to academic applications, while the HLBM has proven its worth in the simulation of realistic compressible case. Based on a 39-velocity model coupled with a primitive entropy equation, Nie \textit{et al.}~\cite{Nie2009} were among the firsts to perform compressible bi-dimensional computations, later applied to three-dimensional turbulent cases by Fares \textit{et al.}~\cite{fares2014validation}. The relevance of this method has been successfully evaluated in the aeronautical context for turbofan broadband noise prediction~\cite{casalino2018turbofan}, aeroacoustics of subsonic and supersonic cavities~\cite{mancini2019very,singh2017lattice} or high subsonic single flow hot jet computation~\cite{nickerson2019simulations}. 
Most of this work is based on the industrial $\mathrm{PowerFLOW}^{\textregistered}$ solver, which has been the only one, so far, showing mature results for compressible cases. Indeed, other HLBM models introduced in the literature are mostly restricted to academic configurations within the framework of the Boussinesq approximation~\cite{feng2019hybrid,ahrar2017novel,bettaibi2014hybrid}. Under this assumption, the thermal coupling methodology between the momentum and the energy equation boils down to a single buoyancy force term, which seems to help recovering similar stability properties as in the athermal case. However, for a more realistic equation of state such as the ideal gas one, severe instabilities~\cite{lallemand2003hybrid, lallemand2003theory} due to modal interactions can occur, making the coupling between the LBM and an extra energy equation much more delicate. The fact that the adopted equation of state can be responsible for a lack of robustness can be supplied by observations of DDF approaches, where most of the Boussinesq-based models show correct stability properties~\cite{guo2002coupled, guo2007thermal, karlin2013consistent}, while an ideal gas one was more unstable~\cite{li2012coupling}. Finding a stable compressible LBM scheme for ideal gases is an important matter of academic research. 

In this context, Feng \textit{et al.}~\cite{feng2018regularized} recently developed a stable ideal gas HLBM on a standard D2Q9 lattice. Using a second-order regularized collision operator\cite{Latt2006}, along with a body-force term addressing the standard lattice Galilean invariance~\cite{dellar2014lattice}, they were able to simulate natural convection phenomena with large temperature differences. This model was later applied to the simulation of low Mach reactive flows~\cite{Feng2018}. However, although ruled by the perfect gas law, this first model did not account for the pressure work term in the (internal) energy equation, boiling down to a basic advection-diffusion equation on temperature. Thus, even though it can model the temperature gas expansion phenomenon which makes it more advanced than a Boussinesq approach, it turns out to be  limited to simulations of nearly incompressible flows.
In order to extend their model to compressible flows, Feng \textit{et al.}~\cite{Feng2019} later proposed an new version based on the hybrid recursive regularized (HRR) collision operator~\cite{Jacob2019}. The consequent gain in stability of this operator allowed the authors to hybridize the LBM system with an entropy equation, thus recovering the full set of compressible Navier-Stokes equation. This model was successfully assessed on compressible high-subsonic test cases and later extended to supersonic regimes~\cite{renard2020improved} and to three-dimensional cases~\cite{guo2020efficient}.
Up to the authors' knowledge, this model is the only stable HLBM one based on standard lattices, among the present literature, that is able to recover the full set of compressible Navier-Stokes equations. However, it relies on a large number of parameters such as a reference temperature, a numerical criterion in the HRR model or the discretization scheme adopted in the computation of spatial gradients. Up to now, an optimal choice of parameters has only been possible by empirical conclusions drawn on several test cases, and no clear reason that makes a model more robust that another has been proposed yet. As an example, no convincing explanation has been provided so far regarding the variable adopted in the energy equation (temperature, entropy, internal or total energy,...), as well as its form (primitive or conservative). There is some evidence that the effect of any of these parameters is driven by the numerics, which makes the understanding of these phenomena very difficult. 

Precisely, the linear stability analysis (LSA), initially proposed by von Neumann~\cite{vonneumann1950method}, can be systematically employed to exhibit the numerical properties of a given scheme. It relies on the resolution of an eigenvalue problem to obtain the dispersion and dissipation rates of plane monochromatic waves in the linear approximation. This methodology was first applied to the LBM by Sterling and Chen~\cite{sterling1996stability}.
Their work paved the way to numerous other studies aiming to shed light on the impact of collision operators~\cite{lallemand2000theory}, to exhibit the low-dissipative properties of the LBM for the aeroacoustics~\cite{marie2009comparison}, to optimize a choice of numerical parameters~\cite{lallemand2000theory,ginzburg2010optimal,xu2011optimal,chavez2018improving,hosseini2019stability}, to exhibit the spurious noise generated at grid refinement interfaces~\cite{Astoul2019}, or to investigate the effect of the reference temperature for isothermal flows~\cite{hosseini2019compressibility}.  
All these publications illustrate the interests of performing a spectral study of the LBM, which strips off the scheme and exposes it numerical properties. Nevertheless, up the the authors' knowledge, only Siebert \textit{et. al}~\cite{siebert2008lattice} applied this methodology to exhibit the stability of thermal LBM, and for MS models only.

The aim of the present article is to perform such analyses of several HLBM schemes including the aforementioned models of Feng \textit{et al.}~\cite{feng2019hybrid}. For this purpose, it is proposed to extend the standard methodology to a coupled system including an energy equation and an ideal gas equation of state. Furthermore, the improved von Neumann analysis of Wissocq \textit{et al.}~\cite{wissocq2019extended} will be employed in order to systematically identify the modes of a given scheme by the knowledge of the macroscopic information they carry. A similar analysis will be applied to several types of energy equations, depending on the considered  variable or on its form itself (primitive or conservative). The objective of such studies is manifold: (1) objectively conclude on the ability of each model to handle compressible flows, (2) find optimal values for the parameters of the HRR collision model, and (3) exhibit the numerical dispersion and dissipation. Furthermore, unintended numerical phenomena induced by a mode coupling will be evidenced thanks to the extended analysis. Most of all, this work aims at retaining one or more model of interest beyond several HLBM classes, based on objective observations from the LSA.

In order to sort all the different models under consideration, the present paper is articulated as follows. 
Sec.~II recalls the governing discrete velocity Boltzmann equation (DVBE), the energy equations and their time and space discrete counterparts. 
Sec.~III is dedicated to linear stability analyses of the DVBE and LBM in the athermal framework, including a corrective term in Mach number. Some notions on the LBM stability are recalled, and discussions regarding the corrective term and stabilization methods are provided. 
Sec.~IV extends the LSA to the hybrid DVBE (HDVBE) and the discrete HLBM. First elements of answer are brought on the choice of the energy variable, and the effects of the numerical parameters of the HRR collision model on the stability are discussed. A parametric study is then performed over the different HLBM classes in Sec. V, which ultimately reduces the choice to only three possible HLBM classes.
Sec.~VI investigates the spectral behavior of the usable classes in terms of modal macroscopic composition. This further put the light on spurious mode couplings and two newly identified phenomena, referred to as \textit{shear-to-entropy} and \textit{entropy-to-shear} production. 
Finally, conclusions are drawn in the last section.


\section{Hybrid Lattice Boltzmann Method}
\label{sec: Governing equations}

The purpose of the so-called HLBM is to solve the mass and momentum equations using a lattice Boltzmann scheme, while an extra energy equation is considered, permitting the account of temperature fluctuations. The continuous space and time equation from which the LBM derives is referred to as the Discrete Velocity Boltzmann Equation (DVBE). This equation is discrete in terms of a finite set of velocities forming the so-called velocity lattice. 
In a first step, the DVBE is introduced on standard lattices with corrective terms so as to get rid of lattice errors~\cite{Karlin2010,Feng2019} and the monatomic limitation~\cite{renard2020improved}. Then, the spatio-temporal discretization of the DVBE is performed, giving rise to the LBM. Different regularized collision operators are introduced, including the hybrid recursive regularization~\cite{Jacob2019} adopted in previous compressible HLBM models~\cite{Feng2019}. In a third step, the different variables and forms of energy equations considered in this work are detailed, followed by their temporal and spatial discretization schemes.

\subsection{Corrected Discrete Velocity Boltzmann Equation}
\label{subsec: DVBE}

In the HLBM framework, the mass and momentum are computed by the moments of a set of DVBE $\left( \mathcal{B}_i \right)_{i \in \llbracket 0, m-1 \rrbracket}$, where $m$ is the number of discrete velocities of the lattice, determining the temporal evolution of the discrete particle distribution function $f_i$. This set of equations reads:
\begin{equation}
\mathcal{B}_i : \frac{\partial f_i}{\partial t} = \underbrace{- c_{i,\alpha}\frac{\partial f_i}{\partial x_\alpha} -\frac{1}{\tau}\left( f_i - f_i^{eq} \right) + \psi_i }_{\mathcal{S}_i^{DVBE}}, \quad \forall i \in \llbracket 0,m-1 \rrbracket ,
\label{eq: DVBE eq}
\end{equation}
where the Bhatnagar-Gross-Krook (BGK) collision operator~\cite{bhatnagar1954model} has been used and where the Greek subscripts $\alpha$ denotes the spatial directions in Cartesian coordinates,. In the present context, the discrete velocities $c_{i,\alpha}$ are built from a Gauss-Hermite quadrature~\cite{shan2006kinetic}. These velocities span the so-called lattice. The one-dimensional D1Q3 and two-dimensional D2Q9 lattices, with respectively three and nine discrete velocities, are considered in this work and can be found in \ref{app: D1Q3 and D2Q9 lattice}.

 $\mathcal{S}_i^{DVBE}$ corresponds to the spatial dependency of the equation and gathers a linear advection term, a non-linear collision and body-force term $\psi_i$. In the collision term, $\tau$ corresponds to the characteristic time for the relaxation of the distribution function $f_i$ toward $f_i^{eq}$, referring to the local thermodynamic equilibrium distribution function. In the present work, this equilibrium is developed in terms of Hermite polynomials~\cite{Shan1998}. This function reads, up to the $N^{th}$ order, as:
\begin{equation}
f_i^{eq,N} = w_i \sum_{n=0}^{N} \frac{1}{n!c_s^{2n}}\boldsymbol{a}^{eq,\left(n\right)} : \boldsymbol{\mathcal{H}}_i^{\left(n\right)},
\label{eq:Equilibrium}
\end{equation}
where $:$ corresponds to the Frobenius inner product and $w_i$ are the so-called weights of the lattice (recalled in \ref{app: D1Q3 and D2Q9 lattice}). The expansion order $N$ corresponds to the higher-order equilibrium moment that can be imposed with such an expansion. In this study, it will be set as $N=2$ for the D1Q3 lattice and $N=3$ for the D2Q9 lattice. Tensors $\boldsymbol{a}^{eq,\left(n\right)}$ and $\boldsymbol{\mathcal{H}}_i^{\left(n\right)}$ are respectively the $n^{th}$-order equilibrium moment and Hermite polynomial. They read, up to the third order, as:

\begin{center}
\begin{tabular}{p{8cm}p{8cm}}
{\begin{subequations}\begin{align}
&\mathcal{H}^{(0)}_{i} = 1, \\  
&\mathcal{H}^{(1)}_{i,\alpha} = c_{i,\alpha}, \\ 
&\mathcal{H}^{(2)}_{i,\alpha\beta} = c_{i,\alpha} c_{i,\beta} - c_s^2\delta_{\alpha\beta}, \\ 
&\mathcal{H}^{(3)}_{i,\alpha\beta\gamma} = c_{i,\alpha}c_{i,\beta}c_{i,\gamma} \nonumber \\
&\hspace{1.2cm} - c_s^2 \left( c_{i,\alpha}\delta_{\beta\gamma} + c_{i,\beta}\delta_{\alpha\gamma} + c_{i,\gamma}\delta_{\alpha\beta} \right),
\end{align}\label{eq: Hermite pol}\end{subequations}}
&%
{\begin{subequations} \begin{align}
&a^{eq,(0)}=\rho,  \\    
&a^{eq,(1)}_{\alpha}=\rho u_\alpha,  \\
&a^{eq,(2)}_{\alpha\beta}=\rho u_\alpha u_\beta + \rho c_s^2 \left(\theta -1\right) \delta_{\alpha\beta}, \\
&a^{eq,(3)}_{\alpha\beta\gamma}=\rho u_\alpha u_\beta u_\gamma \nonumber \\
& \hspace{0.9cm} +\rho c_s^2 \left( \theta-1 \right) \left( u_\alpha \delta_{\beta\gamma} + u_\beta \delta_{\alpha\gamma} + u_\gamma \delta_{\alpha\beta} \right),
\end{align}\label{eq: moments}\end{subequations}}
\end{tabular}
\end{center}

\noindent where $\delta$ denotes the Kronecker symbol and $\rho$ and $u_\alpha$ are respectively the density and velocity.  In the present case, $\theta = r_g T / \left(r_rT_r\right)$ where $T$ and $T_r$ respectively stand for the fluid temperature and a reference temperature along with $r_g$ and $r_r$ corresponding to the fluid and the reference molecular gas constant. Finally, $c_s=\sqrt{r_rT_r}$ is the characteristic velocity. In the present work, no laws on the thermodynamic coefficients is applied and allows to choose $r_g=r_r=287.15\ \mathrm{K}.\mathrm{m}^2.\mathrm{s}^{-2}$ which simplifies the expression of $\theta = T/T_r$ and gives $c_s=\sqrt{r_gT_r}$. Finally, for the D2Q9 lattice, even though $N=3$ will be adopted, only the the third-order off-diagonal moments $a^{eq,(3)}_{xxy}$ and $a^{eq,(3)}_{yyx}$ are considered in Eq.~(\ref{eq:Equilibrium}), the diagonal ones being not supported by the lattice~\cite{malaspinas2015increasing}.

Computing the zeroth and first moments of the distribution functions yields respectively the mass and momentum:
\begin{equation}
\rho \equiv \sum_{i=0}^{m-1}f_i=\sum_{i=0}^{m-1}f_i^{eq,N\geq 0} \quad \mathrm{and} \quad \rho u_\alpha\equiv \sum_{i=0}^{m-1} c_{i,\alpha} f_i = \sum_{i=0}^{m-1} c_{i,\alpha} f_i^{eq,N\geq 1},
\label{eq: zero and first moment of f}
\end{equation} 
which are two collision invariants by construction of $f_i^{eq,N}$, provided that $N \geq 1$.

Finally, in Eq.~(\ref{eq: DVBE eq}), $\psi_i$  refers to a correction term which reads:
\begin{equation}
\psi_i = w_i\frac{\mathcal{H}_{i,\alpha\beta}}{2c_s^4} \left( E_{1,\alpha\beta} + E_{2,\alpha\beta} \right),
\label{eq: Psialgo}
\end{equation}
removing both the lattice closure error induced by the standard lattice~\cite{Karlin2010} ($E_{1,\alpha\beta}$ correction) and the polyatomic inconsistency of the equilibrium distribution function in the hybrid framework ($E_{2,\alpha\beta}$ correction)~\cite{renard2020improved}. These terms read respectively:
\begin{equation}
\begin{split}
E_{1,\alpha\beta} =&  - \frac{\partial}{\partial x_\alpha}  \left( a^{eq}_{\alpha\alpha\alpha} \right)  \delta_{\alpha\beta} \qquad \mathrm{and} \qquad E_{2,\alpha\beta} =p\left(\frac{D+2}{D}-\gamma_g\right) \frac{\partial u_\gamma}{\partial x_\gamma}\delta_{\alpha\beta},
\label{eq: E1 E2 expressions}
\end{split}
\end{equation}
and can both be used on a D1Q3 lattice with $f^{eq,2}_i$ or on a D2Q9 lattice with $f^{eq,3}_i$. Here, $\gamma_g$ is the heat capacity ratio of the gas, $D$ the number of spatial dimensions and $a^{eq}_{\alpha\alpha\alpha}$ the third-order Hermite equilibrium moment of Eq.~(\ref{eq: moments}). 

After performing a Chapman-Enskog expansion~\cite{chapman1970mathematical} of the corrected DVBE, its macroscopic hydrodynamic limit is found to be:

\begin{subequations}
\begin{align}
\label{eq: Mass from NS}&\frac{\partial \rho}{\partial t} + \frac{\partial (\rho u_\alpha)}{\partial x_\alpha} = 0 \\
\label{eq: QDM from NS}&\frac{\partial (\rho u_\alpha)}{\partial t} + \frac{\partial }{\partial x_\beta} \left(\rho u_\alpha u_\beta + p \delta_{\alpha\beta} \right)  = -\frac{\partial a^{(1)}_{\alpha\beta} }{\partial x_\beta} 
\end{align}
\label{eq: NS system}
\end{subequations}
with
\begin{equation}
p=\rho c_s^2\theta \qquad \mathrm{and} \qquad a^{(1)}_{\alpha\beta}  = - \tau p \left( \frac{\partial u_\alpha}{\partial x_\beta} + \frac{\partial u_\beta}{\partial x_\alpha} - \frac{2}{D} \frac{\partial u_\gamma}{\partial x_\gamma}\delta_{\alpha\beta} \right),  
\end{equation}
being respectively the pressure and the error-free viscous stress tensor \cite{renard2020improved}. This set of macroscopic equation is recovered for both the D1Q3 and D2Q9 lattices.
 It is worth noting that Eq.~(\ref{eq: Mass from NS}) and the convective part of Eq.~(\ref{eq: QDM from NS}) depend only on the zeroth-, first- and second-order equilibrium moments. On the other side, as the first two moments are collision invariant, the system is altered exclusively $via$ the second-order off-equilibrium moment $a^{(1)}_{\alpha\beta}$ gathering the viscous effects. Finally, by identification with the Navier-Stokes equations, a dynamic viscosity can be defined as $\mu = \tau p$. In the present work and without loss of generality, $\mu$ is considered as a constant fluid property, meaning that no temperature evolution of $\mu$, \textit{e.g.} with the Sutherland's law
~\cite{sutherland1893lii}, will be considered.

Regarding the mass and momentum equations, the way in which $\theta$ is introduced leads to two distinct cases: 
\begin{itemize}
	\item Athermal case: $\theta$ is considered as a constant.  The equation of state only depends on the density, and the athermal Newtonian speed of sound $c_s\sqrt{\theta}$ is recovered~\cite{wissocq2019extended}. It can be freely adjusted with the arbitrary choice of $\theta$ but remains constant over time and space. For $\theta=1$ and if the correction term $\psi_i$ is neglected, the very standard athermal DVBE is recovered. 
    \item Thermal case: $\theta$ is related to the fluid temperature, whose evolution in time and space is governed by an $(m+1)$ \textit{energy} equation. The perfect gas equation of state is recovered as well as the isentropic speed of sound $c_s\sqrt{\gamma_g \theta}$. 
\end{itemize}
In the HLBM framework, the athermal case can be adopted as far as the Boussinesq hypothesis applies~\cite{feng2019hybrid}. Thus, an \textit{ersatz} of the energy equation is sufficient, such as a single advection-diffusion of temperature, along with a buoyancy force term in the momentum equation. Using this hypothesis, only incompressible flows with small temperature variations can be considered.
To recover the compressible Navier-Stokes-Fourier equations, it is mandatory to consider the thermal case, where a fully detailed energy equation is solved and coupled to the DVBE through a perfect gas equation of state. It is the context of the present work, and the energy equation is further detailed in the next subsection.

\subsection{Energy equation}
\label{subsec: Energy equation}

The energy equation can be written in a general conservative form as:
\begin{equation}
\frac{\partial (\rho \phi)}{\partial t} + \frac{\partial (\rho u_\alpha \phi)}{\partial x_\alpha} = \mathcal{P}_\phi + \mathcal{F}_\phi + \mathcal{V}_\phi,
\label{eq: total phi eq cons}
\end{equation}
or, after expanding the left-hand side term of Eq.~(\ref{eq: total phi eq cons}) and using the mass equation~(\ref{eq: Mass from NS}), in a non-conservative form as:
\begin{equation}
\frac{\partial  \phi}{\partial t} + u_\alpha\frac{\partial \phi}{\partial x_\alpha} = \frac{1}{\rho}\left( \mathcal{P}_\phi + \mathcal{F}_\phi + \mathcal{V}_\phi \right).
\label{eq: total phi eq nocons}
\end{equation}
The quantity $\phi \in \left\{ e,s,E \right\}$ corresponds to the energy variable, where $e=c_vT$ refers to the internal energy, $E=e+u_\alpha^2/2$ to the total energy and $s=c_v\ln\left(T/\rho^{\gamma_g-1}\right)$ to the entropy, with  $c_v=r_g/\left(\gamma_g - 1\right)$ the heat capacity at constant volume.  $\mathcal{P}_\phi$ is the pressure work term modeling heat production induced by compressibility effects, $\mathcal{F}_\phi$ is the Fourier term referring to the heat losses by diffusion and $\mathcal{V}_\phi$ is the viscous production term modeling the heat source by friction.
\begin{table}[h]
\centering
\begin{tabular}{|c||c|c|c|}
    \hline
    \diagbox{terms}{variables}   &      $E$     &     $e$     &     $s$ \\ \hline \hline 
    $T$  &   $\dfrac{1}{c_v} \left(E - \dfrac{1}{2} u_\alpha^2 \right)$   &  $\dfrac{e}{c_v}$ & $\rho^{\gamma_g-1} \exp \left(\dfrac{s}{c_v}\right)$   \\ \hline
    $\mathcal{P}_\phi$  &   $-\dfrac{\partial (u_\alpha p)}{\partial x_\alpha}$   &  $-p\dfrac{\partial u_\alpha}{\partial x_\alpha}$ & $\varnothing$   \\ \hline
    $\mathcal{F}_\phi$  &   $  \dfrac{\partial }{\partial x_\alpha} \left(\lambda\dfrac{\partial T}{\partial x_\alpha} \right)$   &  $  \dfrac{\partial }{\partial x_\alpha} \left(\lambda\dfrac{\partial T}{\partial x_\alpha} \right)$ & $  \dfrac{1}{T}\dfrac{\partial }{\partial x_\alpha} \left(\lambda\dfrac{\partial T}{\partial x_\alpha} \right)$   \\ \hline
    $\mathcal{V}_\phi$  &   $-\dfrac{\partial}{\partial x_\beta} \left( a^{(1)}_{\alpha\beta} u_\alpha \right)$   &  $-a^{(1)}_{\alpha\beta} \dfrac{\partial u_\alpha }{\partial x_\beta}$ & $-\dfrac{a^{(1)}_{\alpha\beta}}{T} \dfrac{\partial u_\alpha }{\partial x_\beta}$   \\ \hline
\end{tabular}
\caption{Energy variables and source terms associated to each form of the energy equation. $\lambda=\mu/\mathrm{Pr}$ is the heat conductivity of the fluid.}
\label{table: energy terms}
\end{table}
The expression of the temperature, as well as the source terms associated to each energy variable, are detailed in Table~\ref{table: energy terms}. It is worth noting the non-linear nature of the viscous heat production terms $\mathcal{V}_e$ and  $\mathcal{V}_s$ which only involve product of partial derivatives, contrary to $\mathcal{V}_E$. The Fourier term $\mathcal{F}$, for its part, is similar for the three energy variables. Finally, the pressure work term $\mathcal{P}_\phi$ is different for the three equations, and even does not appear at all for the entropy equation.

 An interesting feature of the entropy equation precisely relies on this fact. The pressure work is indeed implicitly present in the definition of the entropy itself so that it is taken into account in the advection part of the entropy equation. This can be shown by using a chain rule over the advection part of $s$ with respect to $\rho$ and $e$ and using the mass equation Eq.~(\ref{eq: Mass from NS}):
\begin{equation}
\frac{\partial s}{\partial t} + u_\alpha\frac{\partial s}{\partial x_\alpha} = \frac{1}{T} \left[ \frac{\partial e}{\partial t} + u_\alpha \frac{\partial e}{\partial x_\alpha} - \frac{\mathcal{P}_e}{\rho}  \right],
\label{eq: TP dans s}
\end{equation}
so that the non-conservative equation on $e$ can be recovered, including the pressure work. Details on this derivation can be found in \ref{app: PW in s eq}. The second advantage of this formulation yields in the fact that, for isentropic phenomena such as vortex convection or acoustics, and in low viscosity limit, the primitive form of the entropy equation is naturally reduced to $s=constant$. In such thermal cases, the value of the (non constant) temperature can be directly computed from the knowledge of a constant entropy, so that the DVBE is not actually coupled with a $m+1$ partial differential equation, but with a closure relation on the entropy.

\subsection{LBM with corrective terms}
\label{subsec: The Lattice Boltzmann Method}

In this subsection, the spatio-temporal scheme applied to the DVBE, commonly known as lattice Boltzmann method, is introduced. The LBM is a specific discrete scheme obtained by integrating the DVBE between $t$ and $t+\Delta t$ ($\Delta t$ being the time step) along the characteristic line $c_{i,\alpha}$, combined with a trapezoidal rule approximating the collision and body-force terms. This integration links $\Delta t$ to the space step $\Delta x$ through the exact streaming constraint $|c_{i,\alpha}| = \Delta x /\Delta t$. 
 It results in a second-order accuracy in space and time. The discrete probability functions are successively collided and streamed (\textit{i.e.} convected from node to node along the $c_{i,\alpha}$ line) on a Cartesian mesh.
It leads to:
\begin{equation}
\,g_i^+ = g_i \left( x+c_{i,\alpha} \Delta t ,t+ \Delta t \right) = \underbrace{ g_i  -\frac{\Delta t}{\check{\tau}}\left( g_i - f_i^{eq} \right) + \frac{2\check{\tau} - \Delta t}{2\check{\tau}}\Delta t \, \psi_i }_{\mathcal{S}_i^{LBM}},\\
\label{eq: LBM BGK} 
\end{equation}
with:
\begin{equation}
g_i = f_i + \frac{\Delta t}{2\tau} \left( f_i - f_i^{eq} \right) - \frac{\Delta t}{2} \psi_i  \quad \mathrm{and} \quad \check{\tau} = \tau + \frac{\Delta t}{2},
\end{equation}
being respectively the distribution function and relaxation time after a wise change of variable~\cite{he1998discrete,he1998novel,dellar2013interpretation}. This manipulation allows the switchover from implicit to explicit formulation without any change of the conserved quantities, which are now computed as:
\begin{equation}
\rho = \sum_{i=0}^{m-1}g_i = \sum_{i=0}^{m-1}f_i^{eq} \qquad \mathrm{and} \qquad  \rho u_\alpha = \sum_{i=0}^{m-1} c_{i,\alpha} g_i = \sum_{i=0}^{m-1} c_{i,\alpha} f_i^{eq}.
\label{eq: rhoalgo and rhoualgo}
\end{equation} 
Similarly to its continuous counterpart in Eq.~(\ref{eq: DVBE eq}), $\mathcal{S}_i^{LBM}$ gathers all the spatial linear and nonlinear terms. In the present work, the space-derivatives of $\psi_i$ are computed using second-order centered (D1CO2) finite differences. Details on this scheme can be found in Sec.~\ref{subsec: Discrete energy equation}. 

The numerical scheme obtained by Eq.~(\ref{eq: LBM BGK}) is referred to as the corrected collide-and-stream BGK-LBM~\cite{Feng2018}. However, further variants of this scheme can be drawn by making use of advanced collision operators. The present work is focused on the study of two classes of collision operators, namely (i) the BGK model (\textit{cf.} Eq.~(\ref{eq: LBM BGK})) and (ii) regularized collision operators. The latter have several variants and three of them will be considered here: the projected regularized (PR)~\cite{Latt2006}, the recursive regularized (RR)~\cite{malaspinas2015increasing,Coreixas2017} and the hybrid recursive regularized (HRR)~\cite{Jacob2019} collision operators. 
These regularized operators aim at increasing, in some respect, the numerical stability of the LB scheme~\cite{wissocq2020linear}. They are based on a common principle: filtering the distribution functions, before the collision step, by reconstructing them as:
\begin{equation}
\begin{split}
g_i &= f^{eq}_i + g^{(1)}_i - \frac{\Delta t}{2}\psi_i \qquad \mathrm{with} \qquad g_i^{(1)} = w_i \sum_{n=2}^{N_r} \frac{1}{n!\left(c_s^2\right)^n}\boldsymbol{\check{a}}^{1,\left(n\right)} : \boldsymbol{\mathcal{H}}_i^{\left(n\right)}. 
\end{split}
\end{equation}
Here, $\boldsymbol{\check{a}}^{1,\left(n\right)}$ is the $n^{th}$-order off-equilibrium Hermite moment and $N_r$ refers to the reconstruction order. In this study, it it set to $N_r=2$ for the D1Q3 lattice and $N_r=3$ for the D2Q9 one, where only third-order off-diagonal moments ($\check{a}^{1, (3)}_{xxy}$ and $\check{a}^{1, (3)}_{yyx}$) are considered. The regularization procedure is equivalent to re-writing Eq.~(\ref{eq: LBM BGK}) as~\cite{Feng2019}:
\begin{equation}
g_i^+ = f_i^{eq} + \left(1 - \frac{\Delta t}{ \check{\tau}} \right) g_i^{(1)} + \frac{\Delta t }{2}\psi_i,
\label{eq: streamalgo}
\end{equation} 
referred to as the corrected collide-and-stream regularized scheme.\newline

Regarding the regularized schemes themselves, their major difference lies in the way off-equilibrium moments $\check{\boldsymbol{a}}^{1,\left(n\right)}$ are reconstructed.
The original regularized collision model (PR)~\cite{Latt2006} was developed in such way that only the second-order moment is computed by projecting the off-equilibrium part onto the second-order Hermite polynomials, in the present context as:
\begin{equation}
\begin{split}
\check{a}^{1,(2)\ \mathrm{PR}}_{\alpha\beta} =& \sum_{i=0}^{m-1} \mathcal{H}_{i,\alpha\beta} \left(g_i - f_i^{eq} + \frac{\Delta t}{2} \psi_i  \right). \\ 
\label{eq: a1PR}        
\end{split}
\end{equation}
This off-equilibrium moment is indeed sufficient to recover the viscous stress tensor as mentioned in Sec.~\ref{subsec: DVBE}. Higher order moments are then considered null with no impact on the physics.
 
It is possible to extend this method, in a straightforward manner, to higher-order regularization. Based on the Chapmann-Enskog expansion, the recursive regularization~\cite{Coreixas2017} aims at reconstructing higher-order moments thanks to a recursive formula. For instance, with the D2Q9 lattice, third-order regularized off-equilibrium moments can be included, forming the so-called RR3 model, as:
\begin{subequations}
\begin{align}
\check{a}^{1, (3)}_{xxy} &\simeq  u_y \check{a}^{1, (2)}_{xx} + 2u_x \check{a}^{1, (2)}_{xy},  \\
\check{a}^{1, (3)}_{yyx} &\simeq   u_x \check{a}^{1, (2)}_{yy} + 2u_y \check{a}^{1, (2)}_{xy},  \\ \nonumber
\end{align}
\label{eq: a1xxy a1yyx} 
\end{subequations}
where one reminds that $\check{a}^{1, (3)}_{xxx}$ and $\check{a}^{1, (3)}_{yyy}$ contributions are null due to the lattice closure error~\cite{Karlin2010}. 
This operator greatly extends the numerical stability of the scheme~\cite{Coreixas2017,wissocq2020linear} compared to its second-order counterpart, suggesting that the third-order contribution acts as a numerical artifact changing the spectral properties of the scheme in terms of dispersion and dissipation.

Finally the HRR collision operator shares common features with the aforementioned RR operator. The difference lies in the way $\check{a}^{1, (2)}_{\alpha\beta}$ is computed. In this instance, it is partially reconstructed using finite differences, discretizing the viscous stress tensor, and using the standard projection procedure. This reads:
\begin{equation}
\check{a}^1_{\alpha\beta} = \sigma \check{a}^{1,\mathrm{PR}}_{\alpha\beta} + \left(1-\sigma\right) \check{a}^{1,\mathrm{FD}}_{\alpha\beta} \quad \mathrm{with} \quad \check{a}^{1,\mathrm{FD}}_{\alpha\beta} = - \frac{\check{\tau}}{\tau}\mu \left(\frac{\partial^{\mathrm{FD}} u_\alpha}{\partial x_\beta} + \frac{\partial^{\mathrm{FD}} u_\beta}{\partial x_\alpha} - \frac{2}{D} \frac{\partial^{\mathrm{FD}} u_\gamma}{\partial x_\gamma} \delta_{\alpha\beta} \right),
\label{eq: a1sigalgo}
\end{equation}
where $\sigma \in \left[0,1\right]$ and
\begin{equation}
\frac{\partial^{\mathrm{FD}} \Gamma}{\partial x_\alpha} = \frac{ \Gamma_{\boldsymbol{x}+\boldsymbol{e}_\alpha\Delta x } - \Gamma_{\boldsymbol{x}-\boldsymbol{e}_\alpha\Delta x }}{2\Delta x},
\end{equation}
being a second order centered scheme applied on $\Gamma$ at $\boldsymbol{x}$ along the unit vector $\boldsymbol{e}_\alpha$. When $\sigma=1$ this operator is reduced to the BGK model for the D1Q3 lattice, and to the RR3 model for the D2Q9 one.

\subsection{Discrete energy equation}
\label{subsec: Discrete energy equation}

The spatio-temporal discretization of the energy equation is introduced in this section. For the sake of simplicity but without loss of generality, only its conservative form is considered. 
The energy equation can be split into two different parts, namely temporal and spatial ones, such that:
\begin{equation}
\frac{\partial (\rho \phi)}{\partial t} = \mathcal{S}^{En} \left( f_i, \rho \phi \right),
\end{equation}
where $\mathcal{S}^{En} \left(f_i, \rho \phi \right)$ denotes the operator gathering all the spatial derivative terms (convection and source terms). Three types of derivatives have to be discretized in this equation: time derivatives, first-order spatial derivatives (convective and pressure work terms) and second-order spatial derivatives (diffusive terms). \newline

Adopting a semi-discretization procedure in which the equations are firstly discretized in time and then in space, the most classic scheme for the temporal evolution is the first-order forward Euler scheme, which reads:
\begin{equation}
\begin{aligned}
\mathrm{K}_1 &= \mathcal{S}^{En}\left(f_i^t, \phi^t \right),\\
\left(\rho \phi \right)^{t+\Delta t} &= \left(\rho \phi\right)^{t} + \Delta t \mathrm{K}_1.
\end{aligned}
\end{equation}
This Euler scheme can also be considered as a first-order Runge-Kutta (RK1) scheme. A fourth-order four-stage Runge-Kutta (RK4)~\cite{bao2003high} is also considered in this work:
\begin{equation}
\begin{aligned}
&\mathrm{K}_1 = \mathcal{S}^{En}\left(f_i^t, \phi^t \right),
\quad 
\mathrm{K}_2 = \mathcal{S}^{En}\left(  \left(\rho \phi\right)^{t} + \dfrac{\Delta t}{2} \mathrm{K}_1 \right) ,\\
&\mathrm{K}_3 = \mathcal{S}^{En}\left(  \left(\rho \phi\right)^{t} + \dfrac{\Delta t}{2} \mathrm{K}_2 \right),
\quad
\mathrm{K}_4 = \mathcal{S}^{En}\left( \left(\rho \phi\right)^{t} + \Delta t \mathrm{K}_3 \right), \\ 
&\left(\rho \phi \right)^{t+\Delta t} = \left( \rho \phi \right)^{t} + \frac{\Delta t}{6} \left( \mathrm{K}_1 + 2\mathrm{K}_2 + 2\mathrm{K}_3 + \mathrm{K}_4 \right)).
\end{aligned}
\end{equation}

Regarding first-order space derivatives for convective terms, two different schemes will be considered: a first-order upwind scheme, referred to as D1UPO1, and a second-order centered scheme, referred to as D1CO2. They respectively read:
\begin{align}
 \mathrm{D1UPO1}&: \quad  \frac{\partial (\rho u_\alpha \phi)}{\partial x_\alpha} = \frac{\left(\rho u_\alpha \phi\right)_{\boldsymbol{x}+d^+\boldsymbol{e}_\alpha\Delta x} - \left(\rho u_\alpha \phi\right)_{\boldsymbol{x}-d^-\boldsymbol{e}_\alpha\Delta x} }{\Delta x} + \mathcal{O}\left( \Delta x \right) \quad \mathrm{with} \quad    
    \begin{cases}
      d^+ = |sign(u_\alpha)-1|/2  \\
      d^-=|sign(u_\alpha)+1|/2  \\
    \end{cases},  \label{eq: D1UPO1 FD} \\
 \mathrm{D1CO2}&: \quad  \frac{\partial (\rho u_\alpha \phi)}{\partial x_\alpha} =  \frac{\left(\rho u_\alpha \phi\right)_{\boldsymbol{x}+\boldsymbol{e}_\alpha\Delta x} - \left(\rho u_\alpha \phi\right)_{\boldsymbol{x}-\boldsymbol{e}_\alpha\Delta x} }{2 \Delta x} + \mathcal{O}\left( \Delta x^2 \right).
\label{eq: D1CO2 FD}
\end{align}
Regarding second-order space derivatives, a standard second-order scheme has been chosen for the discretization of $\mathcal{F}$ and $\mathcal{V}$. Applied on the variable $T$, it reads:
\begin{equation}
\mathrm{D2CO2}: \quad  \frac{\partial^2 T}{\partial x_\alpha^2} =  \frac{ T_{\boldsymbol{x}+\boldsymbol{e}_\alpha\Delta x} - T_{\boldsymbol{x}} + T_{\boldsymbol{x}-\boldsymbol{e}_\alpha\Delta x}}{\Delta x^2} + \mathcal{O}\left( \Delta x^2 \right).
\end{equation}

In summary, two combinations of these discrete schemes are retained for the present study: the so-called RK1UPO1 (fairly close to the scheme employed in the recent literature~\cite{Feng2019,renard2020improved}) and RK4CO2 schemes. Details of the particular discretizations used in each scheme are provided in Table~\ref{table:energy_disc}. Also note that a completely similar approach is adopted for the non-conservative form of the energy equation.

\begin{table}[h]
\centering
\begin{tabular}{|c||c|c|c|c|c|}
    \hline
      &      $\dfrac{\partial (\rho \phi)}{\partial t}$     &     $\dfrac{\partial (\rho u_\alpha \phi)}{\partial x_\alpha}$     &     $\mathcal{P}_\phi$ & $\mathcal{F}_\phi$ & $\mathcal{V}_\phi$ \\ \hline \hline 
    RK1UPO1  &   RK1   &  D1UPO1 & D1CO2 & D2CO2 & D2CO2 \\ \hline
    RK4CO2  &   RK4   &  D1CO2 & D1CO2 & D2CO2 & D2CO2 \\ \hline
\end{tabular}
\caption{Discretization schemes used for the energy equation.}
\label{table:energy_disc}
\end{table}


\section{The linear stability analysis}
\label{sec: The Linear stability analysis}

In this section, the methodology of the linear stability analysis are recalled, based on the former work of von Neumann~\cite{vonneumann1950method} and later adapted to the LBM formalism by Sterling and Chen~\cite{sterling1996stability}. In this section, the analysis is applied to the standard athermal DVBE and LBM systems. When performed to the continuous system, this analysis allows highlighting the role of each corrective term. Applied to the discrete system, it can evidence the numerical effects of the discretization on both stability and accuracy.

\subsection{Method and concept}
\label{subsec: Method and concept}

For the sake of clarity, the DVBE system of Eq.~(\ref{eq: DVBE eq}) is considered here. As proposed in \cite{sterling1996stability}, the distribution functions $f_i$ can be decomposed into two parts: (i) a mean flow $\overline{f_i}=\overline{f}_i^{eq}\left( \overline{\rho}, \overline{u}_\alpha \right)$ and (ii) a small perturbation $f^{'}_i$. The mean flow is considered constant in time and space, and $\overline{\rho}$ and $\overline{u}_\alpha$ are respectively the mean density and velocity.
The functions $f_i\left(x,t \right)$ then read: 
\begin{subequations}
\begin{align}
\label{eq: f lin }  f_i(x,t) &= \overline{f_i} + f^{'}_i(x,t), \\
\label{eq: f prime} f^{'}_i(x,t)&=\widehat{f_i} e^{ \mathrm{i} \left( k_\alpha  x_\alpha - \omega t \right) },
\end{align}
\label{eq: f lin and prime}
\end{subequations}
where the fluctuating distribution functions have been expressed as complex monochromatic plane waves, as proposed by von Neumann~\cite{vonneumann1950method}.
Here, $\mathrm{i}^2=-1$, $\omega\in \mathbb{C}$ is the complex temporal pulsation of the wave, $k_\alpha$ is the $\alpha$ component of the wavenumber vector and $\widehat{f_i}$ is the complex amplitude of the perturbation. Considering $f'_i \ll \overline{f_i}$, neglecting the nonlinear terms and injecting Eq.~(\ref{eq: f lin }) into Eq.~(\ref{eq: DVBE eq}), the following linearized system is obtained:
\begin{equation}
\omega \widehat{f}_i = \widetilde{\mathrm{M}}^{DVBE}_{ij} \widehat{f}_j,
\end{equation}
where $\widetilde{\mathrm{M}}^{DVBE}_{ij}$ is defined as the time-advance matrix in the spectral space. For the derivation of this matrix, the reader may relate to~\ref{app: Matrix athermal}.

Computing the eigenvalues $\left( \omega_l \right)_{l \in \llbracket 0, m-1 \rrbracket}$ and eigenvectors $\left( \mathrm{F}^l_i \right)_{l\in \llbracket 0, m-1 \rrbracket }$ of this matrix, $m$ linear eigenmodes of $\mathcal{B}_i$ are obtained. The diagonalized system reads:
\begin{equation}
\omega_l \mathrm{F}^l_i = \widetilde{\mathrm{M}}^{DVBE}_{ij} \mathrm{F}^l_i.
\end{equation}
The real and imaginary parts of $\omega_l$, respectively $\mathrm{Re} (\omega_l)$ and $\mathrm{Im}(\omega_l)$, provide information on the propagation and the dissipation of the perturbation. If $\mathrm{Im}(\omega_l) > 0$, the perturbation associated to the $l^{th}$ mode is exponentially amplified in the time, and thus the system is unstable. On the other hand, if $\mathrm{Im}(\omega_l) < 0$ the mode is damped with time. The system is then linearly stable if and only if every mode behaves in this way. Finally, the phase velocity $v^\Phi_l$ and the group velocity $v^g_{l}$ of each mode $l$ are directly related to $\omega_{l,r}$, and commonly defined as: 
\begin{equation} \label{eq:phase_velocity}
	 v^\Phi_{l} = \frac{\mathrm{Re}(\omega_l)}{||k||} \qquad \mathrm{and}
	\qquad v^g_{l} = \frac{\mathrm{d}\mathrm{Re}(\omega_l)}{\mathrm{d}||k||}.
\end{equation}

To make sure that the system behaves properly in the linear approximation, the spectral properties can be compared with a reference. In this way, the same linear stability analysis can be applied to the Navier-Stokes equations~\cite{chu1958non,kovasznay1953turbulence} by solving the eigenvalue problem for $\widetilde{M}_{i,j}^{NS,ath}$, referred to as the time-advance matrix in the athermal case. It provides three linear modes in two dimensions: one shear (or vortical) mode $\omega _{shear}$, one upstream acoustic $\omega_{ac-}$ mode and one downstream acoustic $\omega_{ac+}$ mode. Their eigenvalues are:
\begin{equation} \label{eq:eigenvaluesNS}
	\begin{aligned}
		&\omega _{shear} = k_\alpha  \overline{u}_\alpha - \mathrm{i} \overline{\nu} \left\Vert k_\alpha \right\Vert^2 + O\left(k^3\right), \\
		&\omega^{ath}_{ac\pm} = k_\alpha  \overline{u}_\alpha \pm \left\Vert k_\alpha \right\Vert  \sqrt{c_s^2\theta} - \mathrm{i} \overline{\nu} \left\Vert k_\alpha \right\Vert^2 + O\left(k^3\right), \\
	\end{aligned}
\end{equation}
where $\overline{\nu} = \mu / \overline{\rho}$ is the mean kinematic viscosity. In one dimension, only the two acoustic modes are present. For more details on the derivation of these expression, the interested reader may refer to ~\cite{chu1958non,kovasznay1953turbulence} or to~\ref{app: Eigen value NS}.

Moreover, following the methodology proposed by Wissocq ~et~al.~\cite{wissocq2019extended}, the DVBE eigenvectors $\left( \mathrm{F}^l_i \right)_{i\in \llbracket 0, m-1 \rrbracket }$ can be used to provide an in-depth physical interpretation of the modes resulting from the linear analysis. This is made possible by computing their macroscopic counterpart $\boldsymbol{\mathcal{E}}_l^{LB}$ as hydrodynamic moments of the DVBE eigenvectors:
\begin{equation}
\boldsymbol{\mathcal{E}}_l^{DVBE} = \left[ \widehat{\rho}^{DVBE}=\sum_{i}^{m-1} \mathrm{F}^l_i \quad , \quad \left(\widehat{u}_\alpha^{DVBE} = \frac{ \sum_{i}^{m-1} c_{i,\alpha} \mathrm{F}^l_i - \overline{u}_\alpha \sum_{i}^{m-1} \mathrm{F}^l_i  }{\overline{\rho}} \right) \right]^\mathrm{T}.
\label{eq: eig vector LB ath}
\end{equation}
This macroscopic information is then compared with the (athermal) Navier-Stokes eigenvectors:
\begin{equation}
\boldsymbol{V}^{NS,ath}_{mode} = \left[ \widehat{\rho}^{NS,ath} \quad , \quad \left(\widehat{u}_\alpha^{NS,ath} \right) \right]^\mathrm{T}, \quad \mathrm{mode} \in \left[shear, ac-, ac+ \right].
\end{equation}
For details about the derivation of $\boldsymbol{V}^{NS,ath}_{mode}$, the reader can refer to~\ref{app: Eigen value NS}. The idea is then to access to the contribution of $\boldsymbol{\mathcal{E}}_l^{LB}$ in terms of physical macroscopic modes such as:
\begin{equation}
\boldsymbol{\mathcal{E}}_l^{LB} = a_l \boldsymbol{V}^{NS,ath}_{shear} + b_l \boldsymbol{V}^{NS,ath}_{ac-} + c_l \boldsymbol{V}^{NS,ath}_{ac+},
\end{equation}
where coefficients $a_l$, $b_l$ and $c_l$ then correspond to the complex contributions of $\boldsymbol{\mathcal{E}}_l^{LB}$ on each macroscopic modes. For details on derivation of $\boldsymbol{V}^{NS,ath}_{mode}$, the interested reader can refer to~\ref{app: Eigen value NS}.

 Their value can be obtained thanks to the passage matrix $\boldsymbol{P}_{NS,ath} = \left[ \boldsymbol{V}^{NS,ath}_{ac-}, \boldsymbol{V}^{NS,ath}_{ac+}, \boldsymbol{V}^{NS,ath}_{shear} \right]$ as:
\begin{equation}
	\left(a_l, b_l, c_l\right)^{T} = P^{-1}_{NS,ath} \boldsymbol{\mathcal{E}}_l^{LB}.
\end{equation}
These coefficients are finally normalized, so that $|a_l| + |b_l| + |c_l|=1$. Looking at the normalized modulus of each coefficient provides information on the physical content carried by a DVBE eigenmode, expressed as (athermal) NS waves. For instance, a mode $l$ of the DVBE for which $(|a_l|, |b_l|, |c_l|)^T=(1,0,0)^T$ can be identified a as pure shear mode. However, in practice, no eigenmode is likely to carry a pure physical information and an arbitrary threshold $\eta$ has to be specified in order to systematically identify the information carried by a mode. In this work, and unless otherwise specified, a mode is considered as a Navier-Stokes mode if the value of one of the normalized coefficients exceeds $\eta=99\%$.\newline

The analysis, recalled here for the DVBE, can be similarly applied to its discrete counterpart, namely the LBM system~(\ref{eq: LBM BGK})\cite{sterling1996stability}, without any change in the methodology. Knowing that:
\begin{equation}
g'_i \left( x + c_{i,\alpha}\Delta t, t+\Delta t \right) = \widehat{g}_i e^{\mathrm{i} \left( k_\alpha \left(x_\alpha + c_{i,\alpha}\Delta t \right) - \omega \left(t + \Delta t \right) \right) },
\label{eq: monochro}
\end{equation}
an equivalent dispersion relation of the discrete equations is obtained and reads:
\begin{equation}
 e^{-\mathrm{i} \omega \Delta t}  \widehat{g_i}   =  \mathrm{\widetilde{M}}^{LBM}_{ij} \widehat{g}_j.
 \label{eq: disp LBM}
\end{equation}
Details about $\mathrm{\widetilde{M}}^{LBM}_{ij}$ can be found in~\ref{app: Matrix athermal}. Once Eq.~(\ref{eq: disp LBM}) is diagonalized, the very same linear analysis can be conducted considering the logarithm of the eigenvalue to compute $\omega$, giving access to the spectral properties of the discrete system.

\subsection{Comments on the corrected DVBE LSA}
\label{subsec: LSA of the corrected DVBE, general results}

This section focuses on the corrected athermal DVBE. Thus, the fluid temperature is considered constant as $T=T_0$ and only $E_{1,\alpha\beta}$ is present in the correction term $\psi_i$ of Eq.~(\ref{eq: Psialgo}). Note that $\mu$ is considered as a constant property of the fluid, so that the kinetic viscosity $\nu$ depends on the local fluid density and its mean value can be defined as $\overline{\nu}=\mu/\overline{\rho}$. Similarly, $\tau=\mu/(\rho r_g T_0)$ is a local variable, its mean counterpart can be defined as $\overline{\tau}=\mu/(\overline{\rho} r_g T_0)$. In order to fully characterize the system, a dimensional analysis is first proposed. In one dimension, the system is described by seven independent variables ($\overline{\rho}$, $\overline{u}$, $\overline{\nu}$, $c=\sqrt{r_gT_0}$, $c_s=\sqrt{r_gT_r}$, $k$ and $\omega$) involving three fundamentals units (mass, length and time). Thus, the overall eigenvalue problem is bounded by considering four dimensionless parameters, $e.g.$:

\begin{equation}
\omega_\tau	= \mathcal{L} \left( \overline{\mathrm{Ma}}, \mathrm{\overline{Kn}}, \theta  \right),
\end{equation}
where 
\begin{equation} 
 \omega_\tau = \frac{\omega \overline{\nu}}{r_gT_0} = \omega \overline{\tau}, \quad \overline{\mathrm{Ma}} = \frac{\overline{u}}{\sqrt{r_gT_0}}, \quad \overline{\mathrm{{Kn}}} = \frac{\overline{\nu} k}{\sqrt{r_gT_0}} \quad  \mathrm{and} \quad \theta=\frac{T_0}{T_r}.
 \label{eq:LSA_DVBE_dimensionless_variables}
\end{equation} 
$\mathrm{\overline{Ma}}$ and $\overline{\mathrm{Kn}}$ are respectively the Mach number of the mean flow and the Knudsen number of the considered perturbation. For larger dimensions, for instance a three-dimensional problem, one could 
consider an angle of Mach and Knudsen numbers per additional spatial degree of freedom ($\overline{\mathrm{Ma}_\theta}$, $\overline{\mathrm{Ma}_\phi}$, $\overline{\mathrm{Kn}_\theta}$, $\overline{\mathrm{Kn}_\phi}$).

Exploring this parameter space allows investigating the linear system in all the possible configurations. Since the present work is applied to fluid mechanics, the Knudsen number will be restricted to $\overline{\mathrm{Kn}} \leq \mathrm{Kn}_c = 5 \cdot 10^{-3}$, so as to ensure the validity of the continuous medium assumption~\cite{bird1994molecular}. Furthermore, as underlined in previous investigations~\cite{PAM2019}, the value of $\theta$ has almost no influence on $\omega_\tau$ for the considered Knudsen numbers. For this reason, the following analyses are performed with $\theta=1$, without any effect on the observations.

\begin{figure}
     \centering
     \begin{subfigure}[b]{0.33\textwidth}
         \centering\hspace{0mm}%
         \includegraphics[scale=1.]{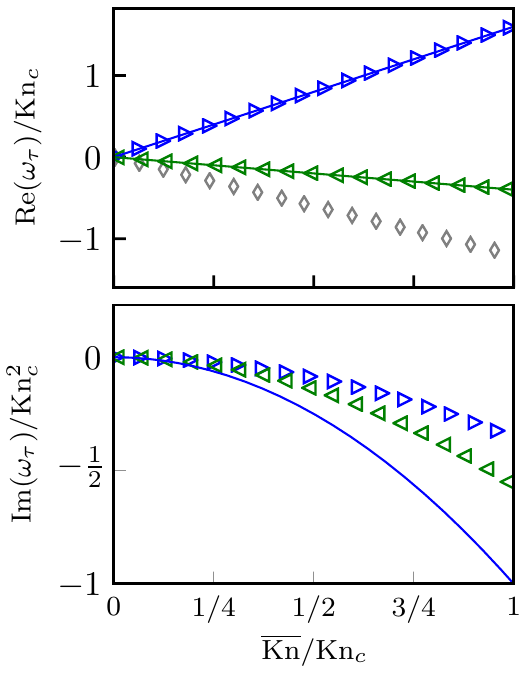}
         \caption{$\overline{\mathrm{Ma}}=0.6$, $E_{1,\alpha\beta} = \varnothing$}
         \label{fig: D1Q3 DVBE no corr stable}
     \end{subfigure}\hspace{0mm}%
     \begin{subfigure}[b]{0.33\textwidth}
         \centering
         \includegraphics[scale=1.]{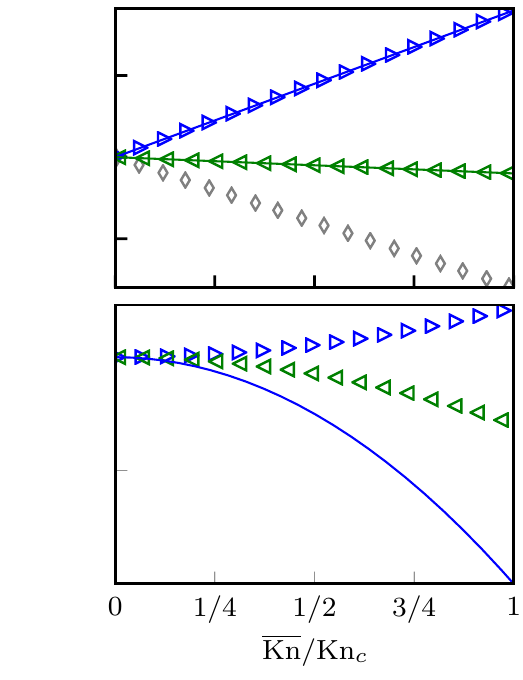}
         \caption{$\overline{\mathrm{Ma}}=0.8$, $E_{1,\alpha\beta} = \varnothing$}
         \label{fig: D1Q3 DVBE no corr unstable}
     \end{subfigure}\hspace{0mm}%
     \begin{subfigure}[b]{0.33\textwidth}
         \centering
         \includegraphics[scale=1.]{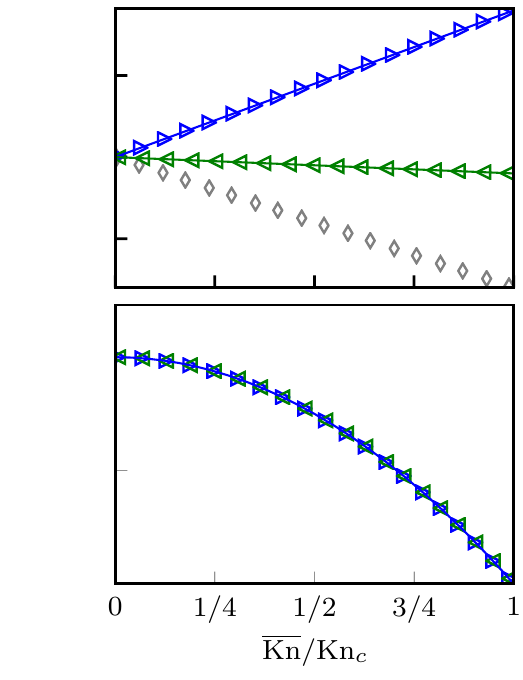}
         \caption{$\overline{\mathrm{Ma}}=0.8$, $E_{1,\alpha\beta} = \checkmark$}
         \label{fig: D1Q3 DVBE corr}
     \end{subfigure}
        \caption{Propagation and dissipation curves of the D1Q3 with $f_i^{eq,2}$ and $\theta=1$ around $\mathrm{\overline{Ma}}=0.732$. Symbols correspond to $\omega_{ac+}$: (\protect\acP) ;  $\omega_{ac-}$: (\protect\acM) ; $\omega_{g}$:(\protect\ghost) and solid lines correspond to the Navier-Stokes solution. }
        \label{fig: DVBE D1Q3 corr/Nocorr}
\end{figure}

Propagation and dissipation curves of the D1Q3 lattice are displayed on Fig.~\ref{fig: D1Q3 DVBE corr} as function of $\mathrm{{Kn}}$. Three modes of different nature are obtained and identified as two physical modes and one ghost mode thanks to the extended analysis. The ghost mode does not carry any macroscopic information and thus cannot be identified as a classical Navier-Stokes modes~\cite{wissocq2019extended}. Note that the latter does not appear on the dissipation curve because its dissipation rate ($\mathrm{Im}(\omega_\tau)=1$) is much larger than the physical one. Regarding the two physical modes, they can be identified as an upstream acoustic mode (\protect\acM) and a downstream one (\protect\acP) and their dispersion and dissipation properties match the Navier-Stokes theoretical values in solid lines.

From this brief modal analysis, the expected athermal Navier-Stokes behavior is recovered. However, close to the specific value $\mathrm{Ma}=0.732$, the DVBE system starts diverging as a consequence of the well-known cubic Mach error present in the viscous stress tensor due to the lattice closure defect. It implies that the linear DVBE systems bearing this defect ($i.e.$ D1Q3, D2Q9, D3Q27 lattices)~\cite{Karlin2010, Li2012} are unstable above $\mathrm{Ma}=0.732$. Such a critical Mach number has already been observed in previous work~\cite{wissocq2019thesis,wilde2019multistep,hosseini2019stability}, its analytical value being found to $\sqrt{3}-1$~\cite{PAM2019}. Thus, before moving to the space discretization, this lattice shortcoming has to be corrected in order to deal with compressible flows, thereby introducing the correction term $E_{1,\alpha\beta}$. Its influence around this critical Mach number can been see on Fig.~\ref{fig: DVBE D1Q3 corr/Nocorr}. In the absence of $E_{1,\alpha\beta}$, the overall system remains stable for $\overline{\mathrm{Ma}}=0.6$ (Fig.~\ref{fig: D1Q3 DVBE no corr stable}), while for $\overline{\mathrm{Ma}}=0.8$ (Fig.~\ref{fig: D1Q3 DVBE no corr unstable}), the downstream acoustic mode is amplified. For the same Mach number, taking $E_{1,\alpha\beta}$ into account, a correct dissipation behavior is recovered in perfect agreement with the theory as shown on Fig.~\ref{fig: D1Q3 DVBE corr}. 
 
\begin{figure}[h!]
\centering
         \includegraphics[scale=1.]{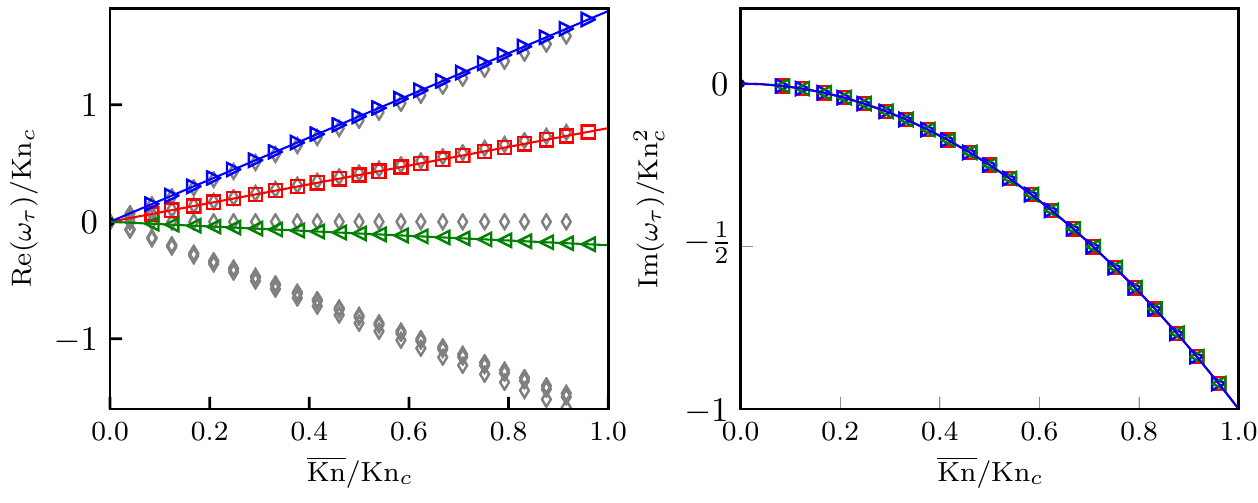}
        \caption{Propagation and dissipation curves of the corrected D2Q9 along $\boldsymbol{e}_x$ at $\overline{\mathrm{Ma}}=0.8$, $\mathrm{Ma}_\theta=0$, $\theta=1$ and $f_i^{eq,3}$. Symbols correspond to $\omega_{ac+}$: (\protect\acP) ;  $\omega_{ac-}$: (\protect\acM) ; $\omega_{shear}$: (\protect\shear) ; $\omega_{g}$:(\protect\ghost) and solid lines correspond to the Navier-Stokes solution.}
        \label{fig: DVBE D2Q9 corr}
\end{figure}

Despite its two-dimensional feature, same conclusions can be drawn for the D2Q9 lattice. On Fig.~\ref{fig: DVBE D2Q9 corr}, propagation and dissipation curves of the corrected D2Q9 lattice are displayed for a mean flow at $\overline{\mathrm{Ma}}=0.8$ along the $x$-direction. As expected and previously observed~\cite{wissocq2019extended}, nine modes are identified: six ghost modes, two acoustic ones and a shear (or vortical) mode, represented by (\protect\shear). The the phase/group velocities and the dissipation rates of the physical modes are found in agreement with the analytical solution across the entire two-dimensional Knudsen space.

\subsection{Von Neumann analysis of the LB scheme}
\label{subsec: LSA of LBM scheme, general results}

In this section, the athermal LBM is considered. In the same way as its continuous counterpart, the temperature is considered constant $T=T_0$ and only $E_{1,\alpha\beta}$ appears in the correction term $\psi_i$. Here again, in order to fully characterize the system, a dimensional analysis is performed. For the one-dimensional discrete case, in addition to the seven independent variables of the continuous case, a numerical variable emerges as the time step $\Delta t$. It is worth noting that the space step $\Delta x$ depends on $c_s$ and $\Delta t$ $via$ the acoustic scaling~\cite{Kruger_Book_2017}, and cannot be considered as an independent variable. The three fundamental units remain unchanged, increasing the number of dimensionless parameters up to five. This linear analysis is then bounded considering:
\begin{equation}
\omega_\tau	= \mathcal{L} \left( \overline{\mathrm{Ma}}, \mathrm{{Kn}} , \theta, \tau^* \right),
\end{equation}
where $\omega_\tau$, $\overline{\mathrm{Ma}}$, $\mathrm{Kn}$ and $\theta$ are still given by Eq.~(\ref{eq:LSA_DVBE_dimensionless_variables}) and
\begin{equation} 
\tau^* = \frac{\overline{\tau}}{\Delta t} = \frac{\mu}{\overline{\rho} c_s^2 \theta \Delta t}.
\end{equation} 
Compared to the continuous case, two parameters are expected to have numerical effects only: $\tau^*$, which appears as a direct consequence of the discrete nature of the system, and  $\theta$, which has no effect on the continuous DVBE in the low-Knudsen regime~\cite{PAM2019}. Regarding the unique number holding the spatial information $k$, \textit{i.e.} $\mathrm{{Kn}}$, a discussion should be held on its relevance in this discrete context, as well as for $\omega_\tau$.
Indeeed, the Nyquist-Shannon sampling theorem~\cite{shannon1949communication} provides here the direct conditions: $k \Delta x \leq \pi$ and $\Re(\omega) \Delta t \leq \pi$. 
This condition is essential for a correct study of a numerical scheme, and for this reason $\mathrm{{Kn}}$ and $\omega^*$ are respectively replaced by $k^*=k \Delta x$ and $\omega^*=\omega \Delta t$ in the present discrete framework. This linear analysis now reads: 
\begin{equation}
\omega^*	= \mathcal{L} \left( \overline{\mathrm{Ma}}, k^* , \theta, \tau^* \right).
\end{equation}
Exploring this parameter space allows assessing the system in all the possible configurations. \newline

An additionnal dimensionless number can be considered as the Courant-Friedrich-Levy (CFL) number~\cite{courant1928partiellen}. It is commonly employed in the stability study of explicit numerical schemes, as a criterion that must be met by the time and the space step. This number corresponds to the ratio of the maximum ``physically expected'' speed over the speed of the numerical information. For the present athermal system, acoustic waves have to be accounted for, which gives:
\begin{equation}
\mathrm{CFL}^{ath}  = \left(\mathrm{\overline{Ma}} + 1 \right)  \sqrt{c_s^{*2}\theta},
\end{equation}
with $c_s^* = c_s \Delta t/\Delta x = 1/\sqrt{3}$ for both the D1Q3 and D2Q9 lattices.
For an explicit scheme, it is recommended to set $\mathrm{CFL}<1$ and even lower, depending on the nature of the scheme and the spatial dimension~\cite{hirsch2007numerical}. In the standard athermal LBM (where $E_{1,\alpha\beta}$ is absent), the value $\theta = 1$ is adopted to minimize the dissipation error in the macroscopic equations~\cite{Karlin2010}. In such a case, the only flexibility over the CFL constraint is to act on $\mathrm{Ma}$, which clearly reduces the physical scope of applications. However, one purpose of the $E_{1,\alpha\beta}$ correction term is precisely to remove this constraint~\cite{renard2020improved}, which now makes it possible to modify the stability or accuracy by varying $\theta$ without any impact on the macroscopic equations. Hence, the analyses below focus on the range $\theta \in [0,1.2]$. 

\begin{figure}[h!]
\centering
         \includegraphics[scale=1.]{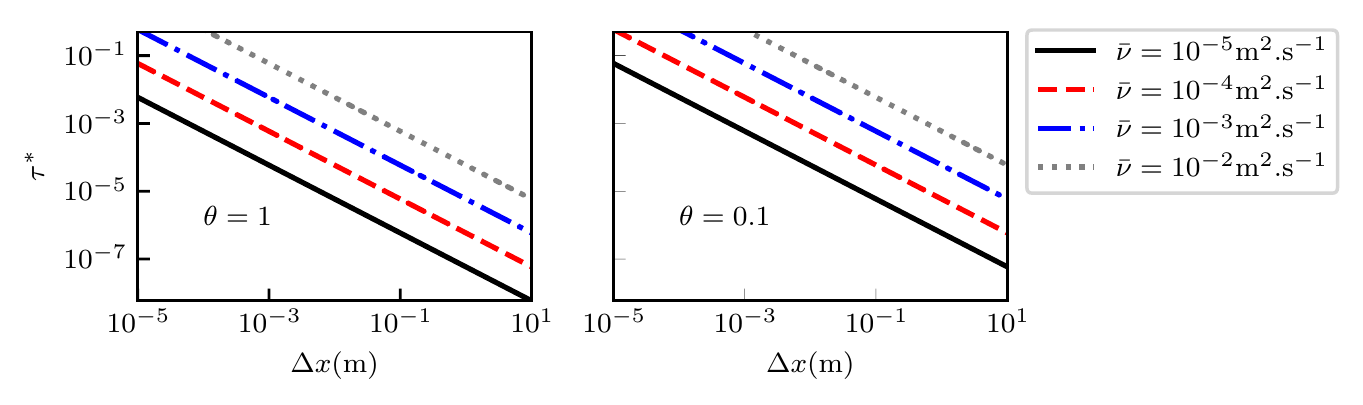}
        \caption{Dimensionless relaxation time $\tau^*=\nu/(\Delta x \theta c_s^* \sqrt{r_gT_0})$ as function of the mesh size $\Delta x$ for $\theta=1$ (left) and $\theta=0.1$ (right). In both cases, different values of viscosity are displayed.}
        \label{fig: tauAdim_funcDx_multiNu}
\end{figure}

Finally, a last discussion has to be conducted regarding the range of interest of the dimensionless relaxation time $\tau^*$. For typical CFD applications, it is fair to admit that the viscosity can vary in time and space, physically or numerically (through turbulence modeling for instance), all along a simulation. Thus, for the targeted applications, a decent range of viscosity ${\nu} \in \left[10^{-5}, 10^{-2} \right] \mathrm{m^2s^{-1}}$ is adopted, as well as a minimal and maximal $\Delta x$ encountered in standard applications $i.e.$ $\Delta x \in \left[10^{-5}, 10^1\right]\mathrm{m}$. This parametric space spanned by ${\nu}$ and $\Delta x$ is displayed on Fig.~\ref{fig: tauAdim_funcDx_multiNu}, which allows considering the lower bound of $\tau^* \simeq 10^{-8}$. 
Regarding the upper bound, $\tau^*=1/2$ is considered. For this value, the collide and stream Eq.~(\ref{eq: LBM BGK}) is reduced to $g^+_i = f_i^{eq}$, which will be considered for its interesting stability properties~\cite{wissocq2020linear}.

\begin{figure}[h!]
     \centering
     \begin{subfigure}[b]{0.33\textwidth}
         \centering\hspace{0mm}%
         \includegraphics[scale=1.]{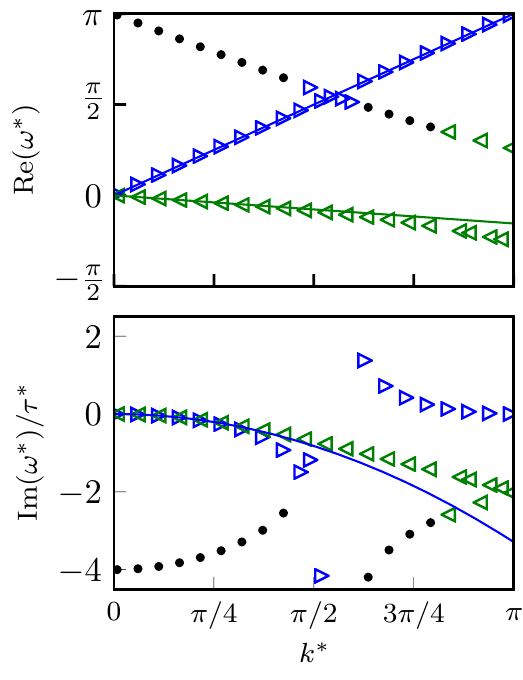}
         \caption{$\theta = 1$, unstable}
         \label{fig: D1Q3 theta 1}
     \end{subfigure}\hspace{0mm}%
     \begin{subfigure}[b]{0.33\textwidth}
         \centering
         \includegraphics[scale=1.]{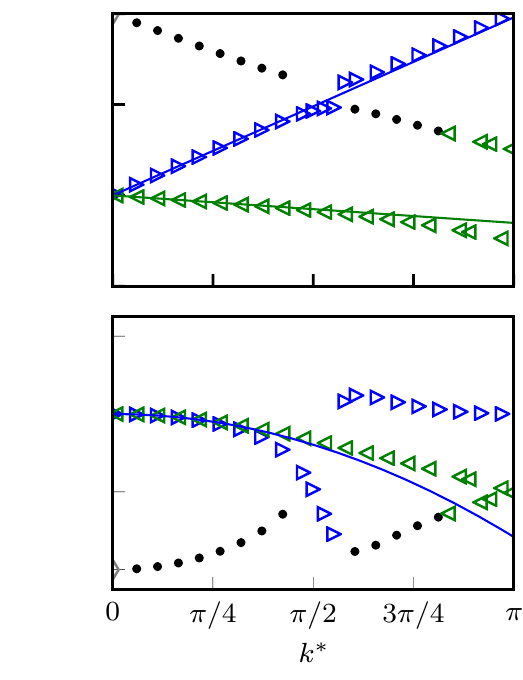}
         \caption{$\theta = 0.96$, unstable}
         \label{fig: D1Q3 theta 0.96}
     \end{subfigure}\hspace{0mm}%
     \begin{subfigure}[b]{0.33\textwidth}
         \centering
         \includegraphics[scale=1.]{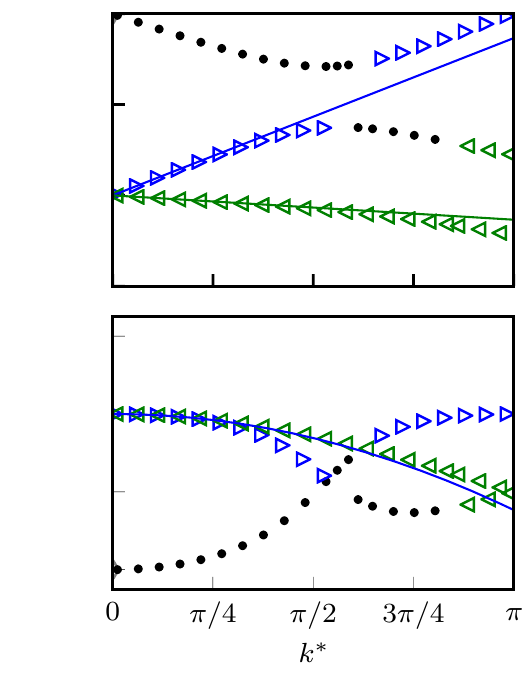}
         \caption{$\theta = 0.75$, stable}
         \label{fig: D1Q3 theta 0.75}
     \end{subfigure}
        \caption{Dispersion and dissipation curves of the D1Q3 BGK-LBM with $f_i^{eq,2}$. In the present case $\overline{\mathrm{Ma}}=0.732$ and $\tau^*=10^{-5}$. Symbols correspond to $\omega_{ac+}$: (\protect\acP) ;  $\omega_{ac-}$: (\protect\acM) ; $\omega_{msp}$:(\protect\macroSpu) and solid lines correspond to the Navier-Stokes solution. }
        \label{fig: LBM D1Q3 effect of theta}
\end{figure}

Fig.~\ref{fig: LBM D1Q3 effect of theta} displays dispersion and dissipation curves of the D1Q3 BGK-LB model.
Due to the space and time discretization, the D1Q3 lattice gives rise to a spurious macroscopic mode $\omega_{msp}$ (black dot), phase shifted by $\pi$ at the origin. The latter carries some macroscopic contribution and can interact with other physical modes~\cite{wissocq2019extended}. For the critical Mach number $\overline{\mathrm{Ma}}= \overline{\mathrm{Ma}^c}=0.732$, with $\theta=1$, Fig.~\ref{fig: D1Q3 theta 1} shows an instability resulting from an eigenvalue collision between the downstream acoustic and the macroscopic spurious mode, which is therefore of a completely different nature than the instability shown in Sec.~\ref{subsec: LSA of the corrected DVBE, general results}. The Mach error being corrected, this instability is uniquely related to numerical aspects as pointed out in~\cite{wissocq2019extended}. \newline

\begin{figure}[h!]
     \centering
     \begin{subfigure}[b]{0.33\textwidth}
         \centering\hspace{0mm}%
         \includegraphics[scale=1.]{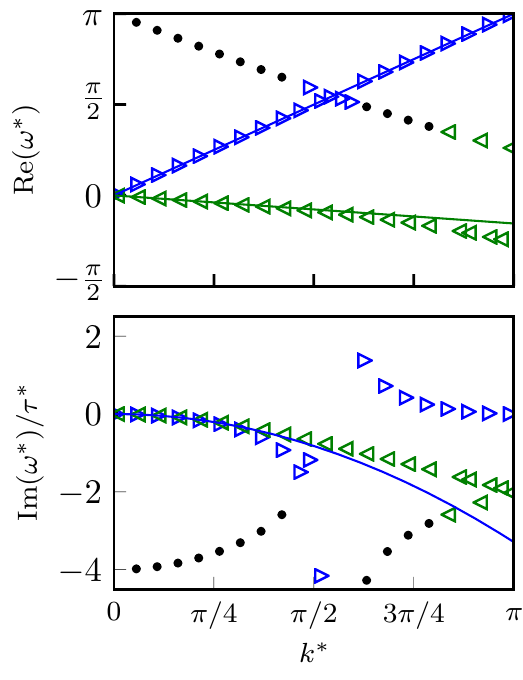}
         \caption{$\sigma = 1$ unstable}
         \label{fig: D1Q3 sigma 1}
     \end{subfigure}\hspace{0mm}%
     \begin{subfigure}[b]{0.33\textwidth}
         \centering
         \includegraphics[scale=1.]{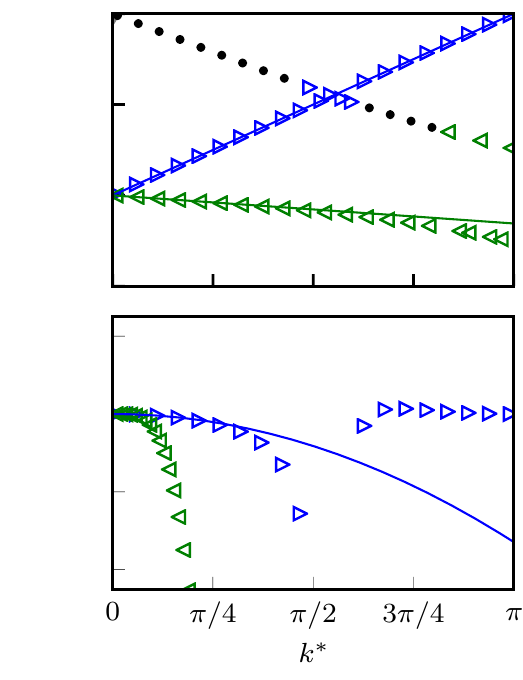}
         \caption{$\sigma = 0.95$ unstable}
         \label{fig: D1Q3 sigma 0.95}
     \end{subfigure}\hspace{0mm}%
     \begin{subfigure}[b]{0.33\textwidth}
         \centering
         \includegraphics[scale=1.]{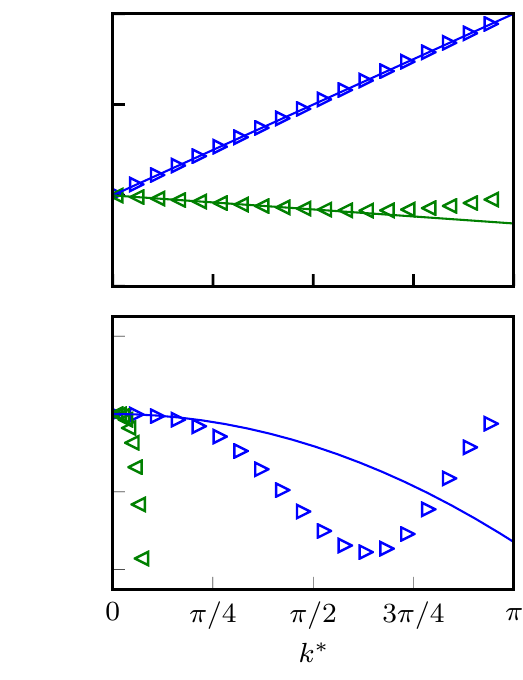}
         \caption{$\sigma = 0$ stable}
         \label{fig: D1Q3 sigma 0}
     \end{subfigure}
        \caption{Dispersion and dissipation curves of the D1Q3 HRR-LBM with $f_i^{eq,2}$. In the present case $\overline{\mathrm{Ma}}=0.732$ and $\tau^*=10^{-5}$. When $100\%$ of finite differences is employed in $g_i^{(1)}$ (ie $\sigma=0$), the spurious macroscopic wave vanishes, suppressing the eigenvalue collision. Symbols correspond to $\omega_{ac+}$: (\protect\acP) ;  $\omega_{ac-}$: (\protect\acM) ; $\omega_{msp}$:(\protect\macroSpu) and solid lines correspond to the Navier-Stokes solutions.}
        \label{fig: LBM D1Q3 effect of sigma}
\end{figure}

In order to circumvent the issue, two strategies can be employed:
\begin{itemize}
\item The first one, showed on Fig.~\ref{fig: LBM D1Q3 effect of theta}, consists in lowering the CFL number \textit{via} $\theta$, acting on the time step. It allows bypassing the eigenvalue collision by narrowing the sonic cone and lowering the group velocity of the spurious mode. For the D1Q3 lattice, the critical Mach number associated to the eigenvalue collision turns out to be obtained when $\mathrm{CFL}^{ath} = 1$, \textit{i.e.}:
\begin{equation}
	\overline{\mathrm{Ma}^c} = \frac{1}{c_s^*\sqrt{\theta}} -1,
\label{eq: Mach c athermal}
\end{equation}
which, for some reason, exactly corresponds to the critical Mach number of the non-corrected athermal DVBE when $\theta=1$.
Lowering $\theta$ thus makes it possible to push back this constraint for the subsonic regime. Other types of instabilities are involved in supersonic configurations, as pointed out in~\ref{app: Supersonic limit}. This strategy allows stabilizing the scheme while preserving correct dissipation properties of well-resolved wavelengths, the drawback being the reduction of the time step increasing the CPU time.
\item Instead of lowering the time step, the second strategy consists in adopting the HRR collision operator~\cite{Jacob2019}. Its particularity is to avoid modal interactions~\cite{astoul2020analysis} by partially reconstructing the second-order non-equilibrium moment with finite differences. This phenomenon is highlighted on Fig.~\ref{fig: LBM D1Q3 effect of sigma}, where the scheme is stabilized by increasing the FD part of $g_i^{(1)}$ in the regularization procedure. In this respect, the dispersion curve of Fig.~\ref{fig: D1Q3 sigma 0.95} shows that the destructive eigenvalue collision can be avoided with $\sigma=0.95$.  Furthermore, note that the spurious mode completely disappears for a full FD reconstruction ($i.e.$ $\sigma=0$) as shown by Fig.~\ref{fig: D1Q3 sigma 0}. However, despite accurate dispersion properties, the FD reconstruction bias the dissipation properties of the scheme acting as a low-pass filter. 
\end{itemize}

\begin{figure}[h!]
\centering
         \includegraphics[scale=1.]{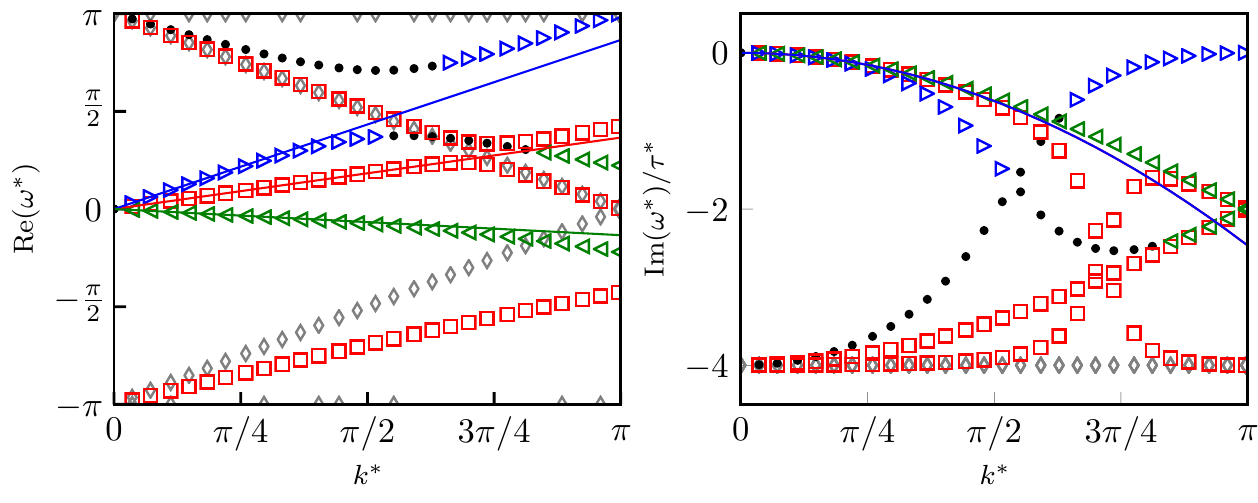}
        \caption{Dispersion and dissipation curves of the corrected D2Q9 BGK-LBM along $\boldsymbol{e}_x$ for $\overline{\mathrm{Ma}}=0.732$, $\mathrm{Ma}_\theta=0$, $\theta=0.75$ and $f_i^{eq,3}$. Symbols correspond to $\omega_{ac+}$: (\protect\acP) ;  $\omega_{ac-}$: (\protect\acM) ; $\omega_{shear}$: (\protect\shear) ; $\omega_{msp}$:(\protect\macroSpu) ; $\omega_{g}$:(\protect\ghost) and solid lines correspond to the Navier-Stokes solution.}
        \label{fig: D2Q9 athermal}
\end{figure}

Fig.~\ref{fig: D2Q9 athermal} displays the spectral properties along $\boldsymbol{e}_x$ of the D2Q9-BGK model for $\overline{\mathrm{Ma}}=0.732$ and $\theta = 0.75$. On these curves, no spurious amplification is obtained and the same stabilization phenomena of Fig.~\ref{fig: D1Q3 theta 0.75} is observed on the acoustics thank to the value of $\theta \leq 1$. However, other kinds of modal interactions, related to the shear waves are underlined. On the dispersion curve, two shear modes emerge, interacting with each other as a curve veering phenomenon~\cite{wissocq2019extended}. This indicates that other instabilities phenomena, directly related to the two-dimensional nature of the scheme, need to be taken into account. In reality, this scheme is found unstable in the spectral domain for wave numbers having a non-zero $k_y$ component. This phenomenon will be further studied in the next subsection. 

The two stabilization techniques introduced in this section ($\theta$ and $\sigma$) will be extensively investigated throughout this article as levers in the aim of stabilizing the HLBM.

\subsection{Two-dimensional LBM instabilities}
\label{subsec: Bi-dimentional lbm}

In the previous section, a similar behavior regarding the acoustic eigenvalues has been observed in the one and two-dimensional cases, \textit{i.e.} the decrease of $\theta$ helps avoiding the eigenvalue collision. However, for the D2Q9 lattice, this observation was made for horizontal plane waves only ($k_y=0$), while nothing ensures that a generalization can be drawn over the whole spectrum. Moreover, the D2Q9 lattice being richer in terms of modes, Fig.~\ref{fig: D2Q9 athermal} highlighted that other kinds of modal interactions can potentially emerge. Therefore, this section is devoted to the behavior of the D2Q9 lattice over its full spectrum, \textit{i.e.} varying both $k^*_x$ and $k^*_y $. 

\begin{figure}[h!]
     \centering
     \begin{subfigure}[b]{1.\textwidth}
         \centering\hspace{0mm}%
         \includegraphics[scale=1.]{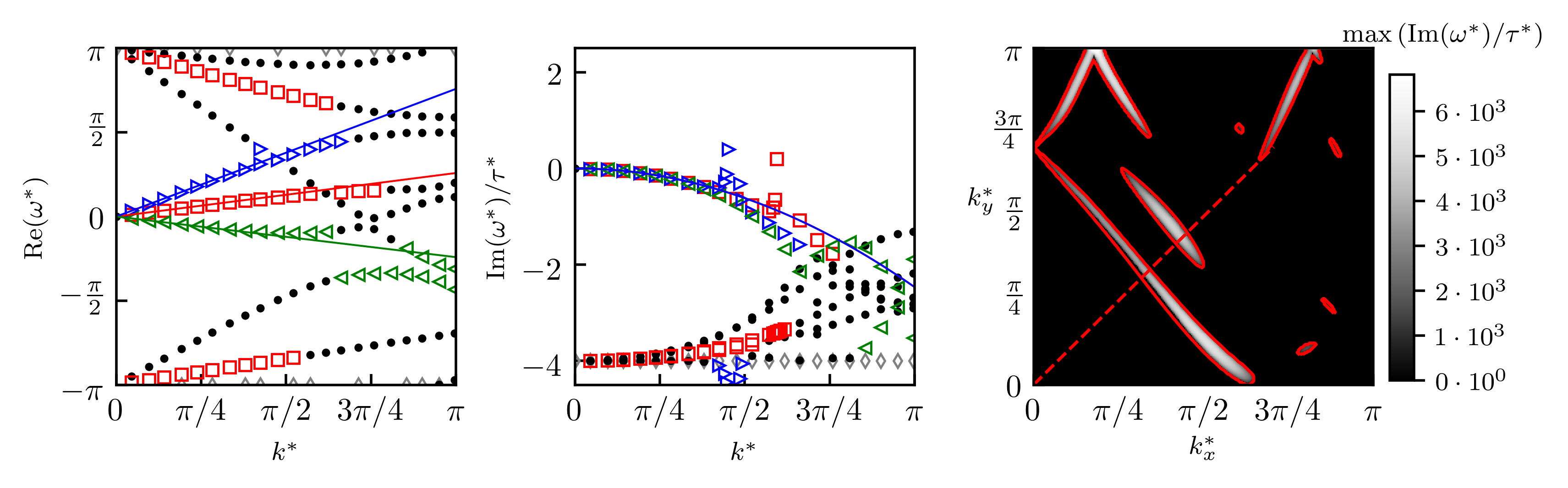}
         \caption{BGK, unstable}
         \label{fig: D2Q9 BGK 3plts}
     \end{subfigure}\hspace{0mm}
     
     \begin{subfigure}[b]{1.\textwidth}
         \centering
         \includegraphics[scale=1.]{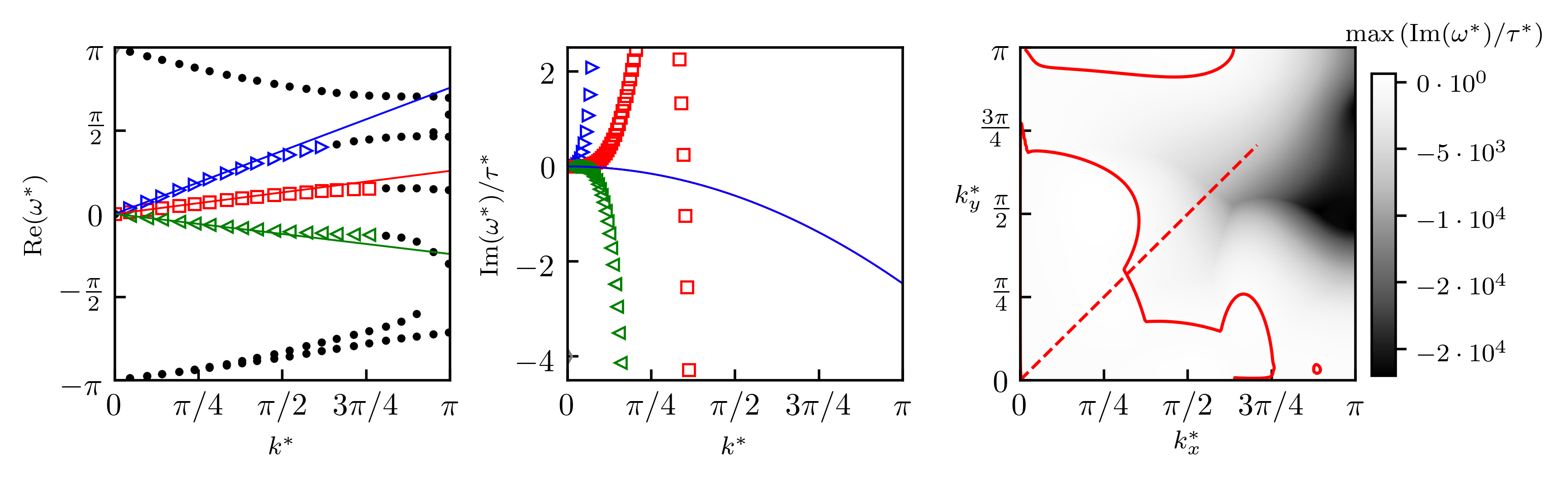}
         \caption{HRR $\sigma = 1$, unstable}
         \label{fig: D2Q9 HRR sig 1 3plts}
     \end{subfigure}\hspace{0mm}

     \begin{subfigure}[b]{1.\textwidth}
         \centering
         \includegraphics[scale=1.]{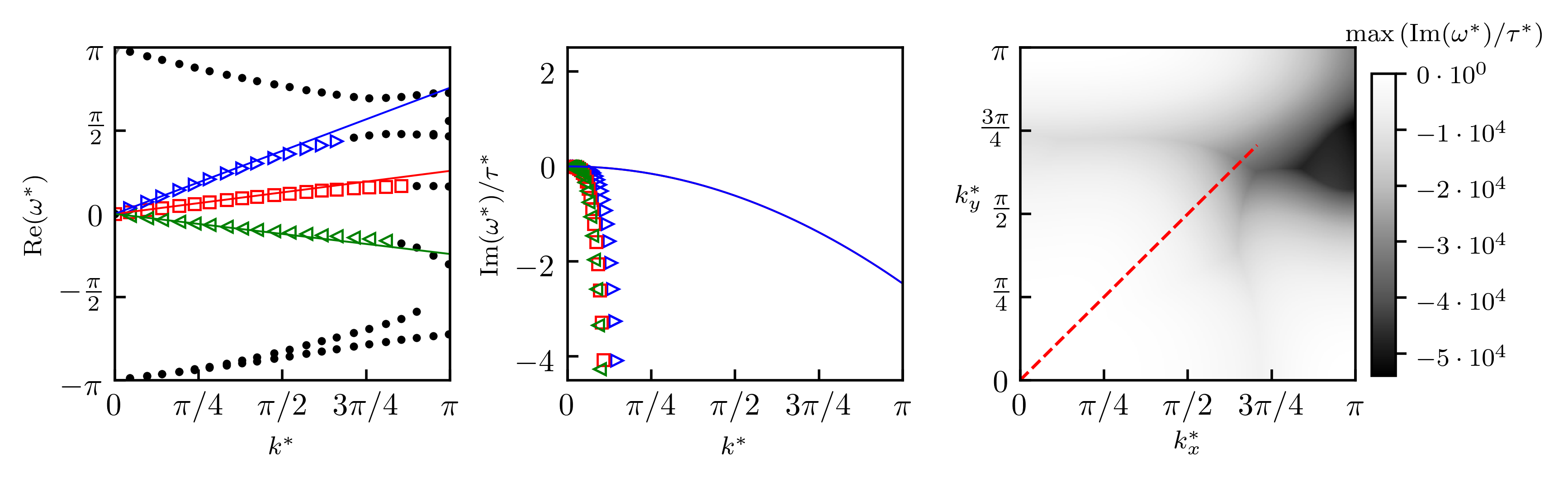}
         \caption{HRR $\sigma = 0.7$, stable}
         \label{fig: D2Q9 HRR sig 0.7 3plts}
     \end{subfigure}\hspace{0mm}
     
        \caption{Spectral properties of the corrected LBM D2Q9 for three collision models : BGK (a), HRR $\sigma=1$ (b) and HRR $\sigma=0.7$. Here, $\overline{\mathrm{Ma}}=0.732$, $\mathrm{Ma}_\theta=0$, $\theta=0.75$, $\tau^*=1.10^{-5}$ and $f_i^{eq,3}$. From left to right : Dispersion graph, dissipation graph and absolute stability map. On this map, the dashed line represent the line along the dispersion and dissipation graphs are plotted ; the solid line represents the unstable regions and therefore corresponds and iso-contour of $\mathrm{max}(\mathrm{Im}(\omega^*)/\tau^* = 0)$.  Symbols correspond to $\omega_{ac+}$: (\protect\acP) ;  $\omega_{ac-}$: (\protect\acM) ; $\omega_{shear}$: (\protect\shear) ; $\omega_{msp}$:(\protect\macroSpu) ; $\omega_{g}$:(\protect\ghost) and solid lines correspond to theory.}
        \label{fig: comp mod D2Q9 athermal}
\end{figure}

Fig.~\ref{fig: D2Q9 BGK 3plts} displays the spectral properties of the D2Q9 BGK model with the same physical parameters as in the previous sub-section, namely $\overline{\mathrm{Ma}} = 0.732$, $\tau^*=10^{-5}$ and $\theta=0.75$, this time plotted for $45^\circ$-inclided waves with the $e_x$ axis. It can be observed that this model is unstable in this direction of the spectrum. Two eigenvalue collisions are highlighted, one of acoustic nature for $k^*<\pi/2$, and one of shear nature for $k^*>\pi/2$. 
On the stability map of the model (Fig.~\ref{fig: D2Q9 BGK 3plts}-right), several instability bubbles (characterized by a very large value of $\mathrm{Im}(\omega^*)/\tau^*$) are observed, consequent to these eigenvalue collisions. It is important to note that this kind of instability has also been detected at lower Mach numbers~\cite{wissocq2019thesis}. Unfortunately, and contrary to the one-dimensional D1Q3 case, these instabilities cannot be addressed by changing the $\theta$ parameter, suggesting the use of another collision operator. 

For this purpose, Fig.~\ref{fig: D2Q9 HRR sig 1 3plts} displays the spectral properties of the D2Q9 HRR model with $\sigma=1$ in similar conditions as Fig.~\ref{fig: D2Q9 BGK 3plts}. Note that for the particular value $\sigma=1$, this operator is reduced to the recursive regularized operator with $N_r=3$ (RR3). On the dispersion curve, it can be observed that the eigenvalue collision disappears. This is made possible thanks to the mode filtering properties of the regularized operators~\cite{wissocq2020linear}. 
Unfortunately, this scheme remains unstable due to large areas of instability as shown on the $\mathrm{max}(\mathrm{Im}(\omega^*)/\tau^*)$ map, also exhibited in previous work~\cite{wissocq2020linear}. Even though these instabilities are much less brutal than that caused by the eigenvalue collision phenomenon, it is noticed that their amplitude is increased by a decrease of $\theta$.  
This observation suggests that decreasing the value of $\theta$ should not be advised with this model. However, for $\overline{\mathrm{Ma}} \geq 0.732$ the decrease of $\theta$ is a necessary ingredient to avoid acoustic eigenvalue collision, as illsutrated in the one-dimensional cases of Sec.~\ref{subsec: LSA of LBM scheme, general results}. The spurious wave responsible for this phenomenon remains present despite the use of the RR3 collision model. To summarize the observations drawn on the D2Q9-RR3 model:
\begin{itemize}
    \item a standard value $\theta \approx 1$ leads to an acoustic eigenvalue collision for $\overline{\mathrm{Ma}}\approx 0.7$,
    \item decreasing $\theta$ yields an amplification of isolated modes, responsible for an unstable scheme,
\end{itemize}
Thus, unlike in one-dimensional cases, the $\theta$ parameter is not sufficient to increase the robustness of 2D compressible LB schemes. It leads to the use of the second stabilization lever, \textit{i.e.} the finite-differences reconstruction parameter $\sigma$. This stabilization phenomenon is illustrated on Fig.~\ref{fig: D2Q9 HRR sig 0.7 3plts}, where the HRR model is used with $\sigma=0.7$. As observed previously, this parameter leads to a dissipation of the physical waves, restoring the numerical stability of the model whatever the wavenumber of the perturbation.\newline

\begin{figure}[h!]
\centering
         \includegraphics[scale=1.]{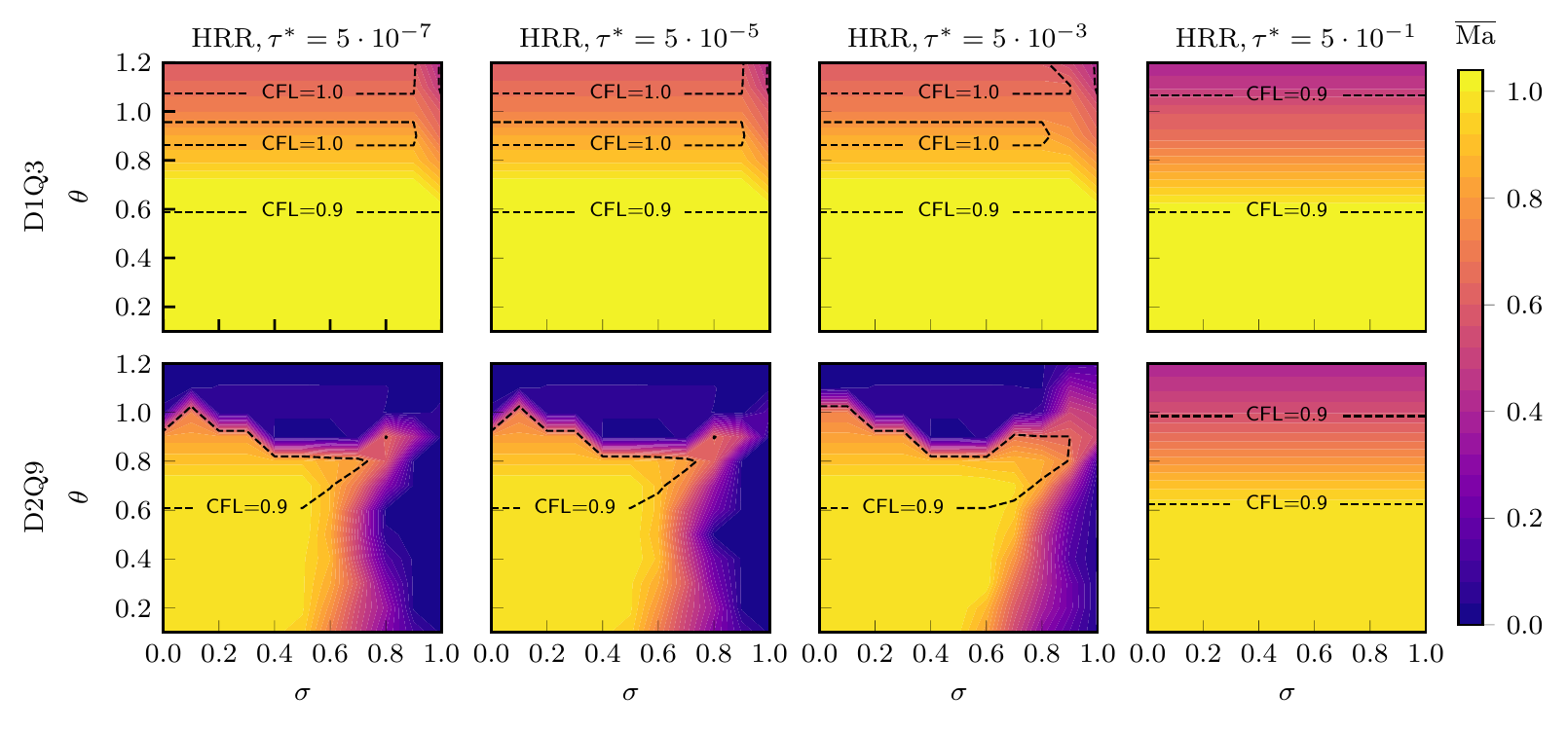}
        \caption{Maximum achievable Mach number acting on both $\theta$ and $\sigma$ stabilization levers. D1Q3 lattice (top)  D2Q9 lattice (bottom). Iso level of CFL number are displayed in dashed lines for the single value of $\mathrm{CFL} = 0.9$ and for ten values evenly distributed for $\mathrm{CFL} \in [1.0, 1.1]$.}
        \label{fig: D1Q3 D2Q9 maps}
\end{figure}

Thus, in order to ensure numerical stability for relatively large Mach numbers, there is a competition between the destabilizing effects of the $\theta$ parameter and the dissipative effects induced by the $\sigma$ parameter. To shed light on this phenomenon, a stability study on the parametric space is carried out for the D1Q3 and the D2Q9 lattices. For four values of the dimensionless relaxation time $\tau^* \in [5.10^{-7}, 5.10^{-5}, 5.10^{-3}, 5.10^{-1}]$, the space formed by $[\theta, \sigma]$ is scanned to find the maximum Mach number allowed by the model. In this case, twelve values are considered for $\theta \in[0.1, 1.2]$ and ten values for $\sigma \in [0,1]$. The maximum Mach number is found using a bisection method with a convergence criterion taken as $\Delta \overline{\mathrm{Ma}} < 0.05$, where the delta represents the difference of the values found between two iterations. For the D2Q9 case, the angle of the average velocity field $\mathrm{Ma}_\theta$ is scanned for 13 values evenly distributed between $0$ and $45$ degrees. The wavenumber space is defined such as $k_x^* \in [-\pi, \pi]$ and $k_y^* \in [0, \pi]$, and discretized by $160$ and $80$ points respectively. For the D1Q3 lattice, $k^* \in [0, \pi]$ and is discretized by $80$ points. The results of this study are shown on Fig.~\ref{fig: D1Q3 D2Q9 maps} for both lattices. As previously mentioned, decreasing $\theta$ has stabilizing effects on the D1Q3 lattice. The latter directly drives the CFL constraint which, according to the iso-countour, never exceeds $\mathrm{CFL} = 1$. On the contrary, $\sigma$ has very little influence on the stability.

Regarding the D2Q9 case, it should be noted that any value $\theta > 1$ has been found unstable whatever the Mach number. As for the 1D case, $\mathrm{CFL}>1$ cannot be reached for the D2Q9 lattice. Moreover it is clearly visible that for values $\theta < 1$, the maximum Mach number is dependent on $\sigma$, which confirms the previous observations. It is worth noting that both schemes have a limit close to $\overline{\mathrm{Ma}} = $1. This limit can be addressed by changing the type of discretization of the correction term $\psi_i$, as illustrated in~\ref{app: Supersonic limit}, which is beyond the purpose of this work. Finally one can see that the two schemes have a nearly similar behavior for the particular value of $\tau^* = 0.5$, where the collide and stream algorithm simply reduces to $g^+_i = f_i^{eq}$, a singular configuration as referred in~\cite{wissocq2019thesis} . In any event, the closer $\tau^*$ is to one-half, the less dissipation needs to be added \textit{via} $\sigma$ to keep the model stable.


\section{Extended linear analyses of the hybrid systems}
\label{subsec: Extended analysis to HDVBE scheme}

The main numerical properties of the LB models of interest in the present work (especially the role of $\theta$ and $\sigma$) have been investigated, up to now, in the athermal framework. The aim of the present section is now to perform an original extended analysis of the hybrid method, so as to better understand the numerical behavior of the HLBM for compressible flows.
The influence of the choice of the thermodynamic variable for the energy equation is also investigated, since this point has been proved to have a strong influence on the properties of the global numerical methods for discretized Euler and Navier-Stokes equations, e.g. \cite{honein2004higher,coppola2019numerically}.

\subsection{Method and concepts}
\label{subsec: Method and concepts HDVBE}

Without loss of generality, the methodology will be detailed here for a set of continuous equations only, knowing that it can be directly transposed to discrete numerical schemes. In this case, the hybrid system is composed of the DVBE coupled with the energy equation. It can be written as:
\begin{equation}
\left\{
\begin{aligned}
\frac{\partial f_i}{\partial t} = \mathcal{S}_i^{DVBE}(f_j, \rho\phi),\\
\frac{\partial (\rho \phi)}{\partial t} = \mathcal{S}^{En}(f_j, \rho\phi),
\end{aligned}
\right.
\label{eq: system equation hybride}
\end{equation}
where in the present case, only the conservative form of the energy equation is considered, without loss of generality for its primitive form. Note that, like $\mu$, the heat conductivity $\lambda$ will be considered as a constant property of the fluid in what follows, as well as $c_p$ and $c_v$. Once the system linearized (with $f_i = \overline{f_i} + f_i'$ and $\rho\phi = \overline{\rho\phi} + \left(\rho\phi\right)'$ ), a time-advance matrix $\mathrm{M}^{HDVBE}_{ij}$ is obtained. It can be decomposed into four distinct blocks: two diagonal and two off-diagonal ones. Fig.~\ref{Fig: multi bloc matrix} details the composition of each block according to the choice of the energy variable. The diagonal parts model the influence of an equation (resp. DVBE or energy equation) on its involved variable (resp. $f_i$ and $\rho \phi$). The off-diagonal parts are responsible for the coupling between the DVBE and the energy equation. The upper-right block of the matrix is the interaction of the aerodynamic variables $\rho$ and $u$ (computed by the DVBE) onto the energy, while the lower-left block is the thermal feedback of the energy equation onto the DBVE system.

Following Eq.~(\ref{eq: f prime}), the fluctuations $f_i'$ and $\left(\rho\phi\right)'$ are expressed under the form of plane monochromatic waves leading to the following eigenvalue problem:
\begin{equation}
\omega \widehat{\mathcal{X}}_i = \widetilde{\mathrm{M}}^{HDVBE}_{ij}  \widehat{\mathcal{X}}_i.
\end{equation}
Here, $\boldsymbol{\mathcal{\widehat{X}}} = \left( \widehat{f}_0, ...,\widehat{f}_{m-1}, \widehat{\rho\phi} \right)^T $ is the vector gathering the amplitude fluctuations of both distribution functions and energy variable, and $\widetilde{\mathrm{M}}^{HDVBE}_{ij}$ is the time advance matrix in the Fourier space. For more details on the derivation of this matrix, one can relate to~\ref{app: Matrix thermal}.

\begin{figure}
\centering
    \begin{tikzpicture}

\matrix[matrix of math nodes,row sep=0.125em,column sep=0.125em, nodes in empty cells,
        left delimiter=(,
        right delimiter=),
        nodes in empty cells] (mymatrix) at (0,0)
{
     \dfrac{\partial \mathcal{S}_i^{DVBE}}{\partial f_j}\bigg|_{\overline{f}_j, \overline{\phi}} &  \dfrac{\partial \mathcal{S}_i^{DVBE}}{\partial \phi}\bigg|_{\overline{f}_j, \overline{\phi}}    \\
    \dfrac{\partial\mathcal{S}^{En}}{\partial f_j}\bigg|_{\overline{f}_j, \overline{\phi}}  & \dfrac{\partial \mathcal{S}^{En}}{\partial \phi}\bigg|_{\overline{f}_k, \overline{\phi}}     \\
};

\draw[ultra thick, draw=black, fill=blue, opacity=0.2, rounded corners=5] 
     (mymatrix-1-1.north west)++(0,0) 
     rectangle 
     (mymatrix-1-1.south east)++(-5,0.6);

\draw[ultra thick, draw=black, fill=blue, opacity=0.2, rounded corners=5] 
     (mymatrix-2-2.north west) 
     rectangle 
     (mymatrix-2-2.south east);

\draw[ultra thick, draw=black, fill=green, opacity=0.2, rounded corners=5] 
     (mymatrix-1-2.north west) 
     rectangle 
     (mymatrix-1-2.south east);

\draw[ultra thick, draw=black, fill=red, opacity=0.2, rounded corners=5] 
     (mymatrix-2-1.north west) 
     rectangle 
     (mymatrix-2-1.south east);
%
%
%


\matrix[matrix of math nodes,row sep=0.125em,column sep=0.125em, nodes in empty cells,
        left delimiter=(,
        right delimiter=),
        nodes in empty cells] (mymatrixTwo) at (3.75,0)
{
 \bf{I} &\bf{II} \\
 \bf{III} & \bf{IV} \\
};

\draw[ultra thick, draw=black, fill=blue, opacity=0.2, rounded corners=5] 
     (mymatrixTwo-1-1.north west)++(0,0) 
     rectangle 
     (mymatrixTwo-1-1.south east)++(0,0);

\draw[ultra thick, draw=black, fill=blue, opacity=0.2, rounded corners=5] 
     (mymatrixTwo-2-2.north west) 
     rectangle 
     (mymatrixTwo-2-2.south east);

\draw[ultra thick, draw=black, fill=green, opacity=0.2, rounded corners=5] 
     (mymatrixTwo-1-2.north west) 
     rectangle 
     (mymatrixTwo-1-2.south east);

\draw[ultra thick, draw=black, fill=red, opacity=0.2, rounded corners=5] 
     (mymatrixTwo-2-1.north west) 
     rectangle 
     (mymatrixTwo-2-1.south east);


\draw (mymatrix-1-1)++(-2,-0.5) node[rotate=0]{$\mathrm{M}^{HDVBE}_{ij}=\,\,\,$};

\draw (mymatrix-1-2)++(1.5,-0.5) node[rotate=0]{$=$};

\draw (mymatrix-1-1)++(0,+0.75) node[rotate=0]{$\boldsymbol{m\times m}$};
\draw (mymatrix-1-2)++(0,+0.75) node[rotate=0]{$\boldsymbol{m\times 1}$};
\draw (mymatrix-2-1)++(0,-0.75) node[rotate=0]{$\boldsymbol{1\times m}$};
\draw (mymatrix-2-2)++(0,-0.75) node[rotate=0]{$\boldsymbol{1\times 1}$};


\draw (-4,-2.0) node[rotate=0]{Block \textbf{III}: LBM on Hybrid};
\draw (-4,-3.5) node[rotate=0]{$ 
        \begin{tabular}{|c|c|c|c|c|c|c|}
            \hline
            &\multicolumn{3}{c|}{{conservative}}&\multicolumn{3}{c|}{{primitive}} \\
             & $E$ & $e$ & $s$ & $E$ & $e$ & $s$ \\
            \hline
            convection    &\checkmark &\checkmark &\checkmark & \ding{55} & \ding{55} & \ding{55} \\
            $\mathcal{P}$ & \checkmark  & \checkmark  & $\varnothing$ & \checkmark  & \checkmark  & $\varnothing$  \\
            $\mathcal{F}$ & \ding{55} & \ding{55} & \checkmark  & \ding{55} & \ding{55} & \checkmark \\
            $\mathcal{V}$ & \checkmark & $\varnothing$ & $\varnothing$  & \checkmark & $\varnothing$ & $\varnothing$ \\
            \hline
        \end{tabular}
$};

\draw (4,-2.0) node[rotate=0]{Block \textbf{IV}: Hybrid on Hybrid};
\draw (4,-3.5) node[rotate=0]{$ 
        \begin{tabular}{|c|c|c|c|c|c|c|}
        \hline
            &\multicolumn{3}{c|}{{conservative}}&\multicolumn{3}{c|}{{primitive}} \\
             & $E$ & $e$ & $s$ & $E$ & $e$ & $s$ \\
            \hline
            convection    & \checkmark  & \checkmark  & \checkmark  & \checkmark  & \checkmark  & \checkmark  \\
            $\mathcal{P}$ & \ding{55} & \ding{55} & $\varnothing$ & \ding{55} & \ding{55} & $\varnothing$ \\
            $\mathcal{F}$ & \checkmark  & \checkmark  & \checkmark  & \checkmark  & \checkmark  & \checkmark \\
            $\mathcal{V}$ & \ding{55} & $\varnothing$ & $\varnothing$  & \ding{55} & $\varnothing$ & $\varnothing$ \\
        \hline
        \end{tabular}
$};

    \end{tikzpicture}
    \caption{Multiblock arrangement of the linearized HDVBE Matrix and their contributions. Block I models the DVBE system acting on itself similarly to block IV for the energy. Block II is responsible for the perfect gas coupling through $f_i^{eq}$ and $\psi_i$ the thermal correction. The contribution of the different energy terms in blocks III and IV are given in the tables in function of the various configurations; the term is present: $\checkmark$, missing: $\ding{55}$ or does not exist in the linearized equation: $\varnothing$.}
    \label{Fig: multi bloc matrix}
\end{figure}

\noindent Diagonalizing the matrix $\widetilde{\mathrm{M}}^{HDVBE}_{ij}$ provides $(m+1)$ eigenvalues $\left( \omega_l \right)_{l \in \llbracket 0, m \rrbracket}$ and eigenvectors $\left( \mathrm{Y}^l_i \right)_{l\in \llbracket 0, m \rrbracket }$. Thus, $m+1$ linear eigen-modes of the system~(\ref{eq: system equation hybride}) are obtained. Once diagonalized it reads:
\begin{equation}
\omega_l \mathrm{Y}^l_i = \widetilde{\mathrm{M}}^{HDVBE}_{ij} \mathrm{Y}^l_i.
\end{equation}

Like in the athermal case of Sec.~\ref{sec: The Linear stability analysis}, the spectral behavior of the system has to be compared with a reference. Eventually linearizing the Navier-Stokes-Fourier equations, an entropy mode $\omega _{entr}$ is obtained in addition to those present in the athermal case. The eigenvalues of these modes are:
\begin{equation} \label{eq:eigenvaluesNS}
	\begin{aligned}
		&\omega _{shear} = k_\alpha  \overline{u}_\alpha - \mathrm{i} \overline{\nu} \left\Vert k_\alpha \right\Vert^2 + O\left(k^3\right), \\
		&\omega _{entr} = k_\alpha  \overline{u}_\alpha - \mathrm{i} \overline{\kappa} \left\Vert k_\alpha \right\Vert^2 + O\left(k^3\right), \\
		&\omega _{ac\pm} = k_\alpha  \overline{u}_\alpha \pm \left\Vert k_\alpha \right\Vert  \sqrt{\gamma c_s^2 \overline{\theta}} - \mathrm{i} \left( \frac{D-1}{D}\overline{\nu} + \frac{\gamma_g -1}{2} \overline{\kappa}  \right) \left\Vert k_\alpha \right\Vert^2 + O\left(k^3\right), \\
	\end{aligned}
\end{equation}
with $\overline{\theta} = \overline{T}/T_r$ the dimensionless temperature and $\overline{\kappa}=\lambda/(\overline{\rho}c_p)$ the mean heat diffusivity. Compared to the athermal case, the shear mode $\omega _{shear}$ (only present in 2D) remains unchanged while the two acoustic modes are altered in terms of dispersion and dissipation, which is a direct consequence of taking account the energy fluctuations. The speed of sound is now isentropic and the thermal diffusivity is now involved in the acoustic dissipation.

Similarly to the athermal study of Sec.~\ref{sec: The Linear stability analysis}, a physical interpretation of the modes obtained by the linear analysis can be done. In that respect, an extension of the methodology of Wissocq~et~al.~\cite{wissocq2019extended} applied to the hybrid LB models is proposed. First, a macroscopic counterpart $\boldsymbol{\mathcal{E}}_l^{HLB}$ of the hybrid LB eigenvectors is defined such as:
\begin{equation}
\boldsymbol{\mathcal{E}}_l^{HLB} = \left[ \widehat{\rho}^{HLB}=\sum_{i=0}^{m-1} \mathrm{Y}^l_i, \quad \left(\widehat{u}_\alpha^{HLB} = \frac{1}{\overline{\rho}} \left( \sum_{i=0}^{m-1} c_{i,\alpha} \mathrm{Y}^l_i - \overline{u}_\alpha \sum_{i=0}^{m-1} \mathrm{Y}^l_i \right) \right), \qquad \widehat{T}_{\rho \phi}^{HLB} \right]^\mathrm{T},
\end{equation}
where the expression of $\widehat{T}_{\rho \phi}^{HLB}$ can be found in Table~\ref{Table: T transfo for NsProj} for the conservative and primitive forms.

\begin{table}[H]
\centering
  \begin{tabular}{|c|c|c|}
    \cline{2-3} \multicolumn{1}{c|}{} & Convervative & Primitive \\
    \hline
    $\widehat{T}_{\rho E,E}^{HLB}$ & $\dfrac{1}{\overline{\rho} c_v} \left( \widehat{\rho E}^{HLB} - \left( c_v \overline{T} + \overline{u}_\alpha^2/2 \right) \widehat{\rho}^{HLB} - \overline{\rho} \overline{u}_\alpha \widehat{u}_\alpha^{HLB} \right)$ & $\dfrac{1}{c_v} \left(\widehat{ E}^{HLB} -  \overline{u}_\alpha \widehat{u}_\alpha^{HLB}\right)$ \\
    \hline
    $\widehat{T}_{\rho e,e}^{HLB}$ & $\dfrac{1}{\overline{\rho}c_v} \left( \widehat{\rho e}^{HLB} - c_v \overline{T} \widehat{\rho}^{HLB}  \right)$ &  $\dfrac{\widehat{ e}^{HLB} }{c_v}$ \\
    \hline
    $\widehat{T}_{\rho s,s}^{HLB}$ & $\dfrac{\overline{T}}{\overline{\rho}c_v} \left\{ \widehat{\rho s}^{HLB} - c_v \left[ \ln\left(\overline{T}\right) - \left( \gamma-1 \right) \left(\ln\left(\overline{\rho}\right)+1\right) \right] \widehat{\rho}^{HLB} \right\}$ & $\overline{T} \left( \dfrac{\widehat{s}^{HLB}}{c_v} + \dfrac{\gamma - 1 }{\overline{\rho}} \widehat{\rho}^{HLB} \right)$ \\

    \hline
  \end{tabular}
\caption{Expression of the temperature fluctuations depending on the HLB energy variables and formulation.}
\label{Table: T transfo for NsProj}
\end{table}

A similar linear stability analysis of the continuous Navier-Stokes-Fourier equations yields the following eigenvectors, written here according to the primitive variables:
\begin{equation}
\boldsymbol{V}^{NS}_{mode} = \left[ \widehat{\rho}^{NS}, \quad \left( \widehat{u}_\alpha^{NS} \right), \qquad \widehat{T}^{NS} \right]^\mathrm{T}, \quad \mathrm{mode} \in \left[shear, ac-, ac+, entr \right].
\end{equation}
Similarly to the athermal analysis of Sec.~\ref{subsec: Method and concept}, the contribution of each HLB mode $\boldsymbol{\mathcal{E}}_l^{LB}$ in terms of physical macroscopic modes can then be expressed as:
\begin{equation}
\boldsymbol{\mathcal{E}}_l^{HLB} = a_l \boldsymbol{V}^{NS}_{shear} + b_l \boldsymbol{V}^{NS}_{ac-} + c_l \boldsymbol{V}^{NS}_{ac+} +  d_l \boldsymbol{V}^{NS}_{entr}, 
\end{equation}
where the new coefficient $d_l$ corresponds to the complex contributions of $\boldsymbol{\mathcal{E}}_l^{HLB}$ on the entropy mode. The passage matrix is now built such as $\boldsymbol{P}_{NS} = \left[ \boldsymbol{V}^{NS}_{ac-}, \boldsymbol{V}^{NS}_{ac+}, \boldsymbol{V}^{NS}_{shear}, \boldsymbol{V}^{NS}_{entr} \right]$ and allows computing each of these coefficients:
\begin{equation}
	\left(a_l, b_l, c_l, d_l\right)^{T} = P^{-1}_{NS} \boldsymbol{\mathcal{E}}_l^{HLB}.
\label{eq: coef NS Fourier}
\end{equation}
These coefficients are finally normalized so that $|a_l|+|b_l|+|c_l|+|d_l|=1$. In this way, the macroscopic nature of the modes resulting of the hybrid LB linear analysis is directly accessible by looking too the module of each coefficient. For details about the derivation of $\boldsymbol{V}^{NS,ath}_{mode}$, the interested reader can refer to~\ref{app: Eigen value NS}. \newline

The methodology introduced here with a set of continuous equations can directly be applied to a discrete system such as the HLBM. In this case, the matrix system reads:
\begin{equation}
e^{-\mathrm{i}\omega_l \Delta t} \mathrm{Y}^l_i = \widetilde{\mathrm{M}}^{HLBM}_{ij} \mathrm{Y}^l_i.
\end{equation}
where details on $\widetilde{\mathrm{M}}^{HLBM}_{ij}$ can be found in~\ref{app: Matrix thermal}. Like in the athermal case, the logarithm of the eigenvalues has to be computed in order to obtain the complex temporal pulsations $\omega_l$.

\subsection{Linear analysis of the HDVBE}
\label{subsec: Analysis of HDVBE scheme}

Here again, a dimensional analysis is first conducted to clearly underline the number of independent parameters of the HDVBE. Compared to the athermal DVBE system, adding an energy equation implies the consideration of the temperature as a new fundamental unit. Furthermore, four additional independent variables naturally emerge from the energy equation, namely the heat capacity ratio at constant pressure $c_p$ and volume $c_v$, the mean fluid temperature $\overline{T}$ and the mean heat diffusivity $\overline{\kappa}=\lambda/(\overline{\rho} c_p)$. It gives rise to two additional dimensionless numbers to fully characterize the system, which now reads in one dimension:
\begin{equation}
\omega_\tau	= \mathcal{L} \left( \overline{\mathrm{Ma}}, \mathrm{\overline{Kn}}, \overline{\theta}, \gamma_g, \mathrm{Pr}  \right),
\end{equation}
with
\begin{equation} 
\begin{split}
&\omega_\tau = \frac{\omega \overline{\nu}}{r_g\overline{T}}, \quad \mathrm{\overline{Ma}} = \frac{\overline{u}}{\sqrt{\gamma_g r_g\overline{T}}}, \quad \mathrm{\overline{Kn}} = \frac{\overline{\nu} k}{\sqrt{\gamma_g r_g\overline{T}}}, \quad \overline{\theta} = \frac{\overline{T}}{T_r}, \\ 
&\gamma_g = \frac{c_p}{c_v} \quad \mathrm{and} \quad \mathrm{Pr} = \frac{\mu c_p}{\lambda}.
\end{split}
\end{equation}
Here, $\mathrm{Pr}$ is the Prandtl number of the considered fluid and $\gamma_g$ its heat capacity ratio. In this work, the value of these parameters, which are related to the fluid characteristics, are set constant as $\mathrm{Pr}=0.71$ and $\gamma_g=1.4$. Similarly to the DVBE system in Sec~(\ref{subsec: LSA of the corrected DVBE, general results}), the same argument stands regarding $\mathrm{\overline{Kn}}\ll 1$. Exploring this parameter space allows studying the system in all the possible configurations.

\begin{figure}
     \centering
     \begin{subfigure}[b]{0.33\textwidth}
         \centering\hspace{0mm}%
         \includegraphics[scale=1.]{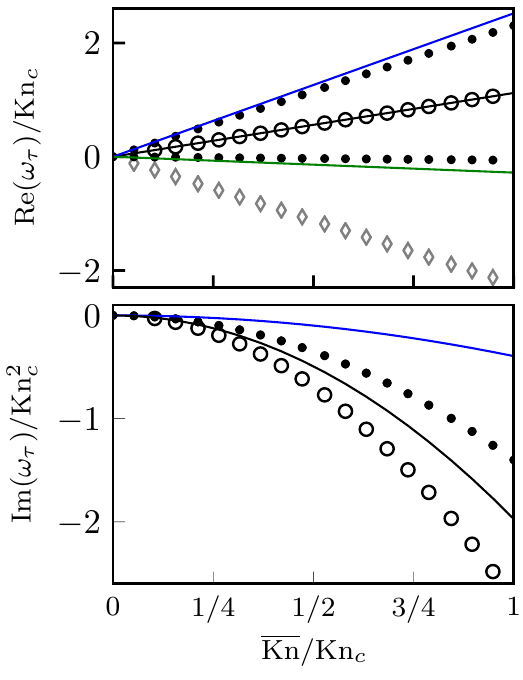}
         \caption{$\mathcal{P}=\varnothing$, $E_{2,\alpha\beta}=\varnothing$.}
         \label{fig: HD1Q3 DVBE no TP no E2}
     \end{subfigure}\hspace{0mm}%
     \begin{subfigure}[b]{0.33\textwidth}
         \centering
         \includegraphics[scale=1.]{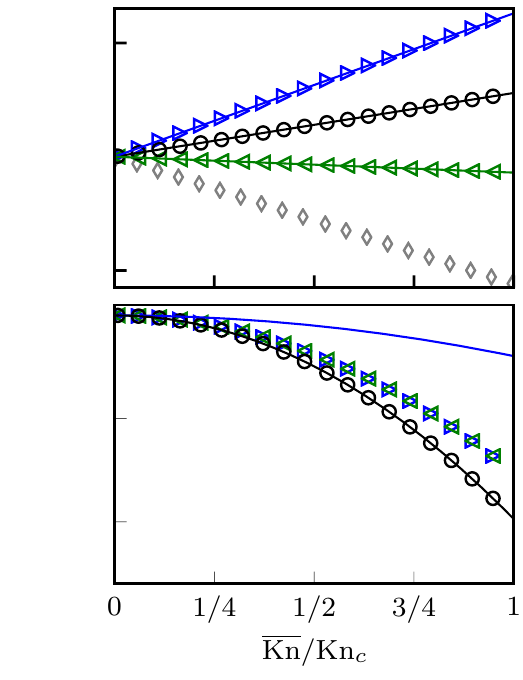}
         \caption{$\mathcal{P}=\checkmark$, $E_{2,\alpha\beta}=\varnothing$.}
         \label{fig: HD1Q3 DVBE with TP no E2}
     \end{subfigure}\hspace{0mm}%
     \begin{subfigure}[b]{0.33\textwidth}
         \centering
         \includegraphics[scale=1.]{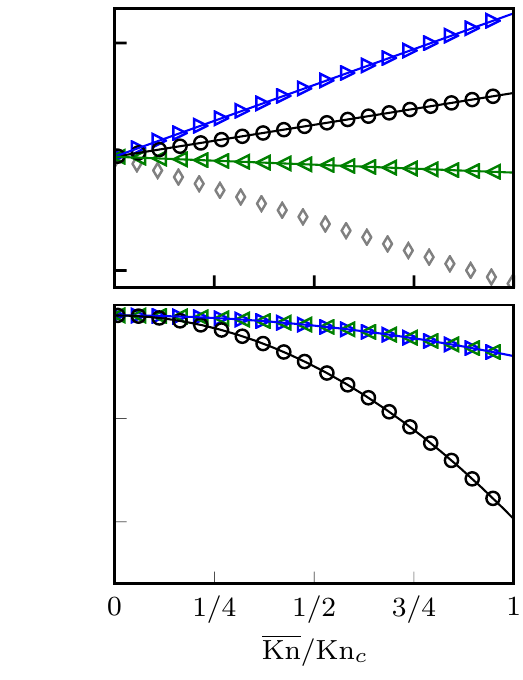}
         \caption{$\mathcal{P}=\checkmark$, $E_{2,\alpha\beta}=\checkmark$.}
         \label{fig: HD1Q3 DVBE with TP with E2}
     \end{subfigure}
        \caption{Propagation (top) and dissipation (bottom) curves of the Hybrid $D1Q3$ on $\rho e$ with $f_i^{eq,2}$ and for $\overline{\mathrm{Ma}}=0.8$. Fig.~(a-b) shows the influence of the pressure work term on the acoustic propagation and Fig.~(b-c) the influence of the $E_{2,\alpha\beta}$ correction acting on the acoustic dissipation. Symbols correspond to $\omega_{ac+}$: (\protect\acP) ;  $\omega_{ac-}$: (\protect\acM) ; $\omega_{entr}$: (\protect\entrop) ; $\omega_{msp}$:(\protect\macroSpu) ; $\omega_{g}$:(\protect\ghost) and solid lines correspond to theory.}
        \label{fig: Hybrid DVBE D1Q3 corr/Nocorr}
\end{figure}

The spectral analysis of the fully corrected D1Q3 HDVBE is displayed on Fig.~\ref{fig: HD1Q3 DVBE with TP with E2}, where the conservative form of the internal energy equation is considered. First, one can notice that this system is linearly stable for $\overline{\mathrm{Ma}}=0.8$, a mean Mach number larger than the critical value without Mach correction. Thus, one can conclude that: (1) the correction term $E_{1,\alpha\beta}$ still owns stabilizing properties in this hybrid context, and (2) coupling the mesoscopic system (DVBE) and the macroscopic energy equation does not lead to singularities. 
Secondly, the propagation curves of Fig.~\ref{fig: HD1Q3 DVBE no TP no E2} and Fig.~\ref{fig: HD1Q3 DVBE with TP no E2} indicate that the pressure work $\mathcal{P}$ is directly involved in the isentropic speed of sound $\sqrt{\gamma c_s^2 \theta}$. On Fig.~\ref{fig: HD1Q3 DVBE no TP no E2}, where $\mathcal{P}$ is missing, two modes of macroscopic nature (black dot) are identified instead of the expected acoustics. They do not propagate the exact acoustic information at the expected isentropic speed of sound. Nevertheless, the propagation of the entropic mode is well recovered, whereas its dissipation rate is overestimated compared to the theoretical value.  
From the two dissipation curves Fig.~\ref{fig: HD1Q3 DVBE with TP no E2} and Fig.~\ref{fig: HD1Q3 DVBE with TP with E2}, it is possible to see the influence of $E_{2,\alpha\beta}$: correcting the polyatomic deficiency of the DVBE affects the trace of the viscous stress tensor~\cite{renard2020improved}. In addition to grant the correct dissipation, modifying this term in a straightforward way also allows adding a potential volume viscosity which is a desirable feature in CFD contex~\cite{dellar2001bulk,cook2007artificial}.

\begin{figure}[h!]
\centering
         \includegraphics[scale=1.]{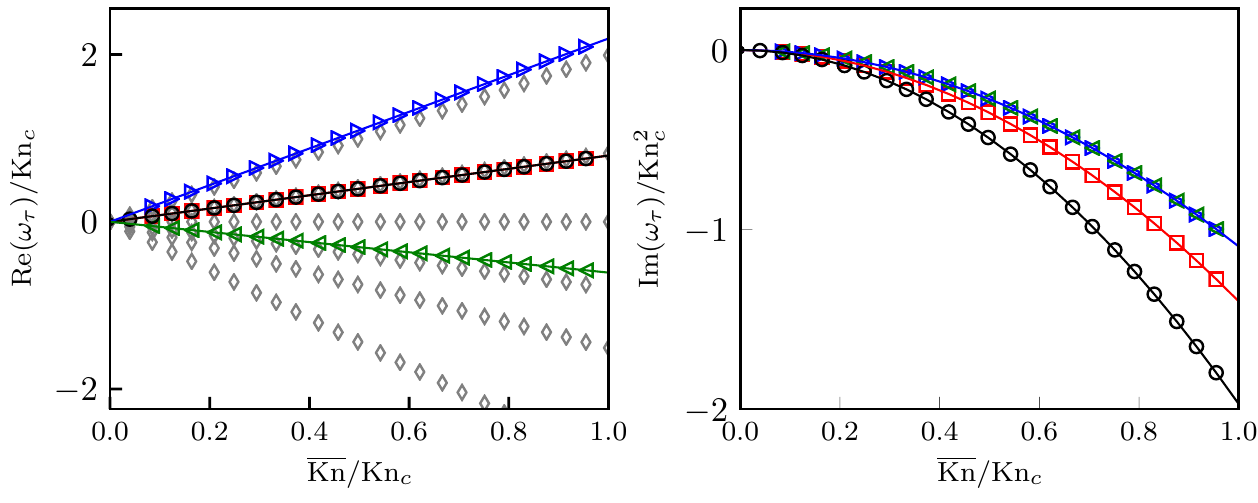}
        \caption{Propagation and dissipation curves of the corrected hybrid D2Q9 along $\boldsymbol{ex}$ at $\overline{\mathrm{Ma}}=0.8$, $\mathrm{Pr=0.71}$, $\gamma_g=1.4$ $\mathrm{Ma}_\theta=0$, $\overline{\theta}=1$ and $f_i^{eq,3}$. Symbols correspond to $\omega_{ac+}$: (\protect\acP) ;  $\omega_{ac-}$: (\protect\acM) ; $\omega_{entr}$: (\protect\entrop) ; $\omega_{shear}$: (\protect\shear) ; $\omega_{g}$:(\protect\ghost) and solid lines correspond to theory.}
        \label{fig: HDVBE D2Q9 corr}
\end{figure}

Finally, the very same behavior is observed for the two-dimensional case on Fig.~\ref{fig: HDVBE D2Q9 corr}, where propagation and dissipation curves of the D2Q9 HDVBE are conducted for $k_y=0$ and a horizontal mean flow at $\overline{\mathrm{Ma}}=0.8$. Here, both $E_{1,\alpha\beta}$ and $E_{2,\alpha\beta}$ are included. The spectral properties of this hybrid system perfectly match the Navier-Stokes theoretical values in terms of dispersion and dissipation, suggesting that the system faithfully recovers the targeted equations. Thus, a numerical method based on this set of equations should converge to the resolution of the compressible Navier-Stokes system. The analysis of such scheme is precisely the subject of the next section.

\subsection{Von Neumann analysis of the HLBM scheme}
\label{subsec: Extended analysis to HLBM scheme}

Compared to the continuous HDVBE and similarly to the athermal LBM, the time step $\Delta t$ has to be considered as an additional independent variable. The number of fundamental units remains unchanged (mass, length, temperature and time), which gives rise to one additional dimensionless number $\tau^*$ to fully characterize the system. The discrete system now reads: 
\begin{equation}
\omega^*	= \mathcal{L} \left( \overline{\mathrm{Ma}}, k^* , \overline{\theta}, \gamma_g, \mathrm{Pr}, \tau^* \right),
\end{equation}
with
\begin{equation} 
\begin{split}
&\omega^* = \omega \Delta t , \quad \mathrm{\overline{Ma}} = \frac{\overline{u}}{\sqrt{\gamma_g r_g\overline{T}}}, \quad k^* = k \Delta x, \quad \overline{\theta} = \frac{\overline{T}}{T_r}, \quad \gamma_g = \frac{c_p}{c_v},  \quad \mathrm{Pr} = \frac{\mu c_p}{\lambda} \quad \mathrm{and} \quad \tau^* =\frac{\overline{\tau}}{\Delta t}.
\end{split}
\end{equation}
Based on the sampling rate arguments detailed in Sec.~\ref{subsec: LSA of LBM scheme, general results}, $k^*$ has been preferred to $\overline{\mathrm{Kn}}$ for this discrete study, along with $\omega^*$ instead of $\omega_\tau$. Moreover, since the isentropic speed of sound is expected to be recovered, $\sqrt{\gamma_g c_s^{*2} \overline{\theta}}$ is the maximal speed allowed by the system for $\mathrm{\overline{Ma}}=0$. Thus, dividing it by the speed of the numerical information ($\Delta x / \Delta t$), the CFL number related to the HLBM reads: 
\begin{equation}
\mathrm{CFL} = \left(\mathrm{\overline{Ma}} + 1 \right)  \sqrt{\gamma c_s^{*2} \overline{\theta}}.
\end{equation}
This number is, by construction, always larger than its athermal counterpart since $\gamma_g>1$.
As previously adopted, $\overline{\theta}$ is set as $\overline{\theta} \in \left[0,1.2\right]$, so that the CFL number remains below, or close to, unity. Finally, the study holds on $\tau^* \in \left[10^{-8},0.5\right]$ for the same reason as mentioned in Sec.~\ref{subsec: LSA of LBM scheme, general results}. Exploring this parameter space allows evaluating the system in all the possible configurations of interest.

\begin{figure}
     \centering
     \begin{subfigure}[b]{0.33\textwidth}
         \centering\hspace{0mm}%
         \includegraphics[scale=1.]{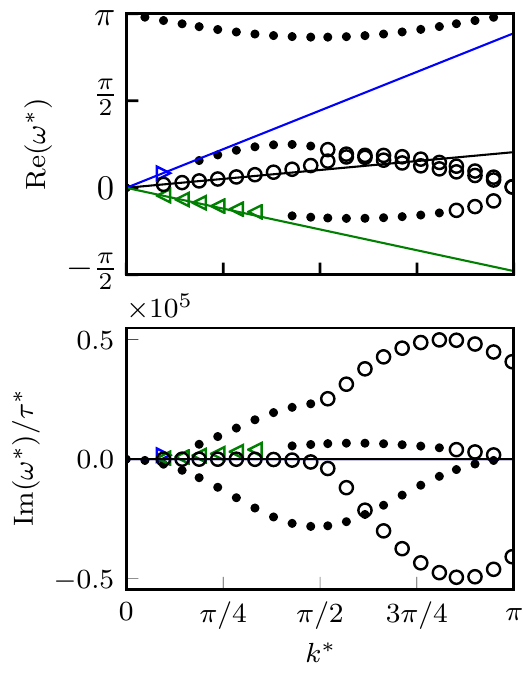}
         \caption{$\rho e$.}
         \label{fig: }
     \end{subfigure}\hspace{0mm}%
     \begin{subfigure}[b]{0.33\textwidth}
         \centering
         \includegraphics[scale=1.]{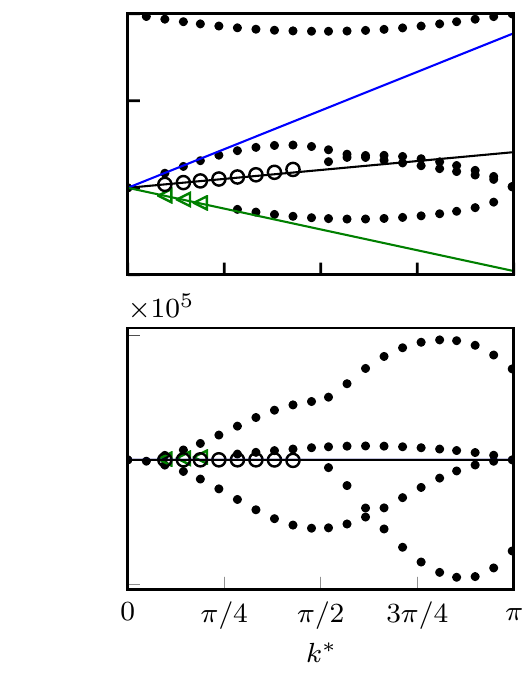}
         \caption{$\rho E$.}
         \label{fig: }
     \end{subfigure}\hspace{0mm}%
     \begin{subfigure}[b]{0.33\textwidth}
         \centering
         \includegraphics[scale=1.]{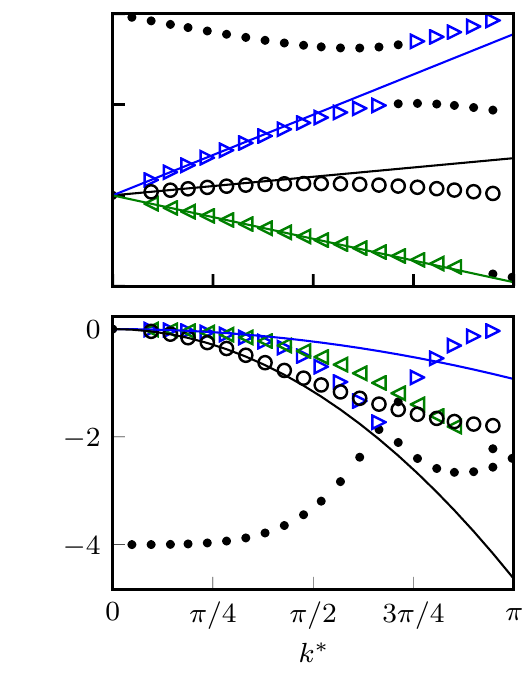}
         \caption{$\rho s$.}
         \label{fig: rhosQ3}
     \end{subfigure}
        \caption{Dispersion (top) and dissipation (bottom) curves of the D1Q3-BGK-RK4C02 scheme for $\overline{\theta}=1$, $\overline{\mathrm{Ma}}=0.3$, $\tau^*=10^{-5}$. Symbols correspond to $\omega_{ac+}$: (\protect\acP) ;  $\omega_{ac-}$: (\protect\acM) ; $\omega_{entr}$: (\protect\entrop) ; $\omega_{msp}$:(\protect\macroSpu) ; $\omega_{g}$:(\protect\ghost) and solid lines correspond to the NS theory.}
        \label{fig: 1D rhoe, rhoE, rhos}
\end{figure}

Fig.~\ref{fig: 1D rhoe, rhoE, rhos} displays the spectral behavior of the D1Q3-BGK-RK4CO2 scheme for three conservative forms of the energy equation, based on: $\rho e$, $\rho E$ and $\rho s$. In the present case $\overline{\theta}=1$, $\overline{\mathrm{Ma}}=0.3$ and $\tau^*=10^{-5}$. First, note that the internal and total energy-based HLBM schemes are found unstable, unlike the entropy-based scheme. The dispersion and dissipation behaviors of the two unstable schemes are rather similar. The dispersion curve highlights a destructive interaction close to $k^*=\pi/2$ between two modes of the HLBM. The nature of this instability appears similar to that observed in Sec.~\ref{subsec: LSA of LBM scheme, general results}, hinting towards the use of the same levers to circumvent this kind of instability. 

The use of the HRR collision operator can be considered, as well as lowering the time step \textit{via} a change of $\overline{\theta}$. However, as observed in Sec.~\ref{subsec: LSA of LBM scheme, general results}, a reduction of $\theta$ is expected to decrease the group velocity of each numerical mode, which might enhance the mode coupling observed in the present case, resulting in lower numerical stability. Yet, the use of $\sigma$ as stabilizing lever will be \textit{a priori} preferred in such a case. Nevertheless, before changing the collision operator, Fig.~\ref{fig: rhosQ3} shows that the entropy-based model is stable and has correct spectral properties. This suggests that the form of the considered equation is at least as important as the type of collision operator, the adopted FD scheme or the value of the $\overline{\theta}$. At last, despite the relatively low Mach number of Fig.~\ref{fig: rhosQ3}, note that a curve veering is observed regarding the downstream acoustic information, contrary to the athermal case, where this only occurs when $\overline{\mathrm{Ma}}\approx 0.6$. This is probably due to the fact that, in the hybrid case, the sound speed is increased by a factor $\sqrt{\gamma}$. As a consequence, the eigenvalue collision phenomenon evoked in Sec.~\ref{subsec: LSA of LBM scheme, general results} for $\theta = 1$ turns out to appear, in the hybrid framework, below $\overline{\mathrm{Ma}^c} = 0.732$. Further observations indicate that this critical value can be, like its athermal counterpart, correlated to the CFL number at unity, so that:
\begin{equation}
	\overline{\mathrm{Ma}^c} = \frac{1}{c_s^*\sqrt{\gamma \theta}} - 1,
	\label{eq: Mach c hybrid}
\end{equation}
which suggests that the $\overline{\theta}$ stabilization lever can play an even more important role in the hybrid context.\newline

\begin{figure}[h!]
     \centering
     \begin{subfigure}[b]{1.\textwidth}
         \centering\hspace{0mm}%
         \includegraphics[scale=1.]{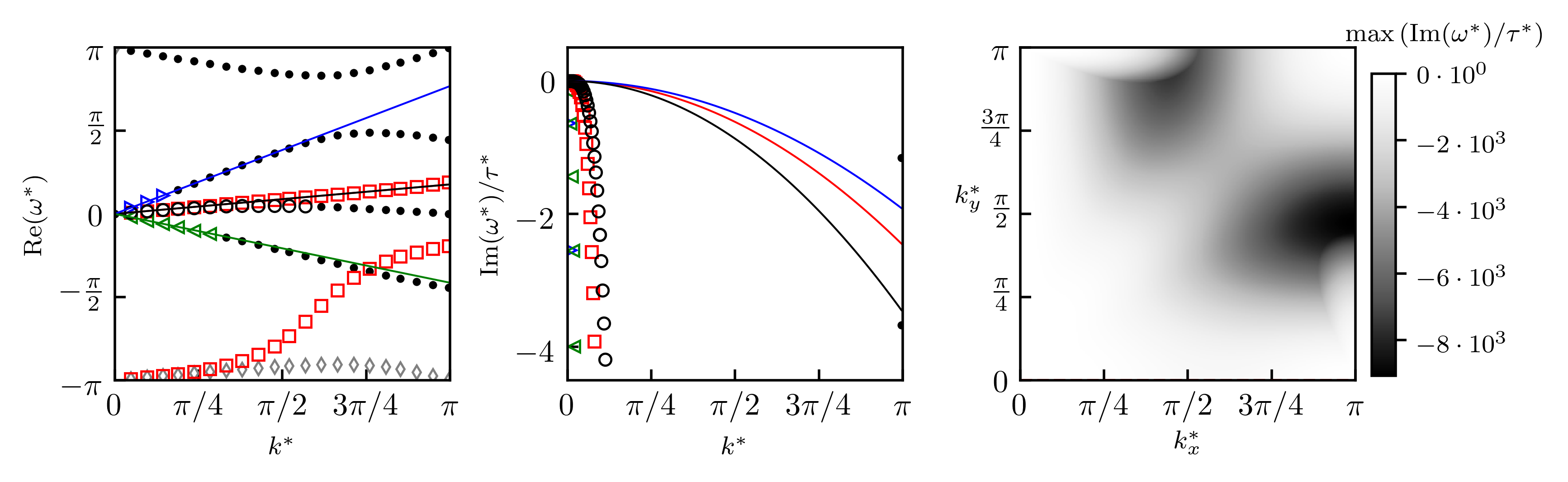}
         \caption{Conservative form, stable}
         \label{fig: HD2Q9 spectrum rho*s}
     \end{subfigure}\hspace{0mm}
     
     \begin{subfigure}[b]{1.\textwidth}
         \centering
         \includegraphics[scale=1.]{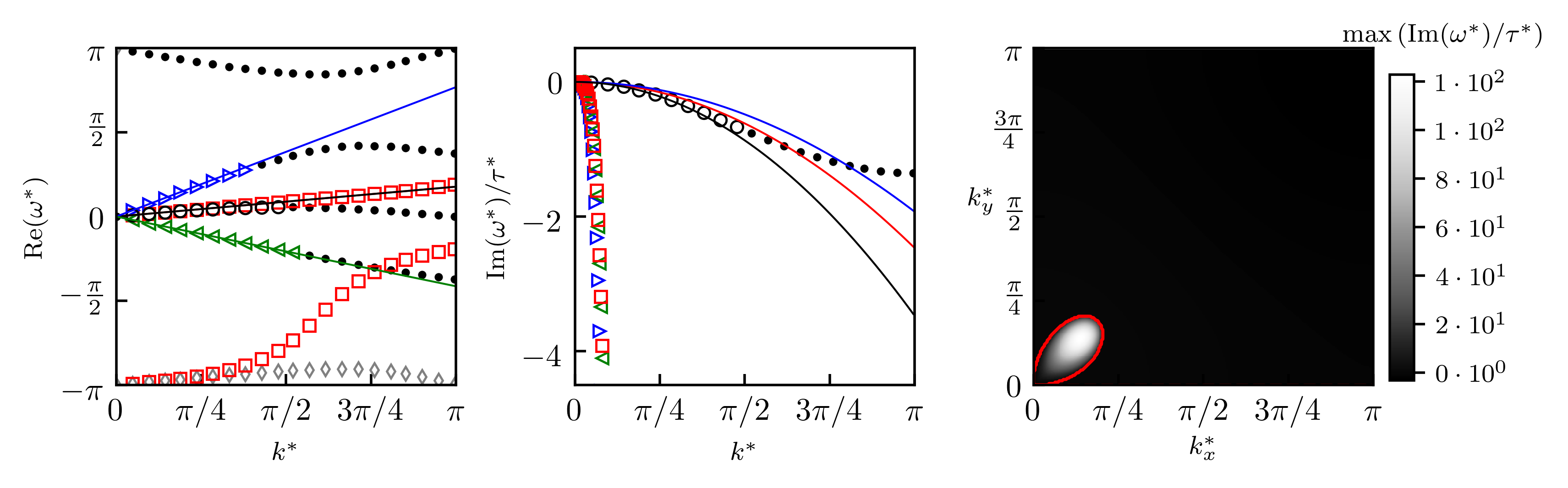}
         \caption{Primitive form, unstable}
         \label{fig: HD2Q9 spectrum s}
     \end{subfigure}\hspace{0mm}
     
        \caption{Spectral properties of the D2Q9 RK2CO2 HRR on entropy for : conservative form (a) and primitive form (b). In the present case, $\overline{\mathrm{Ma}}=0.3$, $\mathrm{Ma}_\theta=0$, $\theta=0.75$, $\tau^*=1.10^{-5}$, $f_i^{eq,3}$ and $\sigma = 0.5$. From left to right : Dispersion graph, dissipation graph (both plotted along $\boldsymbol{e}_x$) and absolute stability map. On this map the solid line represents the unstable regions which corresponds to iso-contour of $\mathrm{max}(\mathrm{Im}(\omega^*)/\tau^* = 0)$.  Symbols correspond to $\omega_{ac+}$: (\protect\acP) ;  $\omega_{ac-}$: (\protect\acM) ; $\omega_{entr}$: (\protect\entrop) ; $\omega_{shear}$: (\protect\shear); $\omega_{msp}$:(\protect\macroSpu) ; $\omega_{g}$:(\protect\ghost) and solid lines correspond to theory.}
        \label{fig: comp s rhos D2Q9 hybrid}
\end{figure}

Let us now focus on the coupled D2Q9 lattice. In Sec.~\ref{subsec: Bi-dimentional lbm}, the two-dimensional athermal models based on the BGK and RR3 collision operators were found to be unstable for Mach numbers of interest in the compressible regime. Among the collision models under consideration, only the use of the HRR one allowed stabilizing the scheme. Consequently, the $\sigma$ parameter turns out to be an essential ingredient in the hybrid context, where compressible applications are targeted. Fig.~\ref{fig: comp s rhos D2Q9 hybrid} displays the spectral properties of the hybrid D2Q9 HRR scheme based on the conservative and primitive forms of the entropy equation for an horizontal mean flow at $\overline{\mathrm{Ma}} = 0.3$. With similar numerical parameters, striking differences can be observed depending on the formulation used. 
First, the $\mathrm{max}(\mathrm{Im}(\omega^*)/\tau^*)$ maps on Fig.~\ref{fig: comp s rhos D2Q9 hybrid} exhibit the stability of the conservative formulation, while the primitive formulation is found unstable. Thus, in addition to the energy variable used, the formulation of the energy equation is a crucial parameter to take into account regarding the stability of the models. 
However, based on these observations, there is no proof that the primitive form is unconditionally unstable. As in Sec.~\ref{subsec: Bi-dimentional lbm}, a wiser choice of the $\overline{\theta}$ or $\sigma$ parameters could lead to a stable scheme. Such an investigation is the purpose of Sec.~\ref{sec: Linear investigation of the HLBM scheme}.
Looking at the dissipation curves, both conservative and primitive models over-dissipate acoustic and shear waves compared to the expected Navier-Stokes behavior. The rate of acoustic dissipation seems to be more impacted in the conservative case than in the primitive one, while the shear dissipation seems to be equivalent. This over-dissipation, also noticed in the athermal cases of Sec.~\ref{subsec: Bi-dimentional lbm}, can be attributed to time and space errors of the regularized scheme~\cite{wissocq2020linear}. However, it is noteworthy that the entropic wave of the primitive case does not seem to suffer from the same over-dissipation and correctly recovers the NS expectations. This suggests that the numerical errors of the LB scheme only impact the entropic wave in the conservative case, while usual low-dissipation properties of the RK4CO2 scheme are recovered in the primitive case. 

Last, the acoustic waves are carried at more than $\eta=99\%$ for very low wavenumbers only in the conservative case (less than 16 points per wavelength for $\omega^*_{r,ac+}$), while in the primitive case, they are identified up to 6 points per wavelength. Thus, if a wave is under-resolved, any combination of macroscopic modes (shear, acoustic and entropic) can be carried at an acoustic group velocity close to the speed of sound, which could be damaging for a simulation.
Despite all these observations, this brief preliminary study does not make it possible to choose between the two formulations, but only to note their main differences. It therefore reinforces the relevance of a parameter study on both primitive and conservative formulations, as proposed in Sec.~\ref{sec: Linear investigation of the HLBM scheme}.

Finally, although the hybrid system has been shown stable for continuous equations regardless the energy variables and the primitive/conservative formulation used, this is far from being the case once the equations are written in a discrete form. 
Throughout this section, extremely different behaviours have been observed only changing these two parameters, reflecting the non-triviality of the problem. 
In addition, it can be expected that the FD scheme applied to the energy equation along with the parameters $\overline{\theta}$ and $\sigma$ can also alter the dispersion, dissipation and the macroscopic content of the waves propagated, further increasing the scope of the study.
Before going deeper into the analysis of these schemes, a first sorting must be done to rule out unusable combinations. 
The following section will therefore be dedicated to a simple stability study of these schemes to establish a certain trend for the possible stable configurations and the possible causes of instability.


\section{Linear investigations of the HLBM scheme}
\label{sec: Linear investigation of the HLBM scheme}

In this section, the potential of different classes of hybrid schemes is discussed. All of the HLBM schemes under consideration are based on the HRR collision operator, and differ by the choice of: (1) the energy variable ($e$, $E$ or $s$), (2) the formulation of the energy equation (primitive or conservative form), (3) the FD scheme applied to the energy equation (RK1UPO1 or RK4CO2) and (4) whether pressure work is considered. This leads to the study of twenty different models in one and two dimensions over the parametric space described in Sec.~\ref{subsec: Extended analysis to HLBM scheme}. The different models are first sorted from a stability viewpoint, then the parameter space of the most linearly stable classes is discussed.

\subsection{Stability results of various HLBM classes}
\label{subsec: Stability of the various HLBM classes}

The stability of the one- and two-dimensional hybrid classes is investigated over the parameter space described in Sec.~\ref{subsec: Extended analysis to HLBM scheme}. 
Linear studies are conducted for: $\tau^* \in [5.10^{-8}, 5.10^{-1}]$ discretized with sixteen logarithmically-spaced values, twelve evenly-spaced values of $\overline{\theta} \in[0.1, 1.2]$ and ten evenly-spaced values of $\sigma \in [0,1]$. The procedure described in Sec.~\ref{subsec: Bi-dimentional lbm} is employed, where the space formed by $[\overline{\theta}, \sigma]$ is scanned to find the maximum Mach number ensuring linear stability. 
For the two-dimensional models, the orientation of the mean velocity field $\mathrm{Ma}_\theta$ is scanned for thirteen evenly distributed values between $0^\circ$ and $45^\circ$. The wavenumber space is defined such as $k_x^* \in [-\pi, \pi]$ and $k_y^* \in [0, \pi]$, respectively discretized using $160$ and $80$ points. For one-dimensional cases, the wavenumber space is composed of $80$ evenly-spaced points for $k^* \in [0, \pi]$.\newline 

\begin{figure}[ht]
     \centering
     \begin{subfigure}[b]{0.49\textwidth}
         \centering\hspace{0mm}%
         \includegraphics[scale=1.]{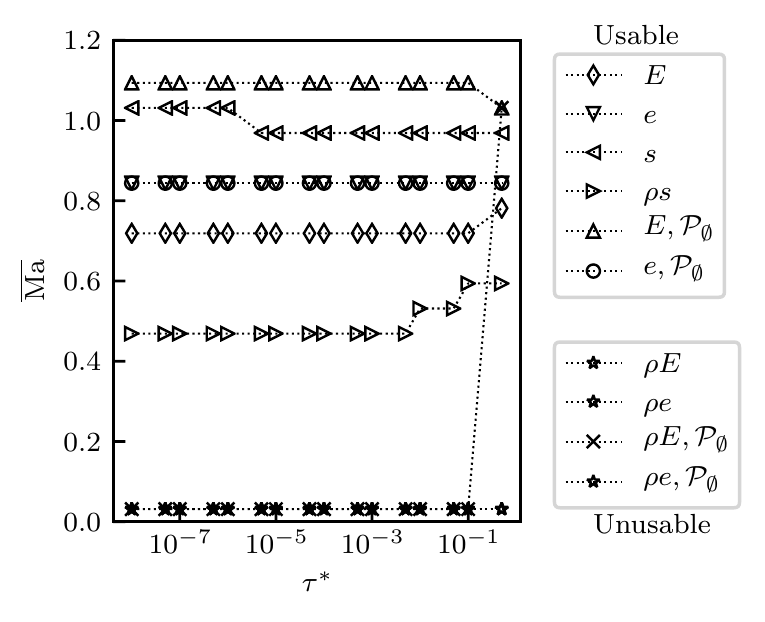}
         \caption{D1Q3, RK1UPO1.}
         \label{fig: D1Q3, RK1UPO1 param}
     \end{subfigure}\hspace{0mm}%
     \begin{subfigure}[b]{0.49\textwidth}
         \centering
         \includegraphics[scale=1.]{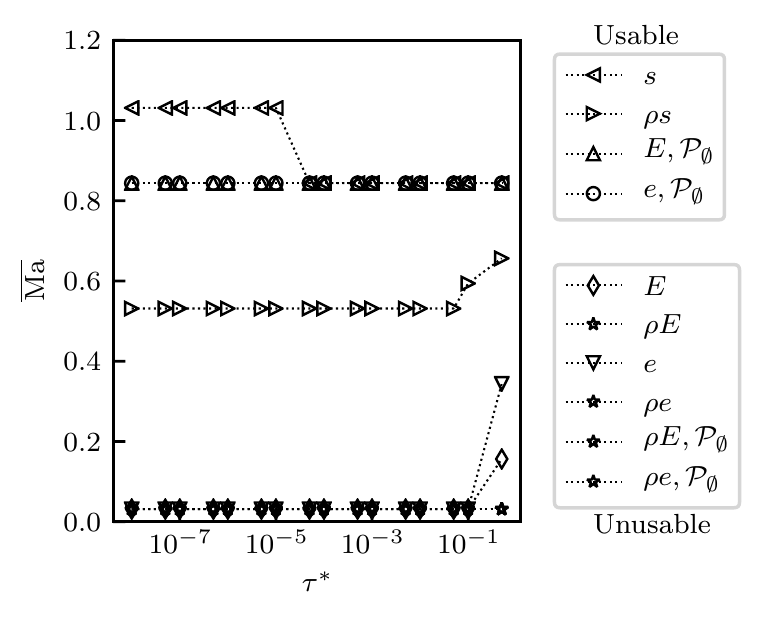}
         \caption{D1Q3, RK4CO2.}
         \label{fig: D1Q3, RK4CO2 param}
     \end{subfigure}\hspace{0mm}
     
     \begin{subfigure}[b]{0.49\textwidth}
         \centering\hspace{0mm}%
         \includegraphics[scale=1.]{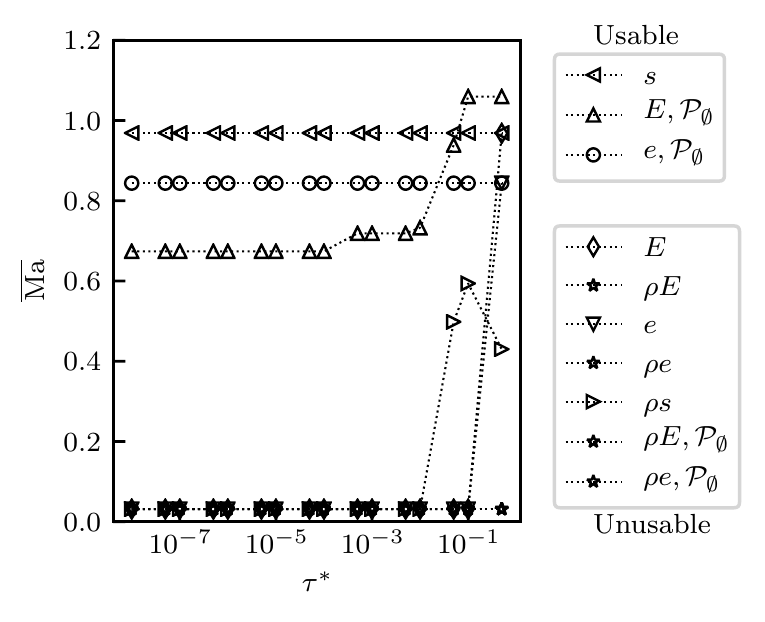}
         \caption{D2Q9, RK1UPO1.}
         \label{fig: D2Q9, RK1UPO1 param}
     \end{subfigure}\hspace{0mm}%
     \begin{subfigure}[b]{0.49\textwidth}
         \centering
         \includegraphics[scale=1.]{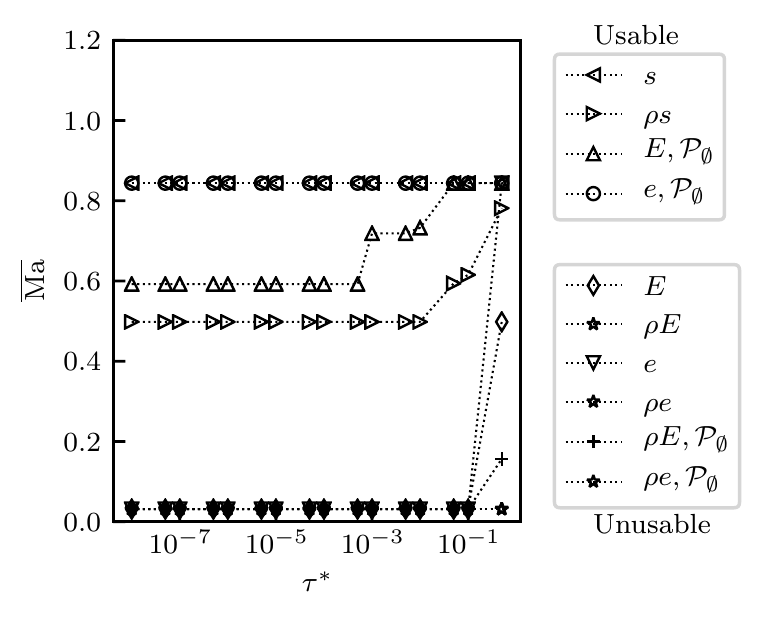}
         \caption{D2Q9, RK4CO2.}
         \label{fig: D2Q9, RK4CO2 param}
     \end{subfigure}\hspace{0mm}%
        \caption{ Maximum stable mean flow Mach number of the hybrid D1Q3 (a)-(b) and D2Q9 (c)-(d) models, investigated over the parametric space $[\overline{\theta}, \sigma ]$ for different relaxation times and energy equations. Here $\mathcal{P}_\varnothing$ denotes the absence of pressure work in the equation. Labels ``Usable'' and ``Unusable'' indicate whether or not the class can be considered for a compressible simulation.}
        \label{fig: All Mach Max param}
\end{figure}

Fig.~\ref{fig: All Mach Max param} displays the maximum reachable Mach number on $[\overline{\theta}, \sigma]$ as function of $\tau^*$ for the twenty different hybrid classes assessed in one and two dimensions. Each class is labelled ``Usable'' if its maximum Mach number exceeds $\overline{\mathrm{Ma}}=0.3$ for all the values of $\tau^*$, meaning that it potentially meets the required standards for compressible simulations, and 
``Unusable'' otherwise.

One-dimensional results are displayed on Fig.~\ref{fig: D1Q3, RK1UPO1 param} and Fig.~\ref{fig: D1Q3, RK4CO2 param} for the RK1UPO1 and RK4CO2 FD schemes respectively. It is worth recalling that the HRR model used with the D1Q3 lattice encompasses the BGK collision model, which is then naturally considered in this study.
First, note that all the classes are restricted to a Mach number close to one. As mentioned in Sec.~\ref{subsec: Bi-dimentional lbm}, this limit can be addressed by modifying the FD scheme of the corrective term~(\cite{renard2020improved}), more details can be found in~\ref{app: Supersonic limit}. 
First, except for the entropy-based classes, all conservative forms are ``Unusable'' whatever the numerical scheme used. It also is noticeable that, working on the conservative form of the entropy equation, the maximum Mach number is found to be lower than its primitive counterpart for the two FD schemes used. This suggests that the conservative formulation may induce stronger instabilities, hinting towards the use of a primitive form in terms of robustness. 
Secondly, classes based on internal and total energy are clearly impacted by the change in the numerical scheme of the energy equation. These forms are found stable for the RK1UPO1 scheme only, which is a first-order scheme, therefore more dissipative. However, once the pressure work is removed from these equations, larger Mach numbers can be reached with both FD schemes. This gain in terms of stability explains the choice made in some previous work to neglect the pressure work in the energy equation~\cite{feng2018regularized,Feng2018}, leading to stable systems with a perfect gas equation of state. 
In addition, no noticeable effect of the discrete scheme used in the pressure work computation has been observed. This indicates that the pressure work is an intrinsic cause of instability. This observation is reinforced by better stability properties of the $\rho s$-based energy equation, which does not require an explicit computation of the pressure work (see Eq.~(\ref{eq: TP dans s})). 

Two-dimensional results are displayed on Fig.~\ref{fig: D2Q9, RK1UPO1 param} and Fig.~\ref{fig: D2Q9, RK4CO2 param}, for the RK1UPO1 and RK4CO2 schemes.
In the athermal case of Sec.~\ref{subsec: Bi-dimentional lbm}, the two-dimensional model was found much more restrictive than the one-dimensional one, which is due to a larger number of degrees of freedom and the existence of an additional physical shear wave. The same conclusions are drawn from this parametric study on the hybrid classes. Only four of them are found ``Usable'' using the D2Q9 lattice: the models based on the variables $(s)$, $(\rho s)$, $(E,\mathcal{P}_\varnothing)$ and $(e,\mathcal{P}_\varnothing)$. Henceforth, no primitive nor conservative classes including the pressure work are found stable. Only the entropy-based classes are found 
``Usable'' in a full compressible context, which remains true for two FD schemes with the primitive formulation. From these results, the same conclusions as the one-dimensional study can be drawn, namely that:
\begin{itemize}
\item classes owning an explicitly pressure work are unstable,
\item classes based on the conservative formulation are less robust than the primitive one.	
\end{itemize}
A conclusion of practical interest can then be drawn: the only ``Usable'' classes for real compressible applications (including pressure work) turn out to be based on the entropy equation.

\subsection{Stability range of HLBM entropic classes}
\label{subsec: Stability range of HLBM entropic classes}

\begin{figure}[h!]
\centering
         \includegraphics[scale=1.]{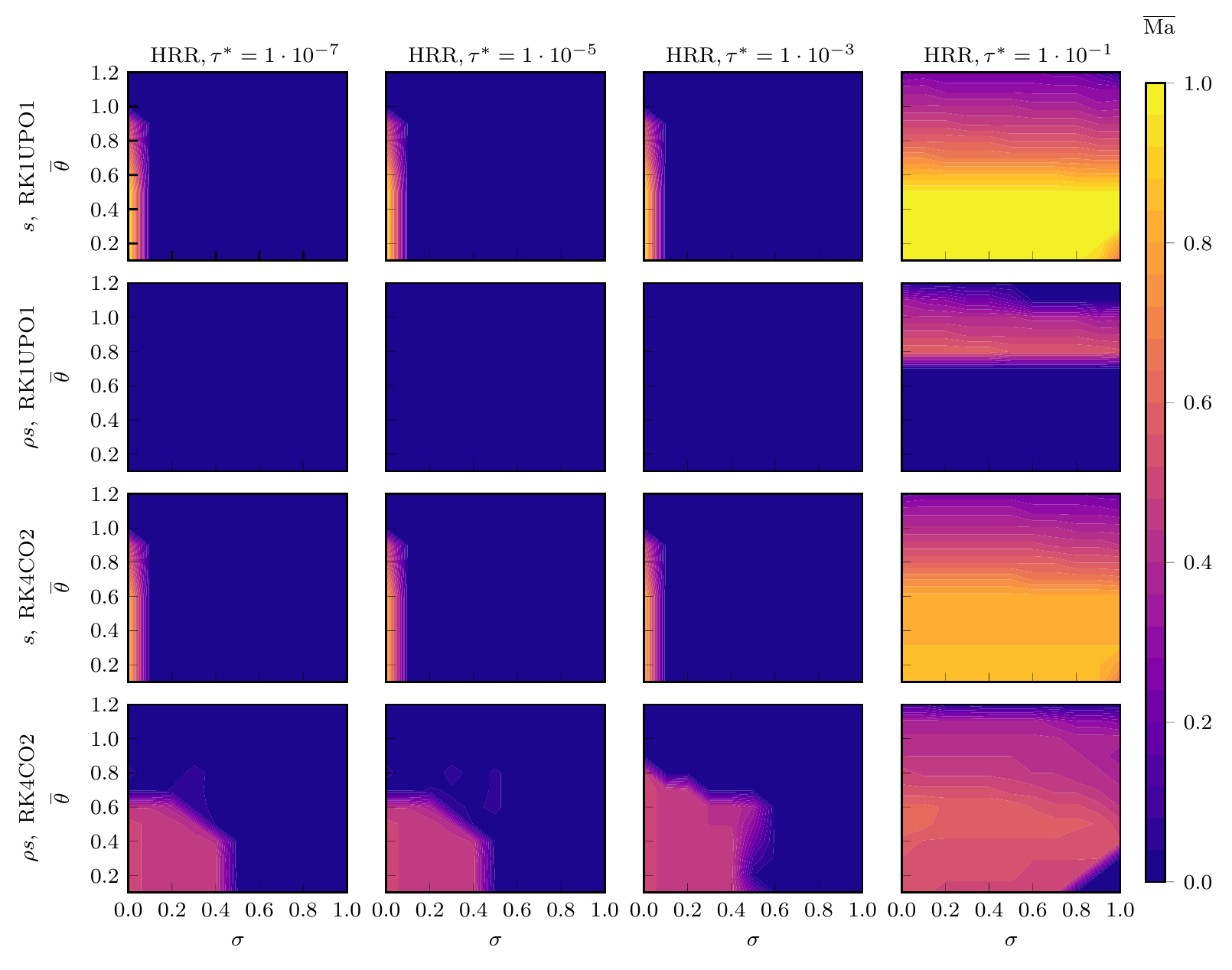}
        \caption{Maximum achievable Mach number of the D2Q9 HRR models based on primitive and conservative entropy equations, acting on both $\overline{\theta}$ and $\sigma$ stabilization levers, and for the RK1UPO1 and RK4CO2 schemes.}
        \label{fig: s, rhos hybrid RK1UPO1 RK4CO2 D2Q9 maps}
\end{figure}
This first sorting being made, the linear analyses of the entropy-based models can be further studied. From now on, only the classes based on $(s)$, $(\rho s)$ are considered. Fig.~\ref{fig: s, rhos hybrid RK1UPO1 RK4CO2 D2Q9 maps} displays the maximum Mach number allowed by each class on the parameter space $[\overline{\theta}, \sigma]$ with the RK1UPO1 and RK4CO2 schemes, for four typical values of $\tau^*$. 

First, the primitive form has the same behavior whatever the FD scheme used. It is found stable for $\overline{\theta} \in [0.1, 0.8]$ yet for the single value $\sigma=0$ only. This point is extremely restrictive since, for low values of $\sigma$, under-resolved shear and acoustic modes are highly attenuated, as shown in Sec.~\ref{subsec: Extended analysis to HLBM scheme}. Nevertheless, these schemes are found stable and ``Usable'' for compressible applications, with a limit of $\overline{\mathrm{Ma}}=0.85$ for the RK4CO2 scheme and $\overline{\mathrm{Ma}}=0.95$ for the RK1UPO1 one. As the RK1UPO1 scheme is more dissipative than the RK4CO2 one, this gain in robustness in the upwind case is however to be paid in terms of precision. 

Regarding the conservative form, the RK1UPO1 scheme is found to be Unusable (as supported by Fig.~\ref{fig: D2Q9, RK1UPO1 param}) except for high values of $\tau^*$ which evinces the non-viability of this class. Concerning the RK4CO2 scheme, it is found to be much more tolerant in permissible values of $\sigma$ compared to the primitive form. This parameter can be set up to $\sigma = 0.4$ even for low values of $\tau^*$. However, it is only Usable for $\overline{\theta} \in [0.1;0.6]$ which is a bit more restrictive than the primitive form since it yields lower usable time steps. This indicates that the conservative formulation is stable for a lower CFL than the primitive form. Furthermore, for a low value of $\tau^*$, the maximum Mach number is restricted to $\overline{\mathrm{Ma}}=0.5$ which is very constraining for compressible applications. 

Finally, for $\tau^*=10^{-1}$ the stability range of the classes is greatly improved, this trend being already observed for the athermal model of Sec.~\ref{subsec: Bi-dimentional lbm}. Thus, for large values of $\tau^*$, both models allow values of $\sigma$ close to unity, which helps recover acceptable dissipation properties. In addition, according to Fig.~\ref{fig: tauAdim_funcDx_multiNu}, it is recalled that for a fixed viscosity, $\tau^*$ is inversely proportional to $\Delta x$. It might then seem wise to adapt the value of $\sigma$ according to the local mesh size: in coarse regions, a value $\sigma \approx 0$ can be advised to enhance numerical stability (yet at the cost of an over-dissipation), while $\sigma \approx 1$ can be adopted in more refined zones. 

To summarize this section, only three ``Usable'' classes, all based on the entropy variable, can be retained for compressible applications, out of the twenty models initially proposed. The first one deals with the conservative equation discretized with the RK4CO2 scheme while the two other ones lie on the primitive form discretized with the RK1UPO1 and RK4CO2 schemes respectively.  
Even though some advantages of one model over the other have been discussed above, a more in-depth study is now mandatory to clearly direct a choice. This task is proposed in Sec.~\ref{sec: HLBM on entropy}.


\section{LBM coupled with an entropy equation}
\label{sec: HLBM on entropy}

In the previous section, an absolute parametric stability study of twenty HLBM classes showed that the RK4CO2 scheme on $\rho s$ and $s$ and the RK1UPO1 scheme on $s$, were the only three stable models for compressible applications. 
In this section, the major flaws of both of these conservative and primitive classes will be investigated thanks to the extended von Neumann Analysis described in Sec.~\ref{subsec: Analysis of HDVBE scheme}.
The observations evidenced by the LSA will be assessed using an in-house C++ code for canonical test-cases.
Note that all the results presented in this section with the primitive form of the entropy equation can be recovered with both the RK1UPO1 and the RK4CO2 schemes. For this reason, and for the sake of clarity, only the RK4CO2 scheme will be detailed hereafter.

\subsection{Conservative form: shear-to-entropy production}
\label{subsec: Conservative instability}

A HLBM involving the conservative form of the entropy equation, discretized with the RK4CO2 scheme, is considered in this section. 
The advanced von Neumann analysis, introduced in~\cite{wissocq2019extended} and adapted to thermal cases in Sec.~\ref{subsec: Method and concepts HDVBE}, makes it possible to investigate the macroscopic behaviour of the HLBM in terms of modal composition.
Without loss of generality, the analyses of this section are performed for $\overline{\mathrm{Ma}}=0.1$, $\tau^*=1.10^{-5}$ and $\sigma=0.5$. Furthermore, note that the properties highlighted below seem to concern this model whatever the considered Mach number and relaxation time.

\begin{figure}[h!]
     \centering
     \begin{subfigure}[b]{1.\textwidth}
         \centering\hspace{0mm}%
         \includegraphics[scale=1.]{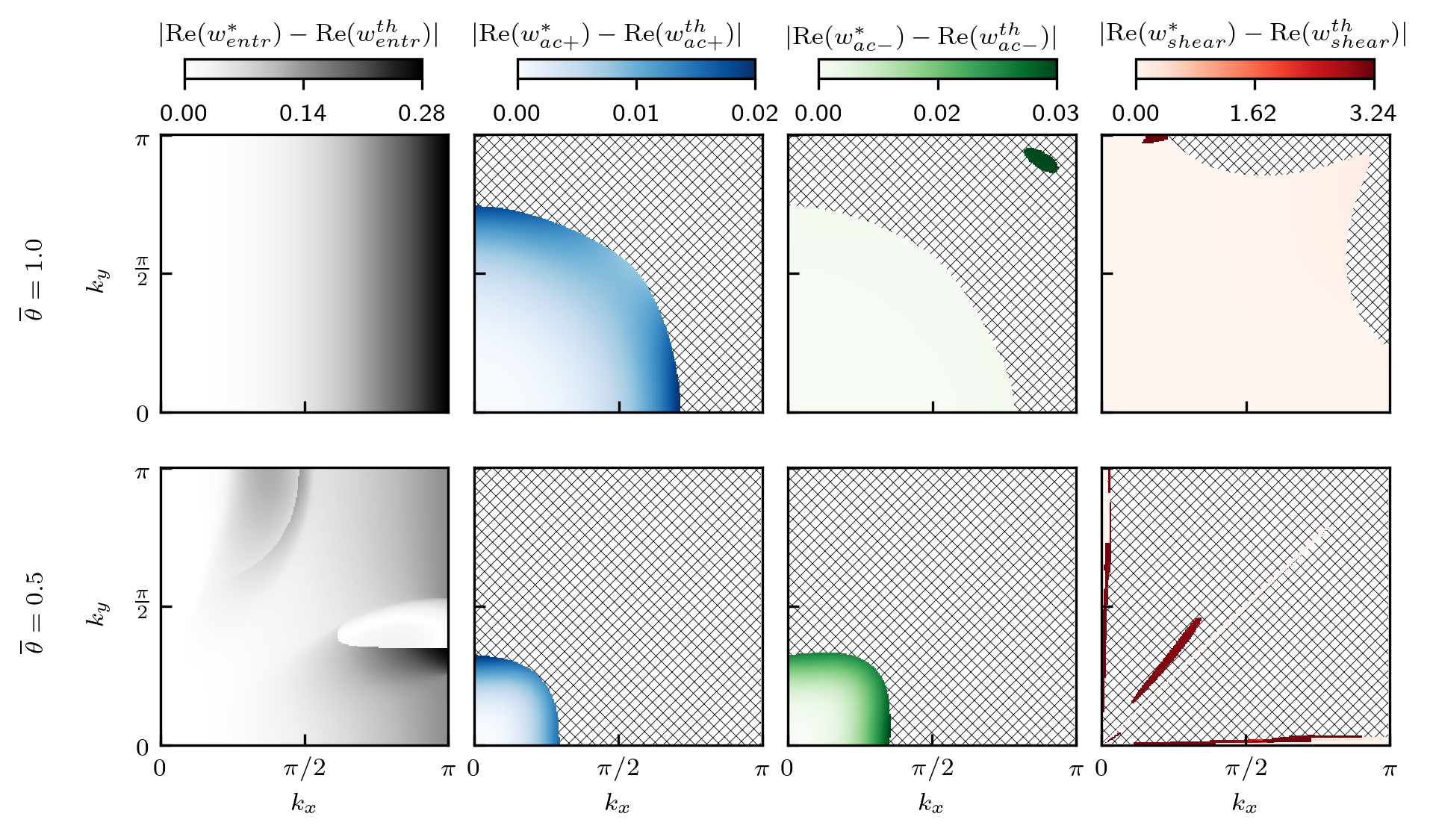}
         \caption{Maps of absolute dispersion absolute error, $\eta=0.9$.}
         \label{fig: rhoS Map2D and graph a}
     \end{subfigure}\hspace{0mm}

     \begin{subfigure}[b]{1.\textwidth}
         \centering\hspace{0mm}%
         \includegraphics[scale=0.95]{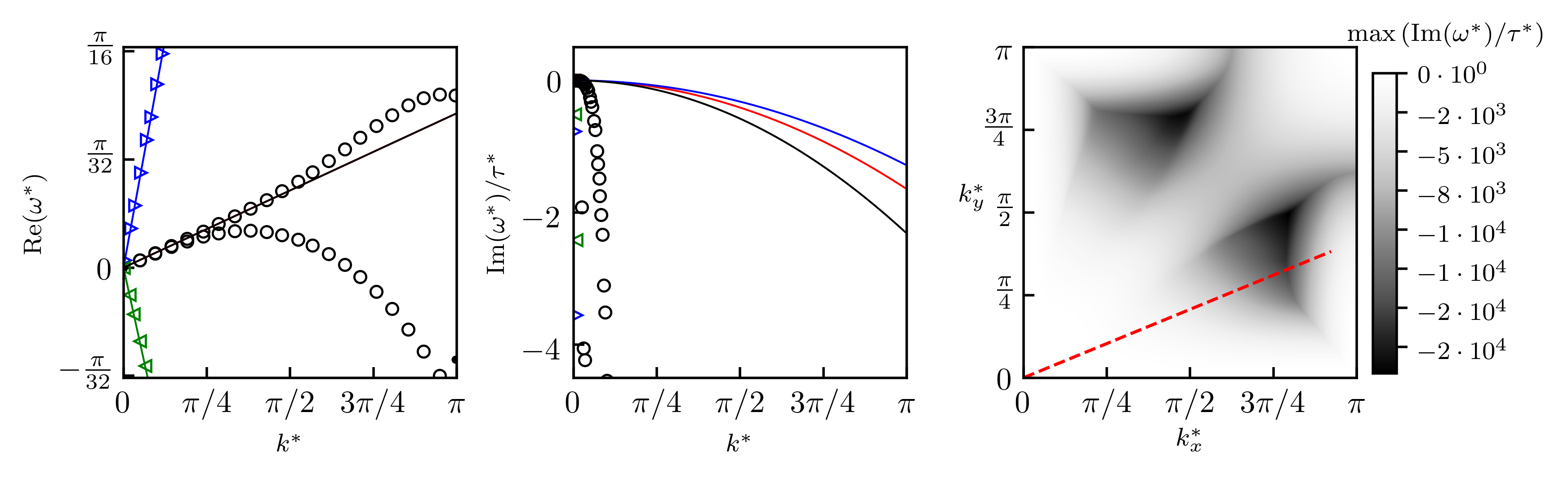}
         \caption{Spectral Properties for $\overline{\theta}=0.5$, $\eta=0.9$}
         \label{fig: rhoS Map2D and graph b}
     \end{subfigure}\hspace{0mm}
        
        \caption{D2Q9 HRR RK4CO2 scheme based on the conservative entropy equation for $\sigma=0.5$, $\overline{\mathrm{Ma}}=0.1$ and $\tau^*=10^{-5}$. Fig.~(a) displays the absolute dispersion error for $\overline{\theta}=1$ (top) and $\overline{\theta} = 1/2$ (bottom). From left to right: entropic, upstream acoustic, downstream acoustic, shear mode ; Dashed regions corresponds to zones where the physical wave is not identified with $\eta=0.9$. Fig.~(b) displays the spectral properties for $\overline{\theta}=1/2$ from left to right : dispersion curve, dissipation curve and absolute stability map. On the latter, the dashed line represents the line along which dispersion and dissipation curves are plotted. Symbols correspond to $\omega_{ac+}$: (\protect\acP) ;  $\omega_{ac-}$: (\protect\acM) ; $\omega_{entr}$: (\protect\entrop) ; $\omega_{shear}$: (\protect\shear); $\omega_{msp}$:(\protect\macroSpu)  and solid lines correspond to the NS theory.}
        \label{fig: rhoS Map2D and graph}
\end{figure}

Fig.~\ref{fig: rhoS Map2D and graph a} displays maps of absolute dispersion error for the four physical modes, $\mathrm{Re}(\omega_{l}^{th})$, $l \in [entr,ac+,ac-,shear]$ being the Navier-Stokes theoretical values. These maps are proposed for $\overline{\theta}=1$ (top) and $\overline{\theta}=0.5$ (bottom). The tolerance parameter is set to $\eta=0.9$, meaning that only modes containing more than $90\%$ of one of the Navier-Stokes waves are identified, a dashed region is represented otherwise. Note that, if several modes of the HLBM system carry the same macroscopic information, only the one with the larger dissipation rate $\omega_i$ is considered. This criterion explains the discontinuities observed on the maps, which are caused by a curve veering phenomenon~\cite{wissocq2019extended} involving the physical and the spurious macroscopic wave located close to $\mathrm{Re}(\omega^*)=\pi$, as illustrated on Fig.~\ref{fig: rhosQ3} for the downstream acoustic mode.
 
At first, for $\overline{\theta}=1$, both acoustic and shear modes are identified over a large part of the spectral space. Regarding the entropy mode, it is detected all over the wavenumber space. 
For this value of $\overline{\theta}$, all Navier-Stokes Fourier modes being recovered from low to moderately large wavenumbers, the ability of this model to recover the intended macroscopic behavior cannot be questioned. 

A radical change can be observed in the case $\overline{\theta}=0.5$. The acoustics mode are identified for much lower wavenumbers than with $\overline{\theta}=1$, \textit{i.e.} for $||\boldsymbol{k}|| < \pi/4$ only. This is, however, not too restrictive since acoustic waves with more than four points per wavelength are still correctly considered. Regarding the entropy wave, some discontinuous areas can be observed, suggesting that an abrupt change of mode with a lower dissipation rate is detected. Finally, and more importantly, no shear wave can be identified with $\eta=0.9$ on the majority of the spectral space, even for well-resolved wavelengths. Only entropy waves travelling in some particular directions, aligned with that of the lattice velocities, seem not to suffer from this apparent strong disease.

To further investigate this deficiency, Fig.~\ref{fig: rhoS Map2D and graph b} displays dispersion and dissipation curves for waves for which $\widehat{(k_x,k_y)}=22.5^\circ$. Instead of one entropy and one shear mode, two entropy modes are identified along this direction. It indicates an inablity of this HLBM system to simulate a pure shear wave in this direction. Indeed, if one would like to initialize such a model with a shear wave, it would unavoidably be converted into two entropy waves, thus generating spurious temperature fluctuations. It is important to note that this phenomenon does not appear in the one-dimensional case and is specific to the multi-dimensional one. Finally, this surprising observation explains the appearance of discontinuous zones on the entropy map: the unintended coexistence of two entropy modes gives birth to abrupt changes of the less attenuated one, represented on Fig.~\ref{fig: rhoS Map2D and graph b}. 

\begin{figure}[h!]
\centering
         \includegraphics[scale=0.8]{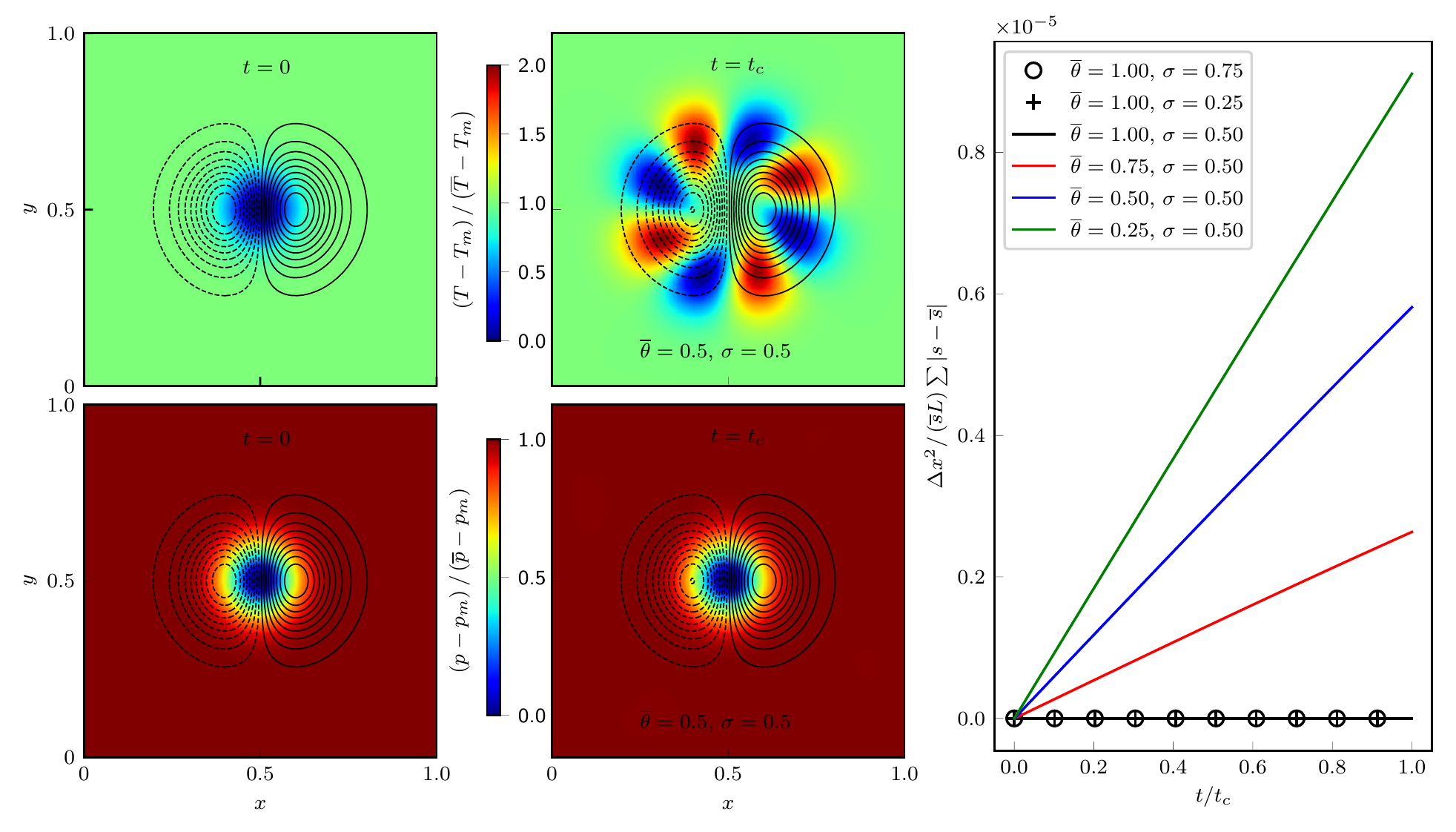}
        \caption{Isentropic vortex for $t=0$ (left) and $t=t_c$ (middle) colored by the reduced temperature (top) and pressure (bottom) with $T_m$ and $p_m$ referring to the minimum temperature and pressure at $t=0$, respectively. Twenty iso-contours ranging from minimum to maximum of $u_y$ are displayed ; solid and dashed contours correspond respectively to positive and negative values. Right: dimensionless entropy variation integrated over the numerical domain over the time, for different simulations parameters $\overline{\theta}$ and $\sigma$. $\overline{s}(\overline{\rho},\overline{T})$ corresponds to the mean flow entropy.}
        \label{fig: Vortex rhos}
\end{figure}

To illustrate the consequences of this spectral defect on the shear wave, a vortical convection test-case is considered with the same parameters as the LSA study; \textit{i.e.} $\overline{\mathrm{Ma}}=0.1$ along $\boldsymbol{e}_x$ and $\tau^*=10^{-5}$.
Even though the convected isentropic vortex is based on the Euler equations, it has been extensively used to assess LBM schemes solving the Navier-Stokes equation~\cite{saadat2019,Frapolli2016a,Feng2019,renard2020improved}. The initialization can be seen as a small perturbation compared to a mean flow $\overline{\rho}, \overline{T}, \overline{u}$ which reads: 

\begin{equation}
u_r(r)=0 \quad \mathrm{and} \quad u_\theta(r)=\overline{c} \, \mathrm{Ma}_v \, r\exp \left( \frac{1-r^2}{2}\right),
\label{eq: V covo}
\end{equation}
where $u_r$ and $u_{\theta}$ are respectively the radial and tangential velocity in polar coordinates and $\overline{c}=\sqrt{\gamma_g r_g \overline{T}}$ the mean flow speed of sound. $\mathrm{Ma}_v$ is the vortex Mach number and $r=\sqrt{ \left(x-x_c\right)^2  + \left(y-y_c\right)^2 } / R$ with $R$ the vortex radius and $(x_c, y_c)$ the initial position of the vortex center. For an isentropic vortex, the density and pressure fields are given by~\cite{wang2013high}: 
\begin{equation}
\rho(r)=\overline{\rho}\left[1-\frac{\gamma_g-1}{2}\mathrm{Ma}_v^2r\exp{(1-r^2)}\right]^{\frac{1}{\gamma_g-1}}.
\label{eq: rho covo}
\end{equation}

\begin{equation}
p(r)=\overline{p} \left( \frac{\rho\left(r\right)}{\overline{\rho}} \right)^{\gamma_g}.
\label{eq: p covo}
\end{equation}
In the present case, the vortex velocity is set so that $\mathrm{Ma}_v = 0.05 \, \overline{\mathrm{Ma}}$, where one recalls that $\overline{\mathrm{Ma}}$ denotes the free-stream Mach number. 
A $\left[L\times L \right]$ periodic domain is considered with $L=1\,\mathrm{m}$, uniformly discretized with 200 points in each direction. 
Finally, $R=20\Delta x$, $\overline{p}= 101325\,\mathrm{Pa}$, $\overline{T} = 300\,\mathrm{K}$, $\overline{\rho} = \overline{p}/(r_g \overline{T})$. The simulation is run for $4140$ iterations, which corresponds to a single convective time $t_c = L / ( \overline{c} \, \overline{\mathrm{Ma}} )$ in the case $\overline{\theta} = 0.5$. 
 
Fig.~\ref{fig: Vortex rhos} displays maps of reduced temperature and pressure, of a simulation run with $\overline{\theta} = 0.5$ and $\sigma=0.5$, for $t=0$ and and after one convective time $t=t_c$. One recalls that the reference solution corresponds to the initialization state for this test-case. Iso-contours of $u_y$ are superimposed on the two fields. At first, they are found in good agreement with the analytical solution, and no noticeable distortion of the velocity field is observed. Same conclusion stands for the pressure field where nor dispersion, dissipation or spurious acoustic waves are observed. 
If one confines oneself to these observations, the model seems to well reproduce the physics of the test-case. 
Regarding the temperature field, one would expect a similar map as in the initial state, only eventually dissipated, which is, however, not recovered by the model. Indeed, large deformations of the temperature field, involving wells and sources, can be observed. A similar phenomenon, in inverse proportion, is also noticed on the density field as a consequence of the perfect gas coupling equation of state. Thus, based on the evolution of thermodynamic variables, one concludes that unintended entropy waves are generated by the system, a totally unexpected feature for an isentropic test case.

This behavior is a direct consequence of the spectral discrepancy of the shear wave mentioned above. Since this HLBM model is not able to properly simulate a shear wave in every direction, an unavoidable conversion to an entropy wave occurs, at the origin of the highlighted non-physical temperature fluctuations. Moreover, one can notice that in the directions of the lattice, no wells nor entropic sources are observed, which supports the previous analysis. 
Finally, the right-hand-side curve of Fig.~\ref{fig: Vortex rhos}, displaying the integrated entropy variation over the time, confirms that the value of $\overline{\theta}$ is directly involved in this phenomenon. On the contrary, changing the values of $\sigma$ does not seem to cause such a disease: as far as $\overline{\theta}=1$, no entropy production is observed. 
Therefore, the lower $\overline{\theta}$ is, the higher the entropy production occurs. As a consequence, the simulated physics is biased, especially if it includes shear flows. 
Unfortunately, in order to keep the model within its stability range (see Sec.~\ref{subsec: Stability range of HLBM entropic classes}), and thus usable for a compressible application, only low $\overline{\theta}$ values are allowed, which will inevitably significantly bias the physics.

Note that all the phenomena described in this section have been observed for many different values of the mean flow Mach number, its orientation and the value of $\tau^*$. In any case, observations done with the extended von Neumann analysis, and corroborated by numerical simulations, showed that this HLBM class, based on the conservative entropy equation, is not recommended for compressible applications, due to:
\begin{itemize}
    \item a linear instability for large values of $\overline{\theta}$,
    \item a shear-to-entropy production for lower values of $\overline{\theta}$.
\end{itemize}

Note that a similar phenomenon as pointed out on Fig.~\ref{fig: Vortex rhos} was observed by Guo~et~al.~\cite{guo2020efficient} for a comparable vortical convection case. Using a D3Q19 HLBM based on a conservative equation on $\rho s$, the authors noticed a large deformation of the density field, whereas the same model using a primitive formulation of the entropy equation did not suffer from this discrepancy. The deformation observed with the conservative formulation is thus explained by the shear-to-entropy transfer exhibited in this section. 
Moreover, spurious mode couplings were also evoked by Lallemand~et~Luo~\cite{lallemand2003hybrid, lallemand2003theory} with a multi-speed D2Q13 model. They noticed an unstable coupling between the shear and entropic modes and proposed a hybrid model to overcome the issue. 
In the light of the present modal study, it would appear that the form of the supplemental equation is at least as important as the choice of dealing with a hybrid method to address this problem. Indeed, such a mode coupling is equally observed on the present hybrid model based on a conservative form of the entropy equation, while Lallemand \& Luo adopted a primitive form of the temperature equation to get rid of this issue.
Note that, in the context of the present work, similar analyses and test case have been performed with the primitive form of the entropy equation discretized with the RK4CO2 scheme, and no such problems have been identified. This point out that such a mode transfer does not regard the primitive form. Nevertheless, the latter is affected by a phenomenon of another nature, which will be presented in the following section.

\subsection{Primitive form: entropy-to-shear production}
\label{subsec: Entropic transport issue}

The primitive form of the entropy equation, discretized with the RK4CO2 scheme, is now considered in this section. The same study as in Sec.~\ref{subsec: Conservative instability} is conducted, where the scheme is analyzed in terms of modal composition. 
Using the advanced von Neumann analysis on several points of the parametric space, a rather similar issue as that observed with the conservative form occurs, here referred to as an entropy-to-shear production. 
To illustrate this phenomenon, and without loss of generality, the analysis is proposed for the same flow parameters as in Sec.\ref{subsec: Conservative instability}, \textit{i.e.} $\overline{\mathrm{Ma}}=0.1$, $\tau^*=10^{-5}$ and $\overline{\theta}=0.5$.

\begin{figure}[h!]
\centering
\includegraphics[scale=1.]{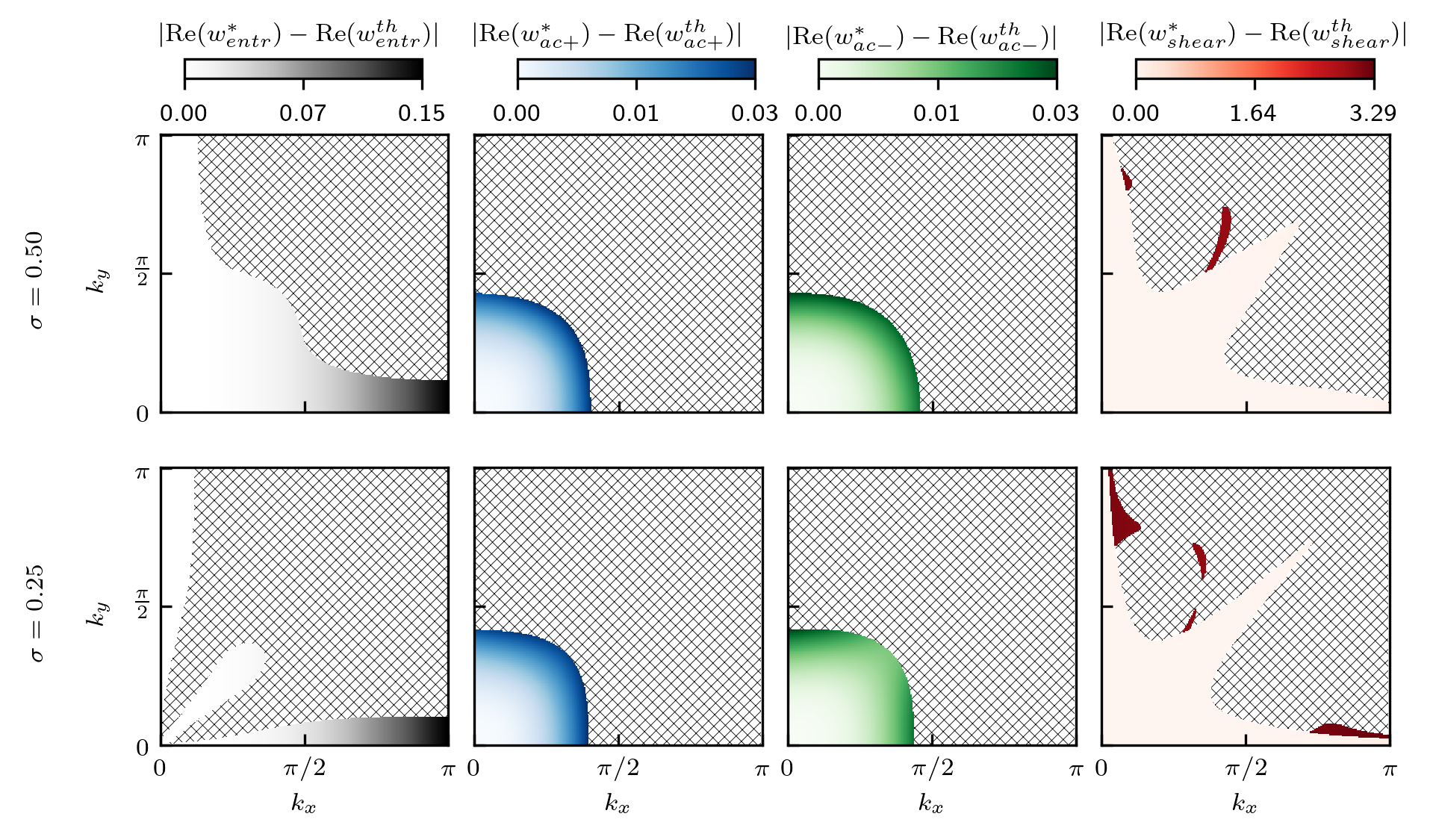}
\caption{D2Q9 HRR RK4CO2 on primitive entropy equation for $\overline{\theta}=0.5$, $\overline{\mathrm{Ma}}=0.1$ and $\tau^*=10^{-5}$. Maps of absolute dispersion error for $\sigma=0.5$ (top) and $\sigma = 0.25$ (bottom). From left to right: entropy, downstream acoustic, upstream acoustic and shear mode ; Dashed regions indicate zones where the physical wave is not identified with $\eta=0.99$.}         
\label{fig: S Map2D and graph a}
\end{figure}

Fig.~\ref{fig: S Map2D and graph a} displays maps of absolute dispersion error for $\sigma=0.5$ (top) and $\sigma=0.25$ (bottom). In the present case, the tolerance parameter has been set to $\eta=0.99$ in order to evidence the phenomenon of interest. 
With $\sigma=0.5$, entropic, acoustic and shear modes are identified over the majority of the spectral space. Large error areas (as observed in Sec.~\ref{subsec: Conservative instability}) are present for the shear modes, but for high wavenumbers only.
Essentially, the ability of the present configuration to correctly simulate a Navier-Stokes Fourier behaviour cannot be questioned. 

Different conclusions can be drawn from the case $\sigma=0.25$. Even though the acoustic and shear modes show little change, no entropy mode is identified on the majority of the spectral space, even for low wavelengths, \textit{i.e.} well resolved fluctuations. In fact, only entropy waves travelling along the directions of the lattice can be correctly considered by this model. This phenomenon is very reminiscent of the shear mode behavior in the primitive case of Sec.~\ref{subsec: Conservative instability}. 

\begin{figure}[h!]
\centering
\includegraphics[scale=1.]{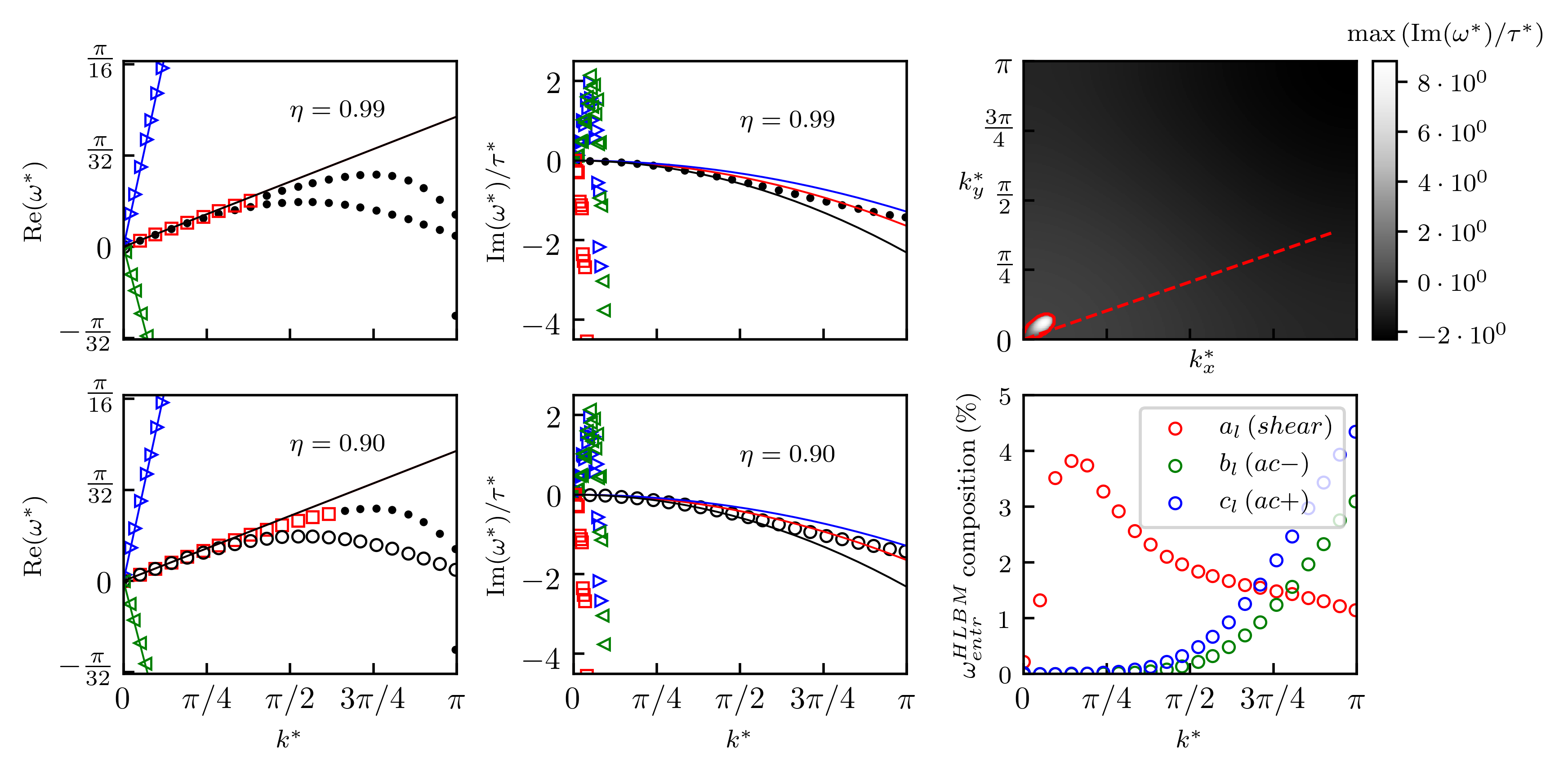}
\caption{D2Q9 HRR RK4CO2 on primitive entropy equation for $\overline{\theta}=0.5$, $\overline{\mathrm{Ma}}=0.1$, $\tau^*=10^{-5}$ and $\sigma=0.25$. Left: dispersion and dissipation curves for two values of $\eta$. Symbols correspond to $\omega_{ac+}$: (\protect\acP) ;  $\omega_{ac-}$: (\protect\acM) ; $\omega_{entr}$: (\protect\entrop) ; $\omega_{shear}$: (\protect\shear); $\omega_{msp}$:(\protect\macroSpu)  and solid lines correspond to the NS theory. Top right: absolute stability map, the dashed line represents the line along which dispersion and dissipation curves are plotted. Bottom right: Entropy HLBM mode decomposition in terms of NS macroscopic contribution Eq.~(\ref{eq: coef NS Fourier}). }
\label{fig: S Map2D and graph b}
\end{figure}

To investigate the deficiency in the hashed area of Fig.~\ref{fig: S Map2D and graph a}, dispersion and dissipation curves for an angle $\widehat{(k_x, k_y)}=22.5^\circ$ are drawn on Fig.~\ref{fig: S Map2D and graph b}. These curves are displayed for two values of the tolerance parameter $\eta=0.99$ and $\eta=0.90$. As expected, acoustic and shear waves are well identified, although unstable in the low $k
^*$ limit, and over-dissipated above. On the contrary, an entropy wave can only be identified by lowering $\eta$ to $90\%$. 
Hence, despite the fact that this mode carries more than $90\%$ percent of an expected entropy wave, a significant percentage of parasitic macroscopic information is also transported.

By construction of the analysis, the macroscopic composition of this mode, in the sense of the compressible NS equations, is directly accessible thanks to Eq.~(\ref{eq: coef NS Fourier}). Normalized coefficients $a_l$, $b_l$ and $c_l$, corresponding to the percentage of shear, upstream and downstream acoustics carried by the entropic HLBM mode respectively, are displayed on the bottom-right curve of Fig.~\ref{fig: S Map2D and graph b}. It can be observed that this mode can carry up to $4\%$ of acoustic and shear information. Its larger values of acoustic contributions are located for $k^*=\pi$, and tend rapidly to zero for well-resolved waves, ensuring the convergence of the mode in terms of entropic composition. However, a more singular behavior is observed regarding the content of shear information ($a_l$). Although it also tends to zero for well-resolved waves, the non-monotony of the curve yields a surprising larger error for some well-resolved waves than under-resolved ones. Its maximum value is found for $k^*=\pi/16$ \textit{i.e.} $16$ points per wavelength.
Similarly to the issue shown in Sec.~\ref{subsec: Conservative instability}, it is important to note that this phenomenon does not appear in the one-dimensional case, and is a specific to the multi-dimensional one. This type of mode transfer is therefore likely to lead to similar adverse effects as those observed with the conservative form.

\begin{figure}[h!]
\centering
         \includegraphics[scale=0.9]{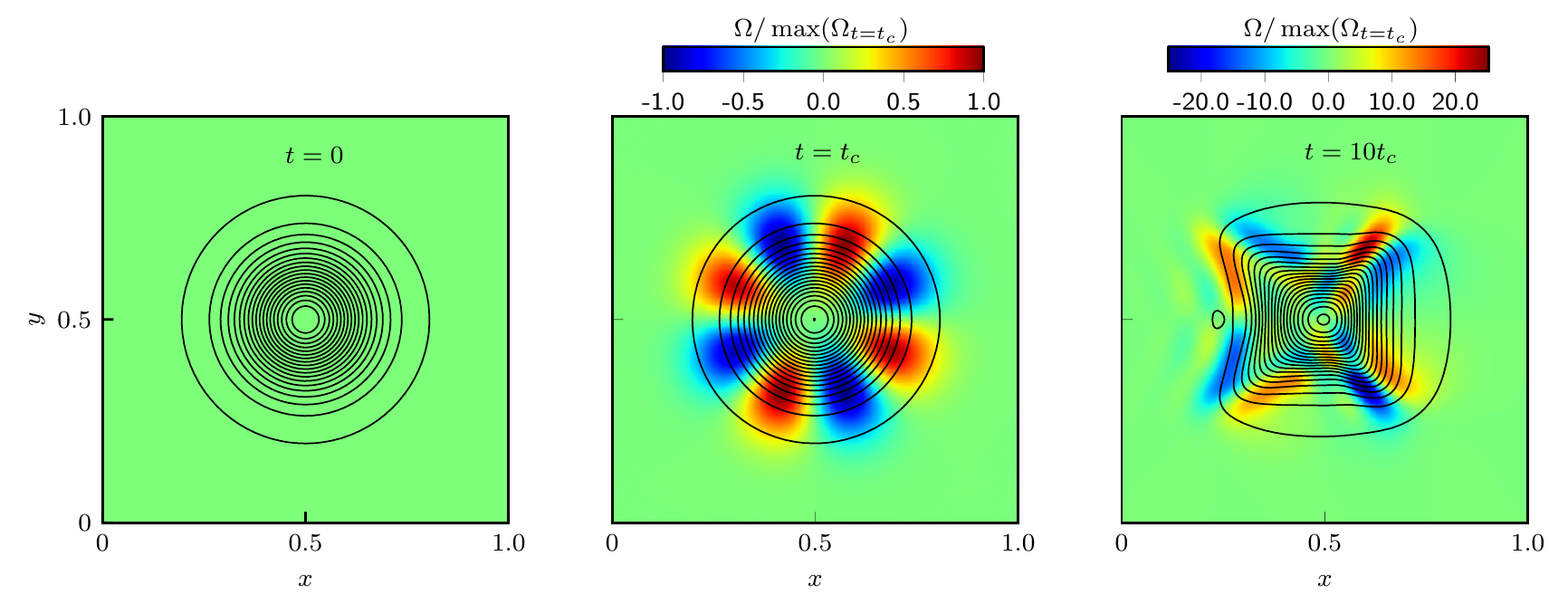}
        \caption{Entropic spot convection for $\theta=0.5$ and $\sigma=0.25$, for $t=0$ (left), after one period of the computational domain (middle) and after ten periods (right). Colormaps of vorticity with twenty iso-contours of temperature ranging from the minimum to the maximum amplitude for $t=0$ are displayed.}
        \label{fig: Entropic spot s}
\end{figure}

To illustrate the consequences of this spectral defect on the HLBM entropy mode, the convection of a Gaussian entropic spot is considered. Its exact solution consists in the simple advection-diffusion of the initial condition over the time. Same parameters as those of the LSA are employed \textit{i.e.} $\overline{\mathrm{Ma}}=0.1$ along $\boldsymbol{e}_x$ and $\tau^*=10^{-5}$.
The fluid is initialized with a small perturbation over the mean flow, convected along $\boldsymbol{e}_x$, which reads: 
\begin{equation}
T(r)= \overline{T} + \Delta T \exp\left(-r^2/2\right) \quad \mathrm{and} \quad \rho(r)= \frac{\overline{p}}{r_g T}.
\label{eq: T, rho entropic spot}
\end{equation}
Here, $\Delta T = 0.05 \overline{T}$ corresponds to the amplitude of the fluctuation. The radius of entropic spot is set to $R=20\Delta x$ and the same mean flow conditions, numerical domain, convective time $t_c$, number of iterations and grid specifications as described in Sec.~\ref{subsec: Conservative instability} are adopted. Given the mesh size and the value of $\tau^*$, the dissipation induced by the heat diffusivity is considered negligible, so that only a hot spot convection is expected.

Vorticity colormaps and temperature iso-contours can be found on Fig.~\ref{fig: Entropic spot s} for the initial state, after one and after ten convective times. One recalls that the vorticity is positive for anti-clockwise rotation and negative for clockwise rotation. 
For $t=t_c$, one can see the emergence of several pairs of azimuthally-distributed counter rotating vortices all around the spot. This phenomenon is similar in all respects to that observed in Sec.~\ref{subsec: Conservative instability}, except that the mode transfer occurs now from entropy to shear production. For such a test case, the entropic spot is supposed to be solely advected, which confirms that the production of vorticity is due to a spurious numerical mode transfer. In addition, along the lattice direction, where the entropic macroscopic information is spectrally recovered, no vorticity is produced, which comforts the results of the LSA. This behavior is a direct experimental observation of the spectral discrepancy of the entropy wave in non lattice-aligned directions. Since this last holds few percents of shear information even for moderate wavenumbers, a spurious shear behavior resulting in the appearance of vortices is observed.  

As a consequence of these spurious vortices, the hot spot is stretched along the diagonal directions and compressed along the $\boldsymbol{e}_x$ and $\boldsymbol{e}_y$ axes. This phenomenon therefore adds errors to both the hydrodynamics and the thermodynamic quantities, where the temperature iso-contours after $t=10t_c$ adopt a square shape. 
Moreover, one can observe that the solution is no longer symmetric with respects to $\boldsymbol{e}_y$. This is directly linked to the dispersion error of the scheme induced by a non zero advection velocity. This dispersion is a common defect for a numerical scheme, and is not related to the observed mode transfer. Nevertheless, it can be mitigated by using a higher-order centered scheme on the convective part of the entropy equation. For a stationary entropic spot, a symmetric solution, but still affected by spurious vortices, would have been observed.

\begin{figure}[h!]
\centering
         \includegraphics[scale=0.9]{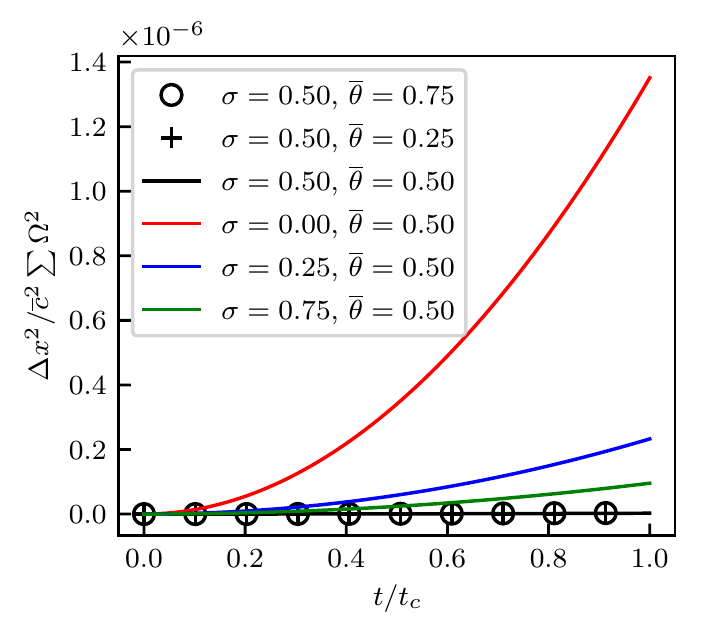}
        \caption{Dimensionless enstrophy variation over the time for different $\overline{\theta}$ and $\sigma$ parameters.}
        \label{fig: Enstrophy entropic spot s}
\end{figure}

Finally, Fig.~\ref{fig: Enstrophy entropic spot s} displays the dimensionless enstrophy variation over the time, up to $t=t_c$. Multiple simulations for several values of $\sigma$ and $\overline{\theta}$ are proposed. The curves confirm that the value of the $\sigma$ parameter is responsible for the entropy-to-shear production, whereas $\overline{\theta}$ has no influence. The independent character of $\overline{\theta}$ on this phenomenon has also been observed through other maps similar to Fig.~\cite{fig: S Map2D and graph a}, but will not be shown for reasons of brevity.

Among the values investigated, the spurious vorticity production is maximum for $\sigma=0$, minimum for $\sigma=0.5$ and non zero for $\sigma=0.75$. This suggests an optimal value $\sigma \neq 1$ ensuring a minimum enstrophy creation together with a stable numerical scheme, which could possibly be found through this analysis for each point of $[\overline{\mathrm{Ma}}, \tau^*]$  space in the spirit of~\cite{lallemand2000theory,xu2011optimal,chavez2018improving}.

At last, a discussion should be held regarding the stability of the model, which is only guaranteed when $\sigma=0$ for low values of $\tau^*$ as evidenced in Sec.~\ref{subsec: Stability range of HLBM entropic classes}. For $\sigma = 0.25$, the dissipation curve and the absolute stability map of Fig.~\ref{fig: S Map2D and graph b}, show that instabilities of the two acoustic modes are responsible for the divergence of the model. 
An instability bubble can be seen for very low wavenumbers, with however a relatively low positive amplification rate of $\omega^*\approx8$. For non-acoustic cases, such unstable low-frequency waves may not be triggered, which makes it possible to perform cases such as vortex or hot-spot convection without any problem~\cite{Feng2019,renard2020improved}. However, this model cannot be used in a straightforward way as is since it is intrinsically unstable. Some improvements will have to be proposed in order to enforce its robustness, especially for the simulation of acoustic cases.

\section{Conclusion}
\label{sec:conclusion}

In this paper, an original von Neumann analysis of the hybrid lattice Boltzmann method under ideal gas thermodynamic closure has been conducted. Inspired by the current state of the art~\cite{Nie2009, feng2018regularized, Feng2019, renard2020improved}, numerous compressible hybrid classes have been derived and investigated. 

A preliminary spectral study on the athermal equations revealed that the Mach error of standard lattices is effectively removed using appropriate corrective terms. Moreover, the latter allows acting on the acoustic CFL constraint by modifying the $\theta$ parameter, without biasing the resolved physics. In the present work, this parameter was one of the two levers that have been introduced to address the eigenvalue collision~\cite{wissocq2019extended} present in the standard LBM. Lowering this parameter reduces the sonic cone and permits to bypass this destabilizing phenomenon, further extending the stability of the LBM to $\mathrm{Ma} > 0.732$. The second lever is the use of the HRR collision operator~\cite{Jacob2019} which makes it possible to alleviate the modal interactions~\cite{Astoul2019} through the FD parameter $\sigma$. In the end, eigenvalue collisions can be avoided if a sufficiently low value of $\sigma$ is considered. Finally, the combination of both levers have been proved to be all the more crucial for bi-dimensional case, ensuring the stability of the LBM system for compressible Mach number ranges.

Once these observations drawn, a von Neumann analysis of the hybrid system corrected in terms of Mach number error and monatomic deficiency~\cite{renard2020improved}, has been conducted. The methodology of Wissocq \textit{et al.}~\cite{wissocq2019extended} adapted in this work to the hybrid system, shown that the compressible Navier-Stokes behavior was spectrally recovered in terms of dispersion, dissipation and modal composition. 
Moreover, in this hybrid context it has been pointed out that the sonic cone is increased by a factor of $\sqrt{\gamma}$, making the eigenvalue collision arising for lower Mach values compared to the athermal case. Thus, it is all the more necessary, at first sight, to use both stabilization levers $\overline{\theta}$ and $\sigma$. 
A basic study of few HLBM classes revealed a completely different spectral behavior compared to the athermal system, which strongly supports the pertinence of this new analysis. Furthermore, variable stability outcomes were obtained depending on the energy variable, type of equation or numerical scheme applied to the energy equation. In that respect, a parametric study of twenty HLBM classes has been conducted to determine "Usable" models for compressible applications, over a space that has been rigorously derived using the Vaschy-Buckingham theorem. It has been observed that classes owning an explicit pressure work are unstable as well as classes based on the conservative formulation, apart from the entropy. Thus, the entropy classes proved to be the only ones "Usable" for compressible applications, and stable up to $\mathrm{Ma}=0.5$ and $\mathrm{Ma}=0.95$ for the conservative and primitive formulations respectively.
This study is therefore in line with the choice of the primitive form of the entropy equation made in the literature~\cite{Nie2009,Feng2019} for the sake of robustness.
It is also worth noting that, in absence of pressure work in the energy equation, other classes turn out to be stable, but they cannot be categorized as "Usable" even though encountered in the literature to models low Mach thermal flow~\cite{feng2018regularized} or reactive flows~\cite{Feng2018}. In the latter, the fluid is only expanded under the effects of strong temperature gradients, modeled by the perfect gas equation of state, and not under compressible effects.

The methodology~\cite{wissocq2019extended} extended to the hybrid method in the present work, makes it possible to access to the macroscopic composition of the modes. In that respect, the two "Usable" entropy classes have been investigated. This paper highlighted original modal transfers from the entropy mode to the shear mode for the primitive formulation, and from the entropy mode to the shear mode for the conservative formulation. These spectral analyses have been qualitatively corroborated by canonical testcases where, because of this mode transfer, large biases relative to the analytical solutions were observed. 
This mode transfer was found to be correlated to $\sigma$ in the primitive case and to $\overline{\theta}$ in the conservative case. As $\overline{\theta}=\overline{T}/T_{ref}$ is directly related to the local temperature (through $\overline{T}$) and to the time step (through $T_{ref}$), improving the conservative form proves more complicated than the primitive form, where $\sigma$ can be freely adjusted in space and time.

This work can be used as a basis for further research, to establish a quantitative link between the prediction of spectral analyses in terms of modal composition, and the experimentally observed phenomena. In future studies, other energy-conservative LBM schemes under a perfect gas equation of state, such as double distribution functions methods~\cite{li2012coupling,saadat2019lattice} or multispeed methods~\cite{shan2006kinetic, philippi2006continuous,Coreixas2017,latt2019efficient}, will have to be investigated to establish whether the mode transfer observed in Sec.~\ref{subsec: Conservative instability} is related to the use of a conservative formulation, or only to the use of the hybrid method.
Moreover,it is important to note that only a fraction of the HLBM classes have been studied in this work. Further research will have to be carried out on classes built using other collision operators such as TRT~\cite{ginzburg2010optimal}, MRT~\cite{DHumieres1992a,lallemand2000theory,MRT_dHumiere_2002} or the entropic collision operator~\cite{karlin1998equilibria,boghosian2001entropic,ansumali2003minimal,karlin2014gibbs,Frapolli2015a} to complete this study.


\appendix

\section{Supersonic limit of athermal and hybrid LBM}
\label{app: Supersonic limit}

In this appendix, the phenomena limiting athermal and hybrid models to the subsonic regime are shown.
Only the D1Q3 lattice will be considered, which is sufficient to exhibit the phenomena involved.
Research~\cite{renard2020improved} has shown that one of the ways to obtain a stable calculation for supersonic flows is based on two ingredients, identified as being:
\begin{itemize}
	\item The HRR collision operator Eq.~(\ref{eq: streamalgo}, \ref{eq: a1PR}, \ref{eq: a1xxy a1yyx}, \ref{eq: a1sigalgo}),
	\item An upwind FD scheme applied on $E_{1,\alpha\beta}$ Eq.~(\ref{eq: E1 E2 expressions}) in the correction term $\psi_i$.
\end{itemize}
Thus, only the HRR collision operator will be considered, gathering for the D1Q3 lattice both BGK ($\sigma=1$) and fully reconstructed operators ($\sigma=0$). Two types of discretizations of $E_{1,\alpha\beta}$ are investigated, the standard CO2 scheme Eq.~(\ref{eq: D1CO2 FD}) and the first order upwind scheme Eq.~(\ref{eq: D1UPO1 FD}).

\begin{figure}[h!]
     \centering
     \begin{subfigure}[b]{0.33\textwidth}
         \centering\hspace{0mm}%
         \includegraphics[scale=1.]{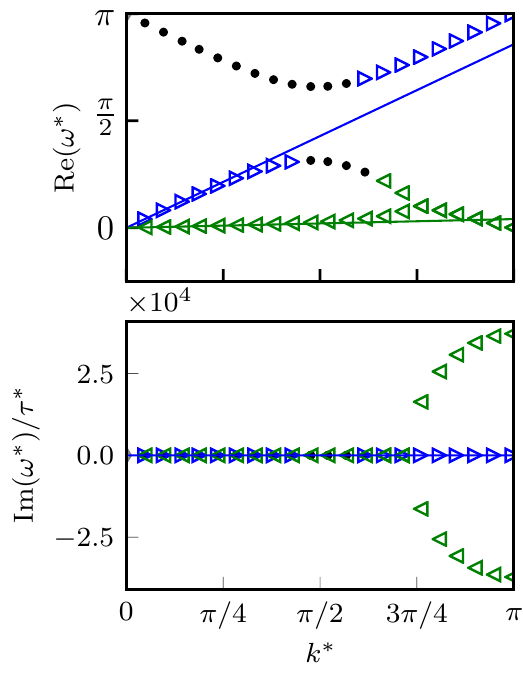}
         \caption{$\sigma = 1$, D1CO2($E_{1,\alpha\beta}$)}
         \label{fig: Athermal superso limit a}
     \end{subfigure}\hspace{0mm}%
     \begin{subfigure}[b]{0.33\textwidth}
         \centering
         \includegraphics[scale=1.]{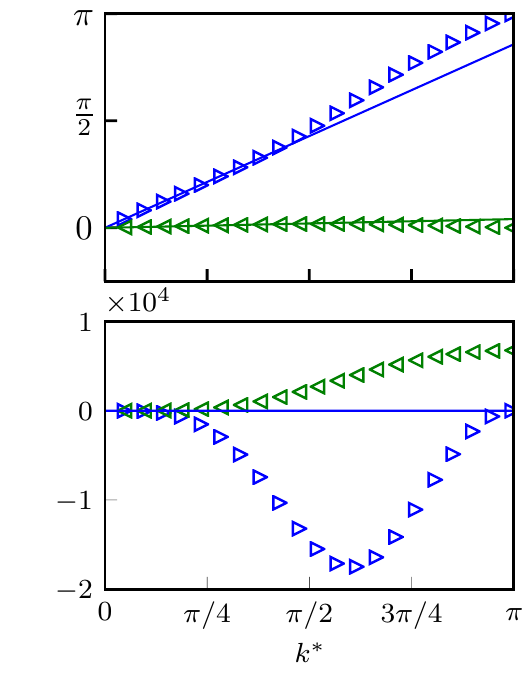}
         \caption{$\sigma = 0$, D1CO2($E_{1,\alpha\beta}$)}
         \label{fig: Athermal superso limit b}
     \end{subfigure}\hspace{0mm}%
     \begin{subfigure}[b]{0.33\textwidth}
         \centering
         \includegraphics[scale=1.]{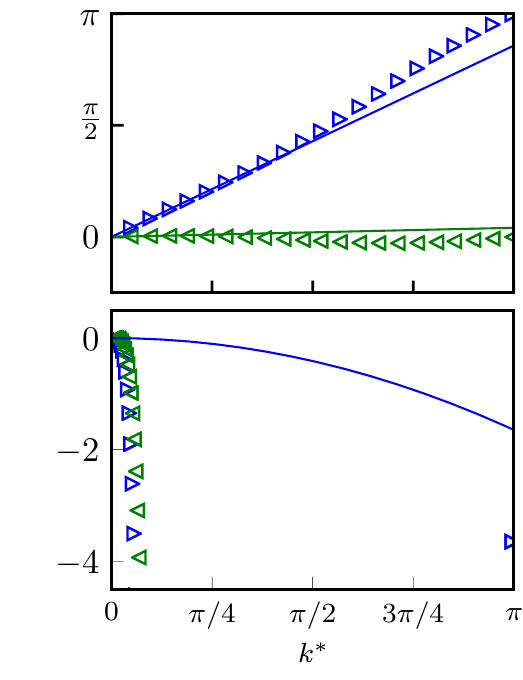}
         \caption{$\sigma = 0$, D1UPO1($E_{1,\alpha\beta}$)}
         \label{fig: Athermal superso limit c}
     \end{subfigure}
        \caption{Dispersion (top) and dissipation (bottom) graph of the LBM D1Q3 HRR model.  In the present case $\overline{\mathrm{Ma}}=1.1$,  $\theta = 0.5$ and $\tau^*=1.10^{-5}$. Symbols correspond to $\omega_{ac+}$: (\protect\acP) ;  $\omega_{ac-}$: (\protect\acM) ; $\omega_{msp}$:(\protect\macroSpu) and solid lines correspond to theory.}
     \label{fig: Athermal superso limit}
\end{figure}

Fig.~\ref{fig: Athermal superso limit} shows the dispersion and dissipation graphs of the athermal model for $\overline{\mathrm{Ma}}=1.1$, $\theta=0.5$ and $\tau^*=1.10^{-5}$. 
On Fig.~\ref{fig: Athermal superso limit a}, the results for the BGK collision operator using a centered discretization of $E_{1,\alpha\beta}$ can be found. According to the dissipation graph, the scheme is unstable. This instability is a direct consequence the eigenvalues collision located around $k^*=3\pi/4$. It is worth noting that this collision is of different kind than the one mentioned in Sec.~\ref{subsec: LSA of LBM scheme, general results} for $\overline{\mathrm{Ma}}= 0 .732$ since, in the present case, it involves the downstream and upstream acoustic modes. 
Indeed, for $\theta=0.5$, the critical Mach number Eq.~(\ref{eq: Mach c athermal}) is $\overline{\mathrm{Ma}^c} \simeq 1.45$ ensuring that the spurious macroscopic mode located at $\mathrm{Re}(\omega^*)=\pi$ (for $k^*=0$ and $k^*=\pi$) do not collide with the downstream acoustic.
Thus, reducing $\theta$ in the expectation of stabilizing the schemes would be pointless since, on the contrary, the sonic cone will be reduced and the two acoustic modes will interact all the more. This feature hints that these modes must be decoupled, and this brings to the second stabilization lever i.e. the $\sigma$ parameter of the HRR.
Fig.~\ref{fig: Athermal superso limit b} shows the results for the HRR at $\sigma=0$. It can be seen that the use of HRR has the desired effect, the acoustic eigenvalue collision is now suppressed. Nevertheless the scheme remains unstable, which is due to the centered discretization  of $E_{1,\alpha\beta}$.
Once this term is discretized using the fist order upwind scheme, the model is found to be stable as shown on Fig.~\ref{fig: Athermal superso limit c}. Test have been done using higher order scheme such as: second and third order upwind and fourth order centered scheme, and the conclusion remains the same: only the upwind class is stable for $\overline{\mathrm{Ma}}>1$. 
This behavior could be certainly explained by an heuristic analysis of the equivalent equation of the hybrid LBM system. It can be assumed that the sign of the coefficients in factor of the even derivatives of error terms changes sign for $\overline{\mathrm{Ma}}=1$ in the centered case. This analysis may be the subject of future work.

\begin{figure}
     \centering
     \begin{subfigure}[b]{0.33\textwidth}
         \centering\hspace{0mm}%
         \includegraphics[scale=1.]{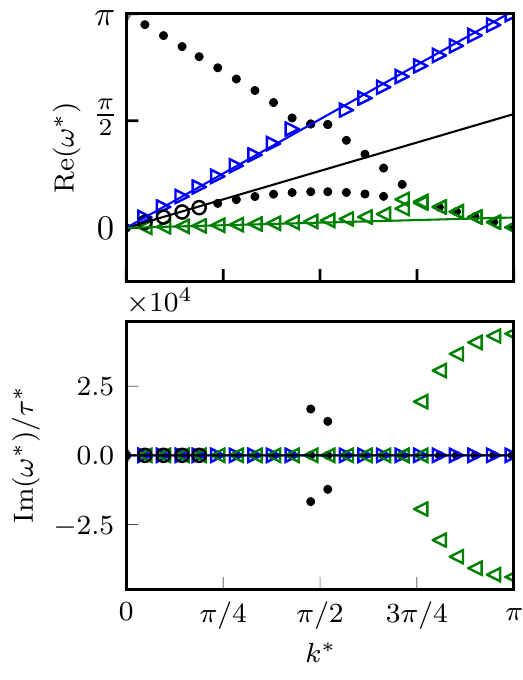}
         \caption{$\sigma = 1$, D1CO2($E_{1,\alpha\beta}$)}
         \label{fig: Hybrid superso limit a}
     \end{subfigure}\hspace{0mm}%
     \begin{subfigure}[b]{0.33\textwidth}
         \centering
         \includegraphics[scale=1.]{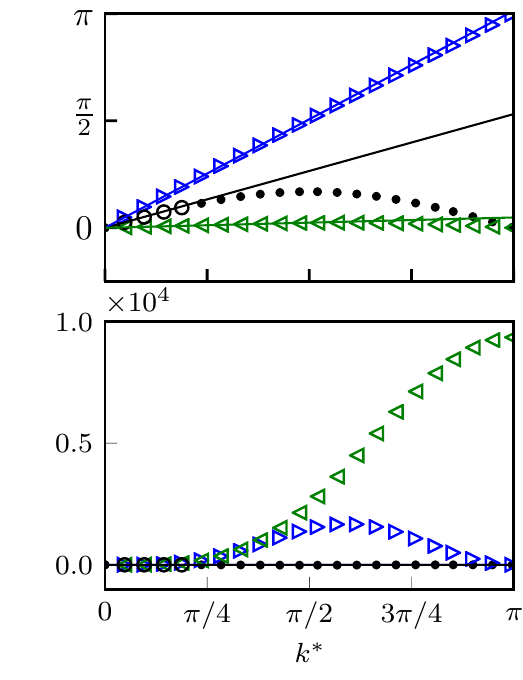}
         \caption{$\sigma = 0$, D1CO2($E_{1,\alpha\beta}$)}
         \label{fig: Hybrid superso limit b}
     \end{subfigure}\hspace{0mm}%
     \begin{subfigure}[b]{0.33\textwidth}
         \centering
         \includegraphics[scale=1.]{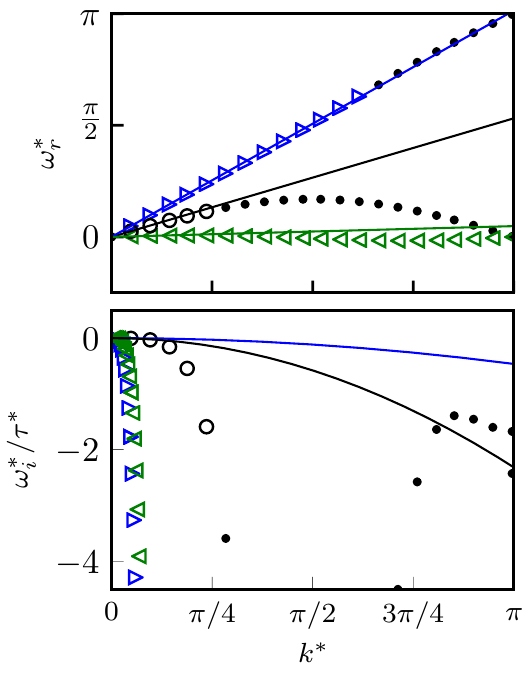}
         \caption{$\sigma = 0$, D1UPO1($E_{1,\alpha\beta}$)}
         \label{fig: Hybrid superso limit c}
     \end{subfigure}
        \caption{Dispersion (top) and dissipation (bottom) graph of the hybrid D1Q3 RK4C02 HRR based on the primitive entropy equation. In this case, $\overline{\theta}=0.5$, $\overline{\mathrm{Ma}}=1.1$ and $\tau^*=1.10^{-5}$. Symbols correspond to $\omega_{ac+}$: (\protect\acP) ;  $\omega_{ac-}$: (\protect\acM) ; $\omega_{entr}$: (\protect\entrop) ; $\omega_{msp}$:(\protect\macroSpu) ; $\omega_{g}$:(\protect\ghost) and solid lines correspond to theory.}
        \label{fig: Hybrid superso limit}
\end{figure}

On Fig.~\ref{fig: Hybrid superso limit}, the same study is conducted on the hybrid scheme based on primitive entropy equation and discretized by RK4CO2 scheme. Results are obtained for the same parameters as the athermal case.
If the BGK operator is employed, (see Fig.~\ref{fig: Hybrid superso limit a}) the same collision between the acoustic eigenvalues is observed. Moreover, one can see an other collision occurring for $k^*=\pi/2$ between the spurious macroscopic and the downstream acoustic mode. As a matter of fact, for $\overline{\theta}=0.5$ and $\gamma=1.4$ the critical Mach number of the hybrid model Eq.~(\ref{eq: Mach c hybrid}) is $\overline{\mathrm{Ma}^c} \simeq 1.07$ and therefore below the current value. Lowering $\overline{\theta}$ would fix this collision, but the other one involving the two acoustic will still be there. 
Similarly to the athermal case, Fig.~\ref{fig: Hybrid superso limit b} shows the results for $\sigma = 0$ where the two collision disappeared but the scheme remains unstable due the the centered discretization of $E_{1,\alpha\beta}$. And finally for an upwind discretization Fig.~\ref{fig: Hybrid superso limit c}, one obtain stable results for supersonic configuration.

In the end, same conclusions than the athermal case are drawn for the hybrid scheme. To obtain a stable model for $\overline{\mathrm{Ma}}>1$:
\begin{itemize}
	\item The physical modes somehow must be decoupled
	\item  An upwind discretization scheme has to be adopted for $E_{1,\alpha\beta}$.
\end{itemize}

\section{D1Q3 and D2Q9 lattices}
\label{app: D1Q3 and D2Q9 lattice}

The one dimensional and two-dimensional lattices~\cite{qian1992lattice} used in this paper are given in the present table:\\
\begin{center}
\begin{tabular}{|c|c|c|c|}
   \hline
    Lattice & velocities $\left(c_{i,\alpha} \right)$ & Velocity factor $\left(r\right)$ & weight $\left(w_i\right)$  \\
    \hline
    \hline
    $D1Q3$ & $0$ & $\sqrt{3r_gT_r}$ & $2/3$ \\
    \cline{2-2}\cline{4-4} 
        & $\pm r$ & \hspace{0.1cm} & $1/6$ \\
    \hline
    \hline
           & $\left(0,0\right)$ &  & $4/9$ \\
    \cline{2-2}\cline{4-4} 
    $D2Q9$   & $\left(\pm r,0\right)$ ; $\left(0,\pm r\right)$ & $\sqrt{3r_gT_r}$ & $1/9$  \\
    \cline{2-2}\cline{4-4} 
        & $\left(\pm r,\pm r\right)$ &  & $1/36$ \hspace{0.1cm} \\
    \hline    
\end{tabular}
\label{table: lattice}
\end{center}

\section{Implicit pressure work in the entropy equation}
\label{app: PW in s eq}

The pressure work term is implicitly contained in the advection part of the entropy equation and can be highlighted by chain rule . Knowing that:
\begin{equation}
\frac{\partial s}{\partial e} \bigg|_\rho = \frac{c_v}{e} \qquad \mathrm{and} \qquad   \frac{\partial s}{\partial \rho}\bigg|_e = - \left(\gamma-1\right) \frac{c_v}{\rho}.
\end{equation}
and using the l.h.s of the non-conservative expression Eq.~(\ref{eq: total phi eq nocons}), with no loss of generalities with respect to the conservative form, one obtain the following equality:
\begin{equation}
\begin{split}
\frac{\partial s}{\partial t} + u_\alpha\frac{\partial s}{\partial x_\alpha} &= \frac{\partial s}{\partial e}\bigg|_\rho \frac{\partial e}{\partial t} + \frac{\partial s}{\partial \rho}\bigg|_e \frac{\partial \rho}{\partial t} + u_\alpha \left( \frac{\partial s}{\partial e}\bigg|_\rho \frac{\partial e}{\partial x_\alpha} + \frac{\partial s}{\partial \rho}\bigg|_e \frac{\partial \rho}{\partial x_\alpha} \right)  \\
&= \frac{c_v}{e} \frac{\partial e}{\partial t} + \left(\gamma-1 \right) \frac{c_v}{\rho} \frac{\partial \rho u_\alpha}{\partial x_\alpha} + u_\alpha \left( \frac{c_v}{e} \frac{\partial e}{\partial x_\alpha} - \left(\gamma-1\right) \frac{c_v}{\rho} \frac{\partial \rho}{\partial x_\alpha} \right)  \\
&= \frac{1}{T}\left[ \frac{\partial e}{\partial t} + u_\alpha \frac{\partial e}{\partial x_\alpha} + r_g T \frac{\partial u_\alpha}{\partial x_\alpha}  \right].
\end{split}
\end{equation}
So by using entropy, it makes it possible to get rid of the discretization of the pressure work term.

\section{Athermal Matrix}
\label{app: Matrix athermal}

In this section, the expression of the time advance matrix of the corrected DVBE and LBM systems are derived in the spectral space. Emphasis will be placed on the spacial derivatives contributions in Fourier space. In the following, Jacobian matrix can be either obtained analytically~\cite{wissocq2020linear}, or using a formal calculation library such as $SymPy$ or $Theano$ as done in \cite{wissocq2019extended,Astoul2019}. The second option using the $Theano$ library was used in this work. 

\subsection{Corrected DVBE}
\label{app: Matrix athermal DVBE}

Starting from Eq.~(\ref{eq: DVBE eq}) and following the linearization procedure around an equilibrium state Eq.~(\ref{eq: f lin }), one get:

\begin{equation}
\frac{\partial f_i' }{\partial t}  = \underbrace{ \frac{\partial \mathcal{S}_i^{DVBE} }{\partial f_j} \bigg|_{\overline{f_j^{eq}}} }_{\mathrm{M}^{DVBE}_{ij}} f'_j + O \left( f_j'^2 \right),
\end{equation}
where $\mathrm{M}^{DVBE}_{ij}$ refers to as the Jacobian matrix of r.h.s. of Eq.~(\ref{eq: DVBE eq}). This matrix is the so-called time advance matrix of the DVBE system. At the moment, it is to be noted that $\mathrm{M}^{DVBE}_{ij}$ is essentially composed of constant terms and space derivative operators applied to $f_i'$.  Once Eq.~(\ref{eq: f prime}) substituted, the derivative operators can be analytically computed in the spectral space changing $\mathrm{M}^{DVBE}_{ij}$ into $\widetilde{\mathrm{M}}^{DVBE}_{ij}$. Keeping only the first-order distribution functions fluctuations, this system reads:
\begin{equation}
\omega \widehat{f}_i = \widetilde{\mathrm{M}}^{DVBE}_{ij} \widehat{f}_j,
\end{equation}
with 
\begin{equation}
\widetilde{\mathrm{M}}^{DVBE}_{ij} = c_{i,\alpha}  k_\alpha \delta_{ij} - \mathrm{i} \frac{1}{\overline{\tau}} \left( \delta_{ij} + \frac{\partial f_i^{eq}}{\partial f_j} \big|_{\overline{f_j^{eq}}} \right) -  \Psi_{ij},
\label{annexEq: M DVBE}
\end{equation}
and
\begin{equation}
\Psi_{ij}= w_i\frac{\mathcal{H}_{i,\alpha\alpha}}{2c_s^4}  k_\alpha \frac{ \partial \left(- a^{eq,(3)}_{\alpha\alpha\alpha} \right) } {\partial f_j} \bigg|_{\overline{f_j^{eq}}}.
\end{equation}
Here, $\Psi_{ij}$ corresponds to the Jacobian of the corrective body-force term expressed in the Fourier space, where the space derivative present in $E_{1,\alpha\beta}$ Eq.~(\ref{eq: E1 E2 expressions}) is now simply $k_\alpha$.

\subsection{Corrected LBM}

In this subsection, the time advance matrix of the LBM-BGK and LBM-HRR respectively Eq.~(\ref{eq: LBM BGK}) and Eq.~(\ref{eq: streamalgo}) will be derived.
Following the linearization procedure around a mean equilibrium state Eq.~(\ref{eq: f lin }), one get for both cases:
\begin{equation}
 g_i'^+  = \underbrace{ \frac{\partial \mathcal{S}_i^{LBM} }{\partial g_j} \bigg|_{\overline{f_j^{eq}}} }_{\mathrm{M}^{LBM}_{ij}} g'_j + O \left( g_j'^2 \right), 
\end{equation}
where $\mathrm{M}^{LBM}_{ij}$ refers to as the Jacobian matrix of the LBM r.h.s discrete system.  Once Eq.~(\ref{eq: monochro}) substituted, the discrete derivative operators can be analytically computed in the spectral space changing $\mathrm{M}^{LBM}_{ij}$ into $\widetilde{\mathrm{M}}^{LBM}_{ij}$. Keeping only the first-order distribution function fluctuations, this system reads:
\begin{equation}
e^{-\mathrm{i}\omega \Delta t } \widehat{g}_i = \widetilde{\mathrm{M}}^{LBM}_{ij} \widehat{g}_j.
\end{equation}
From here, the two collision operator  will be treated independently.\\ 

\noindent \textbf{Time-advance matrix of the corrected LBM-BGK}\\
Starting from the most classical case, the time advance matrix of the LBM-BGK system Eq.~(\ref{eq: LBM BGK}) reads:
\begin{equation}
\mathrm{BGK:} \quad \widetilde{\mathrm{M}}^{LBM}_{ij} = e^{-\mathrm{i} k_\alpha \Delta x } \left[ \delta_{ij}  -\frac{\Delta t}{\check{\tau}}\left( \delta_{ij} - \frac{\partial f_i^{eq}}{\partial g_j} \big|_{\overline{f_j^{eq}}} \right) + \frac{2\check{\tau} - \Delta t}{2\check{\tau}}\Delta t \, \Psi_{ij}  \right],
\label{eqApp: M LBM BGK}
\end{equation}
where $\Psi_{ij}$ refers to as the Jacobian of the corrective term including spacial derivatives expressed in the Fourier space and reads:
\begin{equation}
\Psi_{ij}= w_i\frac{\mathcal{H}_{i,\alpha\alpha}}{2c_s^4}  \mathcal{K}^\mathrm{D1CO2}_\alpha \frac{ \partial \left( -a^{eq,(3)}_{\alpha\alpha\alpha} \right) } {\partial g_j} \bigg|_{\overline{f_j^{eq}}}  .
\label{eqApp: Jd psi athermal}
\end{equation}
Here $\mathcal{K}^\mathrm{D1CO2}_\alpha$ corresponds to the modified wave number of the second order centered finite difference, applied in the present case to $E_{1,\alpha\beta}$ Eq.~(\ref{eq: E1 E2 expressions}). One can relate to~\ref{AnnexSec: mod wave number} for its expression. \\

\noindent \textbf{Time-advance matrix of the corrected LBM-HRR}\\
Regarding LBM-HRR system Eq.~(\ref{eq: streamalgo}), the matrix can be expressed as:
\begin{equation}
\mathrm{HRR:} \quad \widetilde{\mathrm{M}}^{LBM}_{ij} = e^{-\mathrm{i} k_\alpha \Delta x } \left[ \frac{\partial f_i^{eq}}{\partial g_j} \big|_{\overline{f_j^{eq}}} + \left(1 - \frac{\Delta t}{ \check{\tau}} \right)   G^{(1)}_{ij}   + \frac{\Delta t }{2} \Psi_{ij}  \right],
\label{eqApp: M LBM HRR}
\end{equation}
where $\Psi_{ij}$ can be found in Eq.~(\ref{eqApp: Jd psi athermal}), and $G^{(1)}_{ij}$ refers to as the Jacobian of the off-equilibrium moments. 
To take into account the finite difference contributions induced by the HRR collision and the corrective term, $G^{(1)}_{ij}$ is expressed such as: 
\begin{equation}
G^{(1)}_{ij} = \frac{\partial}{\partial f_j} \left( w_i  \frac{\mathcal{H}_{i,\alpha\beta}}{2c_s^4} \left[ \sigma A_{\alpha\beta}^{PR,\left(2\right)} + \left( 1 -\sigma \right)A_{\alpha\beta}^{FD,\left(2\right)} \right] + w_i  \frac{1}{6c_s^6} \left[ \mathcal{H}_{i,xxy} A_{xxy}^{\left(3\right)} + \mathcal{H}_{i,yyx} A_{yyx}^{\left(3\right)} \right]  \right)  \big|_{\overline{f_j^{eq}}}
\end{equation}
with 
\begin{subequations}
\begin{align}
A_{\alpha\beta}^{PR,\left(2\right)} &= \sum_i  \left[ \mathcal{H}_{i,\alpha\beta}\left( f_i - f_i^{eq} \right) \right] + \frac{\Delta t}{2} \mathcal{K}^\mathrm{D1CO2}_\alpha \left(-a^{eq,(3)}_{\alpha\alpha\alpha} \right),\\
A_{\alpha\beta}^{FD,\left(2\right)} &= -\check{\tau} \overline{p} \left( \mathcal{K}^\mathrm{D1CO2}_\beta u_{\alpha} + \mathcal{K}^\mathrm{D1CO2}_\alpha u_{\beta} \right),
\end{align}
\end{subequations}
respectively the second order off-equilibrium moment partially computed using projection procedure and  finite difference scheme modeling the viscous stress tensor. The third order off-equilibrium moments read:
\begin{subequations}
\begin{align}
A_{xxy}^{\left(3\right)} &= 2\overline{u}_x \left( \sigma A_{xy}^{PR,\left(2\right)} + \left(1-\sigma\right) \right) A_{xy}^{FD,\left(2\right)} + \overline{u}_y \left( \sigma A_{xx}^{PR,\left(2\right)} + \left(1-\sigma\right) \right) A_{xx}^{FD,\left(2\right)} ,\\
A_{yyx}^{\left(3\right)} &= 2\overline{u}_y \left( \sigma A_{xy}^{PR,\left(2\right)} + \left(1-\sigma\right) \right) A_{xy}^{FD,\left(2\right)} + \overline{u}_x \left( \sigma A_{yy}^{PR,\left(2\right)} + \left(1-\sigma\right) \right) A_{yy}^{FD,\left(2\right)} .
\end{align}
\label{appEq: third order}
\end{subequations}
where Eq.~(\ref{eq: eig vector LB ath}) impose the mean value for the velocities, since the later contains only first order fluctuation~\cite{Astoul2019}.

\section{Thermal Matrix}
\label{app: Matrix thermal}

In this section, the expression of the time advance matrix of the HDVBE and HLBM systems are given in the spectral space. These matrix are detailed on the form of blocks, presented by Fig.~\ref{Fig: multi bloc matrix}. One recall that, for $\widetilde{\mathrm{M}}^{HDVBE}_{ij}$ (and reciprocally for $\widetilde{\mathrm{M}}^{HLBM}_{ij}$), the blocks are defined such that:
\begin{itemize}
    \item Block I: $i,j\in\left[0,m-1\right]$
    \item Block II: $i\in\left[0,m-1\right]$, $j=m$
    \item Block III: $i=m$, $j\in\left[0,m-1\right]$ 
    \item Block IV: $i=m$, $j=m$ 
\end{itemize}
Emphasis will be placed on the shape of these block in Fourier space, specially in the discrete case where the finite difference contributions are detailed. For the sake of brevity, and without loss of generality, only the total energy equation on conservative form is considered. Similarly to~\ref{app: Matrix athermal}, the Jacobian are computed numerically using the $Theano$ library.

\subsection{HDVBE}
\label{app: HDVBE matrix}

Starting from the system.~(\ref{eq: system equation hybride}) and following the linearization procedure around the mean equilibrium state Eq.~(\ref{eq: f lin }) and around the mean total energy, one get:
\begin{equation}
\left\{
\begin{aligned}
\frac{\partial f_i'}{\partial t} = \overbrace{ \frac{\partial \mathcal{S}_i^{DVBE} }{\partial f_j} \bigg|_{\left( \overline{f_j^{eq}}, \overline{\rho E}\right)} }^{\mathrm{Block\,I}}  f'_j + \overbrace{\frac{\partial \mathcal{S}_i^{DVBE} }{\partial \left(\rho E\right)} \bigg|_{\left( \overline{f_j^{eq}}, \overline{\rho E}\right)}}^{\mathrm{Block\,II}}  \left( \rho E \right)' + O \left( f_j'^2, \left( \rho E \right)'^2 \right),\\
\frac{\partial \left(\rho E \right)'}{\partial t} = \underbrace{ \frac{\partial \mathcal{S}^{En} }{\partial f_j} \bigg|_{\left( \overline{f_j^{eq}}, \overline{\rho E}\right)}}_{\mathrm{Block\,III}}  f'_j + \underbrace{\frac{\partial \mathcal{S}^{En} }{\partial \left(\rho E\right)} \bigg|_{\left( \overline{f_j^{eq}}, \overline{\rho E}\right)}}_{\mathrm{Block\,IV}}  \left( \rho E \right)' + O \left( f_j'^2, \left( \rho E \right)'^2 \right),\\
\end{aligned}
\right. .
\label{eq: system equation hybride linearized}
\end{equation}
where the four blocks appear naturally depending on the nature of the fluctuation in factor of the partial derivatives. Once these blocks are gathered,  the time advance matrix $\mathrm{M}_{ij}^{HDVBE}$ Fig.~\ref{Fig: multi bloc matrix} is obtained. It is worthy to note that $\mathrm{M}^{HDVBE}_{ij}$ is essentially composed of constant terms and space derivative operators applied to $f_i'$ and $\left(\rho E \right)'$.  Once Eq.~(\ref{eq: f prime}) substituted, the derivative operators can be analytically computed in the spectral space changing $\mathrm{M}^{HDVBE}_{ij}$ into $\widetilde{\mathrm{M}}^{HDVBE}_{ij}$. This system reads:
\begin{equation}
\omega \widehat{\mathcal{X}}_i = \widetilde{\mathrm{M}}^{HDVBE}_{ij}  \widehat{\mathcal{X}}_i,
\end{equation}
where $\boldsymbol{\mathcal{\widehat{X}}} = \left( \widehat{f}_0, ...,\widehat{f}_{m-1}, \widehat{\rho E} \right)^T $. Along this section, each blocks of the time advance matrice in the Fourier space $\widetilde{\mathrm{M}}^{HDVBE}_{ij}$ will be detailed.   \\

\textbf{Block I}: $i,j\in\left[0,m-1\right]$\\
The upper left block is similar to the athermal matrix Eq.~(\ref{annexEq: M DVBE}). The only difference lies in the fact that the Jacobian of the equilibrium distribution function is computed holding $\rho E$ constant and is evaluated in $\left(f_j, \rho E \right)=\left( \overline{f_j^{eq}}, \overline{\rho E} \right)$. 
Moreover, for the thermal case the force term also needs to be  modified since it includes $E_{2,\alpha\beta}$ Eq.~(\ref{eq: E1 E2 expressions}). Now it reads:
\begin{equation}
\Psi_{ij}= w_i\frac{\mathcal{H}_{i,\alpha\alpha}}{2c_s^4}  \left(  k_\alpha \frac{ \partial \left(- a^{eq,(3)}_{\alpha\alpha\alpha} \right) } {\partial f_j} \bigg|_{\left(\overline{f_j^{eq}}, \overline{\rho E} \right)}  + \overline{p}\left(\frac{D+2}{D}-\gamma_g\right) k_\gamma \frac{ \partial u_\gamma } {\partial f_j} \bigg|_{\left(\overline{f_j^{eq}}, \overline{\rho E} \right)} \right).
\end{equation}

\textbf{Block II}: $i\in\left[0,m-1\right]$, $j=m$\\
The upper right block corresponds to a column vector and gather all the energy contribution contained in the LB system. It reads:
\begin{equation}
\widetilde{\mathrm{M}}^{HDVBE}_{ij} = \mathrm{i}\frac{1}{\overline{\tau}} \left( \frac{\partial f_i^{eq}}{\partial \rho E} \bigg|_{\left(\overline{f_j^{eq}}, \overline{\rho E} \right)} \right) -  \Psi_{ij}.
\end{equation}
with:
\begin{equation}
\Psi_{ij}= w_i\frac{\mathcal{H}_{i,\alpha\alpha}}{2c_s^4}  \left(  k_\alpha \frac{ \partial \left(- a^{eq,(3)}_{\alpha\alpha\alpha} \right) } {\partial \rho E} \bigg|_{\left(\overline{f_j^{eq}}, \overline{\rho E} \right)} \right),
\end{equation}
the thermal contribution of the corrective term carried through $E_{1,\alpha\beta}$.
It is to be noted that this block is responsible for the perfect gas coupling carried out through the equilibrium distribution function. \\

\textbf{Block III}: $i=m$, $j\in\left[0,m-1\right]$\\
The lower left block corresponds to a line vector and gather all the aero contribution contained in the energy system. It reads:
\begin{equation}
\widetilde{\mathrm{M}}^{HDVBE}_{ij} = k_\alpha \overline{\rho E} \frac{\partial u_\alpha }{\partial f_j} \bigg|_{\left(\overline{f_j^{eq}}, \overline{\rho E} \right)} + k_\alpha \frac{\partial u_\alpha p }{\partial f_j} \bigg|_{\left(\overline{f_j^{eq}}, \overline{\rho E} \right)} - \mathrm{i} k_\alpha^2 \lambda \frac{\partial T }{\partial f_j} \bigg|_{\left(\overline{f_j^{eq}}, \overline{\rho E} \right)} - \mathrm{i} \tau \overline{p} k_\alpha^2 \overline{u}_\beta \frac{\partial u_\beta }{\partial f_j} \bigg|_{\left(\overline{f_j^{eq}}, \overline{\rho E} \right)}.
\end{equation}
From left to right, one has the convective, the pressure work, the heat diffusion and viscous heat productions terms. The presence of these terms depends directly on the variable and the formulation of the energy equation used as shown on Fig.~\ref{Fig: multi bloc matrix}.
Finally, the total energy is the only one that linearly contains the viscous heat production term. For other energy variables this term vanish since it contains only second order contributions.\\

\textbf{Block IV}: $i=m$, $j=m$\\
Finally, the lower right block corresponds to a scalar involving only the energy system. It reads:
\begin{equation}
\widetilde{\mathrm{M}}^{HDVBE}_{ij} = k_\alpha \overline{u}_\alpha  + k_\alpha \frac{\partial u_\alpha p }{\partial \rho E} \bigg|_{\left(\overline{f_j^{eq}}, \overline{\rho E} \right)} - \mathrm{i} k_\alpha^2 \lambda \frac{\partial T }{\partial \rho E} \bigg|_{\left(\overline{f_j^{eq}}, \overline{\rho E} \right)} 
\end{equation}
and its composition depending on the formulation and energy variables can be found on Fig.~\ref{Fig: multi bloc matrix}.

\subsection{HLBM}

Following the same procedure as~\ref{app: HDVBE matrix}, one obtain a linear discrete system. Using Eq.~(\ref{eq: monochro}) to express the finite difference operators in the Fourier space, one obtains the HLBM eigenvalue problem which reads:
\begin{equation}
e^{-\mathrm{i}\omega \Delta t } \widehat{\mathcal{X}}_i = \widetilde{\mathrm{M}}^{HLBM}_{ij}  \widehat{\mathcal{X}}_i.
\end{equation}
Here, $\boldsymbol{\mathcal{\widehat{X}}} = \left( \widehat{g}_0, ...,\widehat{g}_{m-1}, \widehat{\rho E} \right)^T $ and $\widetilde{\mathrm{M}}^{HLBM}_{ij}$ refers to the time advance matrix of the system in the Fourier space. Similarly to it continuous counterpart, this matrix is composed of blocks sketched on Fig.~\ref{Fig: multi bloc matrix}, which will be detailed along this section. \\

\textbf{Block I}: $i,j\in\left[0,m-1\right]$\\
The upper left block is very similar to the athermal matrix for both BGK Eq.~(\ref{eqApp: M LBM BGK}) and HRR Eq.~(\ref{eqApp: M LBM HRR}) collision operators. Nevertheless, some modifications are needed for the thermal matrix. 
First, the Jacobians must be evaluated in $\left(f_j, \rho E \right)=\left( \overline{f_j^{eq}}, \overline{\rho E} \right)$ holding $\rho E$ constant.
Second, the force term Eq.~(\ref{eqApp: Jd psi athermal}) must be modified to includes the corrective term $E_{2,\alpha\beta}$ Eq.~(\ref{eq: E1 E2 expressions}), and reads:
\begin{equation}
\Psi_{ij}= w_i\frac{\mathcal{H}_{i,\alpha\alpha}}{2c_s^4}  \left(  \mathcal{K}^\mathrm{D1CO2}_\alpha \frac{ \partial \left(- a^{eq,(3)}_{\alpha\alpha\alpha} \right) } {\partial g_j} \bigg|_{\left(\overline{f_j^{eq}}, \overline{\rho E} \right)}  + \overline{p}\left(\frac{D+2}{D}-\gamma_g\right) \mathcal{K}^\mathrm{D1CO2}_\gamma \frac{ \partial u_\gamma } {\partial g_j} \bigg|_{\left(\overline{f_j^{eq}}, \overline{\rho E} \right)} \right).
\end{equation}
For the expression of the modified wavenumber $\mathcal{K}^\mathrm{D1CO2}$ one can relate to ~\ref{AnnexSec: mod wave number}.
And finally, for the HRR collision operator, the off-equilibrium moments must be corrected accordingly to the traceless viscous stress tensor. They now read:
\begin{subequations}
\begin{align}
A_{\alpha\beta}^{PR,\left(2\right)} &= \sum_i  \left[ \mathcal{H}_{i,\alpha\beta}\left( f_i - f_i^{eq} \right) \right] + \frac{\Delta t}{2} \mathcal{K}^\mathrm{D1CO2}_\alpha \left(-a^{eq,(3)}_{\alpha\alpha\alpha}\right) + \overline{p}\left(\frac{D+2}{D}-\gamma_g\right) \mathcal{K}^\mathrm{D1CO2}_\gamma  u_\gamma ,\\
A_{\alpha\beta}^{FD,\left(2\right)} &= -\check{\tau} \overline{p} \left( \mathcal{K}^\mathrm{D1CO2}_\beta u_{\alpha} + \mathcal{K}^\mathrm{D1CO2}_\alpha u_{\beta} - \frac{2}{D}\mathcal{K}^\mathrm{D1CO2}_\gamma u_{\gamma} \right),
\end{align}
\end{subequations}
where the higher order moment are computed using Eq.~(\ref{appEq: third order}).\\

\textbf{Block II}: $i\in\left[0,m-1\right]$, $j=m$\\
Similarly to its continuous counterpart, the upper right block is responsible for the perfect gas equation of state $via$ the equilibrium distribution function. It reads for the BGK collision:
\begin{equation}
\mathrm{BGK:} \quad \widetilde{\mathrm{M}}^{LBM}_{ij} = e^{-\mathrm{i} k_\alpha \Delta x } \left[   \frac{\Delta t}{\check{\tau}}  \frac{\partial f_i^{eq}}{\partial \rho E} \big|_{\left(\overline{f_j^{eq}}, \overline{\rho E} \right)} + \frac{2\check{\tau} - \Delta t}{2\check{\tau}}\Delta t \, \Psi_{ij}  \right],
\label{eqApp: M HLBM BGK}
\end{equation}
where 
\begin{equation}
\Psi_{ij}= w_i\frac{\mathcal{H}_{i,\alpha\alpha}}{2c_s^4}   \mathcal{K}^\mathrm{D1CO2}_\alpha \frac{ \partial \left(- a^{eq,(3)}_{\alpha\alpha\alpha} \right) } {\partial \rho E} \bigg|_{\left(\overline{f_j^{eq}}, \overline{\rho E} \right)},
\label{eq: Psi HLBM}
\end{equation}
refers to as the energy dependence of the corrective term.
Finally for the HRR collision, this block reads:
\begin{equation}
\mathrm{HRR:} \quad \widetilde{\mathrm{M}}^{LBM}_{ij} = e^{-\mathrm{i} k_\alpha \Delta x } \left[ \frac{\partial f_i^{eq}}{\partial\rho E} \big|_{\left(\overline{f_j^{eq}}, \overline{\rho E} \right)} + \left(1 - \frac{\Delta t}{ \check{\tau}} \right)   G^{(1)}_{ij}   + \frac{\Delta t }{2} \Psi_{ij}  \right],
\end{equation}
with 
\begin{equation}
G^{(1)}_{ij} = \frac{\partial}{\partial \rho E} \left( w_i  \frac{\mathcal{H}_{i,\alpha\beta}}{2c_s^4} \left[ \sigma A_{\alpha\beta}^{PR,\left(2\right)} + \left( 1 -\sigma \right)A_{\alpha\beta}^{FD,\left(2\right)} \right] + w_i  \frac{1}{6c_s^6} \left[ \mathcal{H}_{i,xxy} A_{xxy}^{\left(3\right)} + \mathcal{H}_{i,yyx} A_{yyx}^{\left(3\right)} \right]  \right)  \big|_{\left(\overline{f_j^{eq}}, \overline{\rho E} \right)},
\end{equation}
and $\Psi_{ij}$ corresponding to Eq.~(\ref{eq: Psi HLBM}). A particular attention must be paid on the second order off-equilibrium moments:  
\begin{subequations}
\begin{align}
A_{\alpha\beta}^{PR,\left(2\right)} &= \sum_i  \left[ \mathcal{H}_{i,\alpha\beta}\left( f_i - f_i^{eq} \right) \right] + \frac{\Delta t}{2} \mathcal{K}^\mathrm{D1CO2}_\alpha \left(-a^{eq,(3)}_{\alpha\alpha\alpha} \right),\\
A_{\alpha\beta}^{FD,\left(2\right)} &= 0,
\end{align}
\end{subequations}
where the thermal dependence is only present in the projection part. Here again, the higher order moment are computed using Eq.~(\ref{appEq: third order}).\\

\textbf{Block III}: $i=m$, $j\in\left[0,m-1\right]$\\
The lower left block of the discrete matrix is given by:
\begin{equation}
\widetilde{\mathrm{M}}^{HDVBE}_{ij} = - \mathcal{K}^\mathrm{D1CO2}_\alpha \overline{\rho E} \frac{\partial u_\alpha }{\partial g_j} \bigg|_{\left(\overline{f_j^{eq}}, \overline{\rho E} \right)} - \mathcal{K}^\mathrm{D1CO2}_\alpha \frac{\partial u_\alpha p }{\partial g_j} \bigg|_{\left(\overline{f_j^{eq}}, \overline{\rho E} \right)} + \mathcal{K}^\mathrm{D2CO2}_\alpha \lambda \frac{\partial T }{\partial g_j} \bigg|_{\left(\overline{f_j^{eq}}, \overline{\rho E} \right)} -  \tau \overline{p} \mathcal{K}^\mathrm{D2CO2}_\alpha \overline{u}_\beta \frac{\partial u_\beta }{\partial g_j} \bigg|_{\left(\overline{f_j^{eq}}, \overline{\rho E} \right)},
\end{equation}
and its composition can be found on Fig.~\ref{Fig: multi bloc matrix} according to the energy equation considered.
In conformity with the  RK1UPO1 and RK4CO2 schemes, the modified wave number in factor of the convective term corresponds either to $\mathcal{K}^{D1UPO1}_\alpha$ or $\mathcal{K}^{D1CO2}_\alpha$. Finally, as only the fluctuations related to the LBM system are considered in this block, the temporal integration of $\rho E$ does not affect this block.\\ 

\textbf{Block IV}: $i=m$, $j=m$\\
Finally, the lower right block corresponds to a scalar involving only the energy system. Depending on the temporal integration, two different expressions are obtained:
\begin{subequations}
\begin{align}
\mathrm{RK1UPO1:}& \quad \widetilde{\mathrm{M}}^{HLBM}_{ij} = 1 + z,\\
\mathrm{RK4CO2:}& \quad \widetilde{\mathrm{M}}^{HLBM}_{ij} = 1 + z + \frac{z^2}{2} + \frac{z^3}{3!} + \frac{z^4}{4!},
\end{align}
\label{appEq: RK spectral space}
\end{subequations}
where
\begin{equation}
z = \Delta t \left( - \mathcal{K}^\mathrm{D1CO2}_\alpha \overline{u}_\alpha  - \mathcal{K}^\mathrm{D1CO2}_\alpha \frac{\partial u_\alpha p }{\partial \rho E} \bigg|_{\left(\overline{f_j^{eq}}, \overline{\rho E} \right)} + \mathcal{K}^\mathrm{D2CO2}_\alpha \lambda \frac{\partial T }{\partial \rho E} \bigg|_{\left(\overline{f_j^{eq}}, \overline{\rho E} \right)} \right).
\end{equation}
For more details on how to obtain Eq.~(\ref{appEq: RK spectral space}), one can refers to~\cite{bao2003high}. 
Here again, the modified wave number in factor of the convective term must be chosen accordingly to the RK1UPO1 or RK4CO2 schemes.

\section{ Dispersion relation of the Navier Stokes systems}
\label{app: Eigen value NS}

Here the dispersion relation of the Navier Stokes system for athermal and thermal cases are derived. In this way, the macroscopic quantities are systematically developed in small perturbations on a base flow:
\begin{equation}
\rho = \overline{\rho} + \rho', \quad u_\alpha = \overline{u}_\alpha + u_\alpha' \quad \mathrm{and} \quad T = \overline{T} + T' 
\label{eqApp: Lin perturb}
\end{equation}
to obtain the linearized equations. Once this linearisation achieved, the perturbation are considered as solution of monochromatic wave:
\begin{equation}
\rho' = \widehat{\rho} \exp\left(\mathrm{i} \left( k_\alpha x_\alpha - \omega t \right)\right), \quad u_\alpha' =\widehat{u}_\alpha \exp\left(\mathrm{i} \left( k_\alpha x_\alpha - \omega t \right)\right)  \quad \mathrm{and} \quad T' = \widehat{T} \exp\left(\mathrm{i} \left( k_\alpha x_\alpha - \omega t \right)\right), 
\label{eqApp: Monochro perturb}
\end{equation}
where $\widehat{\rho}$, $\widehat{u_\alpha}$ and $\widehat{T}$ are the complexe amplitude of the perturbations. Computing the eigen value of the problem gives the dispersion relation of the system.

\subsection{Athermal equations}

Lets consider the athermal Navier-Stokes equations in the following form:
\begin{subequations}
\begin{align}
\frac{\partial \rho}{\partial t} &= - u_\alpha \frac{\partial \rho}{\partial x_\alpha} - \rho \frac{\partial u_\alpha}{\partial x_\alpha} \\
\frac{\partial u_\alpha}{\partial t} &= - u_\beta \frac{\partial u_\alpha }{\partial x_\beta}  - c_s^2 \theta \frac{\partial \rho }{\partial x_\beta} \delta_{\alpha\beta}  + \nu \frac{\partial }{\partial x_\beta} \left( \frac{\partial u_\alpha }{\partial x_\beta} + \frac{\partial u_\beta }{\partial x_\alpha} \right),
\end{align}
\end{subequations}
where $\sqrt{c_s^2 \theta} = \sqrt{r_g T}$ the newtonian sound speed (with $T$ considered as a constant of the fluid) and $\nu=\mu/\rho$ being the kinematic viscosity. Once Eq.~(\ref{eqApp: Lin perturb}) is replaced, the linearized system is obtained: 
\begin{subequations}
\begin{align}
\frac{\partial \rho'}{\partial t} &= - \overline{u}_\alpha \frac{\partial \rho'}{\partial x_\alpha} - \overline{\rho} \frac{\partial u_\alpha'}{\partial x_\alpha} \\
\frac{\partial u_\alpha'}{\partial t} &= - \overline{u}_\beta \frac{\partial u_\alpha' }{\partial x_\beta}  - c_s^2 \theta \frac{\partial \rho' }{\partial x_\beta} \delta_{\alpha\beta}  + \overline{\nu} \frac{\partial }{ \partial x_\beta} \left( \frac{\partial u_\alpha' }{\partial x_\beta} + \frac{\partial u_\beta' }{\partial x_\alpha} \right),
\end{align}
\end{subequations}
with $\overline{\nu}=\mu/\overline{\rho}$. Using Eq.~(\ref{eqApp: Monochro perturb}), the dispersion relation of the system is obtained and reads:
\begin{equation} 
	\begin{aligned}
		&\omega \widehat{\rho}  = k_\alpha  \overline{u}_\alpha \widehat{\rho} + k_\alpha  \overline{\rho} \widehat{u}_\alpha, \\
		&\omega \widehat{u}_\alpha  = k_\beta\overline{u}_\beta \widehat{u}_\alpha + k_\beta \frac{c_s^2\theta}{\overline{\rho}} \delta_{\alpha\beta}\widehat{\rho} -   \mathrm{i} \overline{\nu} k_\beta \left( k_\beta \widehat{u}_\alpha + k_\alpha \widehat{u}_\beta \right).
    \end{aligned}
\label{eqApp: disp rel athermal}
\end{equation}
Considering a two-dimensional problem, one can define a vector ${V}^{NS}_i = \left[ \widehat{\rho}, \widehat{u}_x, \widehat{u}_y \right]^T$ gathering the complex amplitude of the perturbations and allows to write Eq.~(\ref{eqApp: disp rel athermal}) on the form of an eigen-value problem:
\begin{equation}
	\omega V^{NS}_i = \widetilde{M}_{ij}^{NS,ath} V^{NS}_i,
\end{equation}
with the matrix:
\begin{equation}
\widetilde{M}_{ij}^{NS,ath} = 
\begin{pmatrix}
k_\alpha \overline{u}_\alpha & k_x \overline{\rho} & k_y \overline{\rho} \\
k_x \dfrac{c_s^2\theta}{\overline{\rho}} & k_\alpha \overline{u}_\alpha - \mathrm{i}\overline{\nu}\left(k_\alpha^2 + k_x^2\right) & - \mathrm{i}\overline{\nu}k_x k_y \\
k_y \dfrac{c_s^2\theta}{\overline{\rho}} & - \mathrm{i}\overline{\nu}k_y k_x  & k_\alpha \overline{u}_\alpha - \mathrm{i}\overline{\nu}\left(k_\alpha^2 + k_y^2\right) \\
\end{pmatrix}
\end{equation}
being the space and time evolution matrix of the athermal Navier Stokes equations in the Fourier space. One diagonalized, the system gives three eigen-modes: two acoustic ($\omega^{ath}_{ac-}$ upstream and $\omega^{ath}_{ac+}$ downstream ) and one transverse or shear mode $\omega^{ath}_{shear}$. These modes can be either computed numerically using a linear algebra library such as $numpy$ or analytically by expanding the solution in terms of wavenumber $k$, which gives:
\begin{equation}
	\begin{aligned}
		&\omega _{shear} = k_\alpha  \overline{u}_\alpha -\mathrm{i} \nu \left\Vert k_\alpha \right\Vert^2+ O\left(k^3\right), \\
		&\omega^{ath}_{ac\pm} = k_\alpha  \overline{u}_\alpha \pm \left\Vert k_\alpha \right\Vert \sqrt{c_s^2 \theta} -\mathrm{i}\nu \left\Vert k_\alpha \right\Vert^2+ O\left(k^3\right), \\
	\end{aligned}
\label{eqApp: eigenvaluesNS athermal}
\end{equation}
In this work, the eigen-vectors used for the extended analysis of~\cite{wissocq2019extended} are essentially computed numerically.

\subsection{Compressible equations}

Lets now consider the compressible Navier-Stokes Fourier equations in the following form:
\begin{subequations}
\begin{align}
\frac{\partial \rho}{\partial t} &= - u_\alpha \frac{\partial \rho}{\partial x_\alpha} - \rho \frac{\partial u_\alpha}{\partial x_\alpha} \\
\frac{\partial u_\alpha}{\partial t} &= - u_\beta \frac{\partial u_\alpha }{\partial x_\beta}  - c_s^2  \frac{\partial \rho \theta }{\partial x_\beta} \delta_{\alpha\beta}  - \nu \frac{\partial }{\partial x_\beta} \left( \frac{\partial u_\alpha }{\partial x_\beta} + \frac{\partial u_\beta }{\partial x_\alpha} - \frac{2}{D}\frac{\partial u_\gamma }{\partial x_\gamma} \delta_{\alpha\beta} \right) \\
\frac{\partial T}{\partial t} &= - u_\alpha \frac{ \partial T }{\partial x_\alpha}  - \left( \gamma_g - 1\right) T  \frac{\partial u_\alpha }{\partial x_\alpha} + \gamma_g \kappa \frac{\partial^2T}{\partial x_\alpha^2} +   \frac{\nu}{c_v} \frac{\partial u_\alpha }{\partial x_\beta} \left( \frac{\partial u_\alpha }{\partial x_\beta} + \frac{\partial u_\beta }{\partial x_\alpha} - \frac{2}{D}\frac{\partial u_\gamma }{\partial x_\gamma} \delta_{\alpha\beta} \right)
\end{align}
\end{subequations}
where $c_s^2 =r_g T_r$, $\theta = T/T_r$, $\nu=\mu/\rho$, $\gamma_g=c_p/c_v$ and $\kappa = \lambda / (\rho c_p)$.
It is worth noting that the choice of the energy equation do not have any impact on the solution. Here the work is conducted using the internal energy equation in primitive form and assuming that $c_v$ and $r_g$ are constant. Once Eq.~(\ref{eqApp: Lin perturb}) is replaced, the linearized system is obtained: 
\begin{subequations}
\begin{align}
\frac{\partial \rho'}{\partial t} &= - \overline{u}_\alpha \frac{\partial \rho'}{\partial x_\alpha} - \overline{\rho} \frac{\partial u_\alpha'}{\partial x_\alpha} \\
\frac{\partial u_\alpha'}{\partial t} &= - \overline{u}_\beta \frac{\partial u_\alpha' }{\partial x_\beta}  - r_g \frac{\partial T' }{\partial x_\beta} \delta_{\alpha\beta} - \frac{r_g \overline{T}}{ \overline{\rho}} \frac{\partial T' }{\partial x_\beta} \delta_{\alpha\beta}  + \overline{\nu} \frac{\partial }{ \partial x_\beta} \left( \frac{\partial u_\alpha' }{\partial x_\beta} + \frac{\partial u_\beta' }{\partial x_\alpha} - \frac{2}{D} \frac{\partial u_\gamma' }{\partial x_\gamma} \delta_{\alpha\beta} \right) \\
\frac{\partial T'}{\partial t} &= - \overline{u}_\alpha \frac{ \partial T' }{\partial x_\alpha}  - \left( \gamma_g - 1\right) \overline{T}  \frac{\partial u_\alpha' }{\partial x_\alpha} + \gamma_g \overline{\kappa} \frac{\partial^2T'}{\partial x_\alpha^2},
\end{align}
\end{subequations}
with $\overline{\kappa} = \lambda/\overline{\rho}$ the mean heat conductivity. For the sake of clarity, $c_s$ and $\theta$ have been replaced by their expressions. It is important to notice that the equation above do not contain the viscous heat production term in the energy equation. This non-linear term being essentially composed of velocity derivative products vanish once the linearization is conducted. Classically, using Eq.~(\ref{eqApp: Monochro perturb}), the dispersion relation of the compressible system is obtained and reads:
\begin{equation} 
	\begin{aligned}
		&\omega \widehat{\rho}  = \overline{u}_\alpha k_\alpha \widehat{\rho} + \overline{\rho}  k_\alpha \widehat{u}_\alpha, \\
		&\omega \widehat{u}_\alpha  = \overline{u}_\beta k_\beta \widehat{u}_\alpha + r_g k_\beta \delta_{\alpha\beta} \widehat{T} + \frac{r_g \overline{T}}{\overline{\rho}} k_\beta  \delta_{\alpha\beta}\widehat{\rho} -   \mathrm{i} \overline{\nu} k_\beta \left( k_\beta \widehat{u}_\alpha + k_\alpha \widehat{u}_\beta - \frac{2}{D} k_\gamma \delta_{\alpha\beta} \widehat{u}_{\gamma} \right),\\
        &\omega \widehat{T}  =  \overline{u}_\alpha k_\alpha \widehat{T}  + \left( \gamma_g -1 \right) \overline{T}k_\alpha \widehat{u}_\alpha - \mathrm{i}\gamma_g \overline{\kappa} k_\alpha^2 \widehat{T}. 
    \end{aligned}
\label{eqApp: disp compressible}
\end{equation}
Considering a two-dimensional problem, one can define a vector ${V}^{NS}_i = \left[ \widehat{\rho}, \widehat{u}_x, \widehat{u}_y, \widehat{T} \right]^T$ gathering the complex amplitude of the perturbations and allows to write Eq.~(\ref{eqApp: disp compressible}) on the form of an eigen-value problem:
\begin{equation}
	\omega V^{NS}_i = \widetilde{M}_{ij}^{NS} V^{NS}_i,
\end{equation}
with the matrix:
\begin{equation}
\widetilde{M}_{ij}^{NS} = 
\begin{pmatrix}
\overline{u}_\alpha k_\alpha &  \overline{\rho} k_x &  \overline{\rho} k_y & 0 \\
 \dfrac{r_g \overline{T}}{\overline{\rho}} k_x & \overline{u}_\alpha k_\alpha - \mathrm{i}\overline{\nu}\left(k_\alpha^2 + k_x^2 - \dfrac{2}{D}k_x^2\right) & - \mathrm{i}\overline{\nu} \left( k_x k_y - \dfrac{2}{D}k_x k_y \right) & r_g k_x \\
\dfrac{r_g \overline{T}}{\overline{\rho}} k_y & - \mathrm{i}\overline{\nu} \left( k_y k_x - \dfrac{2}{D}k_yk_x \right)  &  \overline{u}_\alpha k_\alpha - \mathrm{i}\overline{\nu}\left(k_\alpha^2 + k_y^2 - \dfrac{2}{D}k_y^2 \right) & r_g k_y \\
0 & \left( \gamma_g - 1\right) \overline{T} k_x  &  \left( \gamma_g - 1\right) \overline{T} k_y & \overline{u}_\alpha k_\alpha - \mathrm{i}\gamma_g \overline{\kappa} k_\alpha^2
\end{pmatrix}
\end{equation}
From the zeros present in the matrix above, in the absence of a perfect gaz coupling ($\widetilde{M}_{13}^{NS} =\varnothing$ and $\widetilde{M}_{23}^{NS} =\varnothing$), the thermal feedback of the energy equation would be impossible. Similarly to the athermal case, these modes can be either computed numerically using a linear algebra library such as $numpy$ or analytically by expanding the solution in terms of wavenumber $k$, which gives:
\begin{equation} \label{eq:eigenvaluesNS}
	\begin{aligned}
		&\omega _{shear} = k_\alpha  \overline{u}_\alpha - i\nu \left\Vert k_\alpha \right\Vert^2+ O\left(k^3\right), \\
		&\omega _{entr} = k_\alpha  \overline{u}_\alpha - \mathrm{i} \overline{\kappa} \left\Vert k_\alpha \right\Vert^2+ O\left(k^3\right), \\
		&\omega _{ac\pm} = k_\alpha  \overline{u}_\alpha \pm \left\Vert k_\alpha \right\Vert \sqrt{\gamma r_g \overline{T}} - \mathrm{i} \left( \frac{D-1}{D}\nu + \frac{\gamma_g -1}{2} \kappa  \right) \left\Vert k_\alpha \right\Vert^2+ O\left(k^3\right) , \\
	\end{aligned}
\end{equation}
Finally in the specific case of the entropic energy variable, the modulus of eigenvectors are found to be very low and the library $mpmath$~\cite{mpmath} allowing an  arbitrary-precision floating-point is preferred as $numpy$.

\section{Modified wave number of finite difference schemes}
\label{AnnexSec: mod wave number}
In this section,the spatial finite difference operators in the Fourier space denoted $\mathcal{K}^{scheme}_\alpha$ are presented. Lets consider a second order centered finite difference scheme of a quantity $\phi$:
\begin{equation}
 \mathrm{D1CO2}: \quad  \frac{\partial \phi}{\partial x_\alpha} =  \frac{\phi_{\boldsymbol{x}+\boldsymbol{e}_\alpha\Delta x} - \phi_{\boldsymbol{x}-\boldsymbol{e}_\alpha\Delta x} }{2 \Delta x} + \mathcal{O}\left( \Delta x^2 \right).
\label{eq: appendix CO2 lin}
\end{equation}
Once this expression is linearized in the manner of Eq~(\ref{eq: f lin }), and once the small perturbation is expressed using Eq~(\ref{eq: f prime}), one get :
\begin{equation}
\mathrm{i} k_\alpha \widehat{\phi} e^{\mathrm{i} \left( k_\alpha x_\alpha -  \omega t \right) } =  \frac{e^{\mathrm{i}k_\alpha\Delta x } - e^{-\mathrm{i}k_\alpha\Delta x } }{2\Delta x} \widehat{\phi} e^{\mathrm{i} \left( k_\alpha x_\alpha -  \omega t \right) } + \mathcal{O}\left( \Delta x^2 \right).
\label{eq:}
\end{equation}
On the l.h.s, the exact wave number modeling the space derivative in the Fourier space is $\mathrm{i} k_\alpha$. By identification, the r.h.s of the equation gives the so-called modified wave number of the scheme:
\begin{equation}
    \mathcal{K}_\alpha^{D1CO2} = \frac{e^{\mathrm{i}k_\alpha\Delta x } - e^{-\mathrm{i}k_\alpha\Delta x } }{2\Delta x} = \mathrm{i} \frac{\sin\left(k_\alpha \Delta x\right)}{\Delta x}.
\end{equation}
This modified wave number characterizes the dispersion and dissipation properties of the scheme. The same manipulation can be done on the other spacial schemes considered in this paper, resulting in:

\begin{subequations}
\begin{align}
&\mathcal{K}^\mathrm{D1UPO1}_\alpha =  \frac{2\sin^2 \left(k\Delta x/2 \right) + \mathrm{i}\sin \left(k\Delta x \right) }{\Delta x}, \\
&\mathcal{K}^\mathrm{D2CO2}_\alpha = 2 \frac{ \cos \left(k\Delta x \right) -1 }{\Delta x^2}.
\label{eq:}
\end{align}
\end{subequations}


\end{document}